\documentclass[a4paper,11pt]{article}
\pdfoutput=1 

\usepackage{jcappub} 
\usepackage{changepage}

\usepackage[T1]{fontenc} 
\usepackage{mathrsfs}
\usepackage{amsfonts,mathtools}
\usepackage{braket}
\usepackage{cleveref}
\usepackage[toc]{appendix}
\usepackage{comment}
\usepackage{cellspace} 
\usepackage{orcidlink}

\newcommand{\Ci}{\text{Ci}}
\newcommand{\DM}{\text{DM}}
\newcommand{\eMD}{\text{eMD}}
\newcommand{\G}{\text{G}}
\newcommand{\GW}{\text{GW}}
\newcommand{\HH}{\text{H}}
\newcommand{\hc}{\text{hc}}
\newcommand{\LS}{\text{LS}}
\newcommand{\NL}{\text{NL}}
\newcommand{\PBH}{\text{PBH}}
\newcommand{\R}{\text{R}}
\newcommand{\RD}{\text{RD}}
\newcommand{\res}{\text{res}}

\newcommand{\secrev}[1]{\textcolor{black}{#1}}
\newcommand{\MD}{\text{MD}}
\newcommand{\Si}{\text{Si}}
\newcommand{\scsc}{\text{ss}}
\newcommand{\st}{\text{st}}
\newcommand{\thh}{\text{th}}

\title{\boldmath Probing scalar-tensor-induced gravitational waves in the nHz band: NANOGrav and SKA}

\author[a]{William Iania\,\orcidlink{0009-0004-1966-9211
}}
\author[b,c]{and Angelo Ricciardone\,\orcidlink{0000-0002-5688-455X
}}
\affiliation[a]{PPGCosmo, Universidade Federal do Esp\'irito Santo, 29075-910, Vit\'oria, ES, Brazil}
\affiliation[b]{Dipartimento di Fisica ``Enrico Fermi'',  Università di Pisa, Largo Bruno Pontecorvo 3, Pisa
I-56127, Italy}
\affiliation[c]{INFN, Sezione di Pisa, Largo Bruno Pontecorvo 3, Pisa I-56127, Italy}

\emailAdd{william.iania@edu.ufes.br}
\emailAdd{angelo.ricciardone@unipi.it}

\abstract{Scalar-induced gravitational waves (SIGWs) have recently attracted considerable interest, both as a possible explanation for the nanohertz signal reported by the Pulsar Timing Array (PTA) collaboration and for their connection with primordial black hole (PBH) physics. In addition to SIGWs, scalar-tensor-induced gravitational waves (STGWs) have emerged as a promising cosmological source of the stochastic gravitational wave background (SGWB). In this paper, we compute the STGWs generated during a generic matter-dominated (MD) era, as well as during an early matter-dominated (eMD) epoch followed by a sudden transition to the standard radiation-dominated (RD) stage, working in the Poisson gauge. We find that, in a purely MD age, the corresponding energy density rapidly dilutes, whereas in the presence of an eMD phase it remains non-vanishing due to the short duration of the eMD period. We then investigate whether the STGW signal could provide a dominant contribution to the \texttt{NANOGrav 15-year} dataset and we forecast the prospects for its detection with future observations by the Square Kilometre Array (SKA). In particular, we consider STGWs generated during both eMD and RD eras, including their linear-order contributions. We further inject the scalar-induced contribution, both as an additional component and as the only signal, and compare the spectra against the same datasets. Our results show that, although the SIGWs still appear to provide the dominant contribution and the posteriors are tightly constrained by the requirement of avoiding PBH overproduction, the GWs induced by scalar-tensor mixing constitute a viable target for future, more sensitive nano-Hz detectors of the SGWB.}

\keywords{cosmological perturbation theory in GR and beyond, physics of the early universe, primordial black holes, primordial gravitational waves (theory)}

\arxivnumber{2604.13012}

\begin{document}
\maketitle
\flushbottom

\section{Introduction}
\label{sec:introduction}

Nanohertz gravitational waves (GWs) have drawn increasing attention, driven by the evidence of a SGWB in the PTA data~\cite{NANOGrav:2023gor,EPTA:2023sfo,EPTA:2023fyk,EPTA:2023xxk,Xu:2023wog,Reardon:2023gzh,Zic:2023gta,Agazie:2026tui}. While supermassive black hole binaries (SMBHBs) are a well-motivated astrophysical candidate~\cite{NANOGrav:2023hfp,NANOGrav:2023pdq,NANOGrav:2025gqp}, a broad range of cosmological sources has also been proposed, which may provide alternative or complementary explanations for the observed PTA signal~\cite{NANOGrav:2023hvm,EPTA:2023xxk}.

Several cosmological scenarios can give rise to a SGWB~\cite{Caprini:2018mtu}. These include vacuum fluctuations generated during inflation~\cite{Grishchuk:1974ny,Starobinsky:1979ty,Rubakov:1982df,Fabbri:1983us}, large primordial scalar perturbations~\cite{Tomita:1967wkp,Matarrese:1994wa,Matarrese:1993zf,Matarrese:1997ay}, first-order phase transitions in the early universe~\cite{Kamionkowski:1993fg,Caprini:2007xq,Hindmarsh:2017gnf,Cutting:2018tjt,Cutting:2019zws,RoperPol:2019wvy,Caprini:2019egz,Cutting:2020nla,Han:2023olf,Ashoorioon:2022raz,Athron:2023mer,Li:2023yaj}, and cosmic defects such as cosmic strings~\cite{Vachaspati:1984gt,Sakellariadou:1990ne,Damour:2000wa,Damour:2001bk,Damour:2004kw,Figueroa:2012kw,Hiramatsu:2013qaa,Blanco-Pillado:2017oxo,Auclair:2019wcv,Gouttenoire:2019kij,Figueroa:2020lvo,Chang:2021afa,Yamada:2022aax,Yamada:2022imq}. The SGWB also gives the opportunity to constrain alternative expansion histories, like periods of kination domination~\cite{Giovannini:1998bp,Giovannini:1999bh,Boyle:2007zx,Li:2016mmc,Li:2021htg,Figueroa:2018twl,Figueroa:2019paj,Gouttenoire:2021wzu,Co:2021lkc,Gouttenoire:2021jhk,Oikonomou:2023qfz}, and high-energy physics beyond the standard model (BSM), for example through particle production during inflation~\cite{Anber:2006xt,Sorbo:2011rz,Pajer:2013fsa,Adshead:2013qp,Adshead:2013nka,Maleknejad:2016qjz,Dimastrogiovanni:2016fuu,Namba:2015gja,Ferreira:2015omg,Peloso:2016gqs,Domcke:2016bkh,Caldwell:2017chz,Guzzetti:2016mkm,Bartolo:2016ami,DAmico:2021fhz,DAmico:2021vka}, which remain otherwise inaccessible to current and foreseeable laboratory experiments. For a recent review, see~\cite{Sesana:2025udx}.

Among the possible cosmological sources proposed to explain the PTA detection, GWs induced at second order in perturbation theory by large scalar fluctuations, known as SIGWs, have emerged as one of the most compelling candidates~\cite{Ananda:2006af,Baumann:2007zm,Domenech:2021ztg}. In this scenario, an enhancement of the scalar power spectrum on small scales generates a GW signal that provides a good fit to the observed GWB among the many proposed cosmological interpretations~\cite{Figueroa:2023zhu,Das:2023nmm}. Furthermore, SIGWs are the gravitational counterpart to a PBH population~\cite{Saito:2008jc,saito2010gravitational}: if the LIGO-Virgo-KAGRA (LVK) collaboration, over its four observing runs~\cite{LIGOScientific:2018mvr,LIGOScientific:2020ibl,KAGRA:2021vkt,LIGOScientific:2025slb}, has observed BHs of PBH origin, the corresponding induced GWB is expected to peak around the PTA frequency band~\cite{Ando:2022tpj,Garcia-Bellido:2017fdg,LISACosmologyWorkingGroup:2023njw}.

It has been shown that the induced GWB can also serve as a powerful probe of the early universe thermal history~\cite{Domenech:2020kqm,Balaji:2023ehk,Liu:2023pau,Liu:2023hpw,Zhu:2023gmx}. In particular, SIGWs produced during a phase of early matter domination prior to the standard RD era have been extensively studied, since such an epoch arises naturally in a variety of physical scenarios. Examples include compact objects such as PBHs~\cite{Hawking:1971ei,Carr:1974nx} and non-topological solitons (i.e. Q-balls)~\cite{Coleman:1985ki}, as well as inflationary or post-inflationary dynamics involving overdamped scalar fields~\cite{Turner:1983he}, oscillons~\cite{Bogolyubsky:1976pw,Lozanov:2022yoy,Sui:2024grm} or rotating axion fields~\cite{Co:2019jts,Co:2019wyp,Co:2022qpr}, all of which can generate an early pressureless component. Remarkably, Ref.~\cite{Ghoshal:2026ros} showed how the nHz GW signal maps onto the decay lengths of sufficiently long-lived particles (LLPs), which could generate an early matter-dominated era, and such lengths will be directly accessible in upcoming LLP experiments. SIGWs in an eMD epoch were first studied in Ref.~\cite{Assadullahi:2009nf}; then, Ref.~\cite{Kohri:2018awv} investigated scenarios with multiple transitions between matter- and radiation-dominated eras. Their potential as a distinctive source of gravitational radiation was clarified in Refs.~\cite{Inomata:2019ivs,Inomata:2019zqy,Inomata:2020lmk}, that analyzed both a gradual and a sudden eMD-RD transition. As the scalar perturbations in matter domination grow proportionally to the scale factor, an eMD phase can lead to a strong SIGW observable energy density, especially when keeping into account the generation right after the transition to radiation domination in the sudden reheating scenario: it is the so-called ``poltergeist mechanism''; see~\cite{Inomata:2025wiv} for a recent pedagogical review. The precision of these predictions has been improved using fully numerical calculations in Ref.~\cite{Pearce:2023kxp}, whereas Ref.~\cite{Pearce:2025ywc} demonstrated that the resulting signal can exhibit characteristic features that help distinguish between different PBH and Q-ball scenarios.

Linear tensor fluctuations coupled to scalar perturbations can similarly induce gravitational radiation, leaving an imprint on the energy density of the GWB~\cite{Gong:2019mui,Chang:2022vlv,Chen:2022dah,Bari:2023rcw,Picard:2023sbz,Yu:2023lmo}. In particular, the STGWs can reach amplitudes comparable to those of the purely scalar-induced component, and may even dominate on small scales. This contribution has only recently been explored in detail~\cite{Bari:2023rcw,Picard:2023sbz,Picard:2024ekd,Picard:2025bwq}; it has also been investigated as a potentially dominant source of the PTA signal~\cite{Wu:2024qdb}, and as a probe of small-scale primordial curvature perturbations as well as primordial gravitational waves (PGWs)~\cite{Wu:2025gwt}. However, its behavior in scenarios with a non-standard thermal history of the universe has not been thoroughly investigated. 

Hence, in this work we compute the STGWs arising from a generic MD epoch working in the Poisson gauge. We extend the results of Ref.~\cite{Gong:2019mui}, where the second-order solutions in the Newtonian gauge during matter domination were first presented, but the corresponding energy density was not derived (see also~\cite{Garoffolo:2022usx} for a discussion in the wave-optics limit). Then, we apply the formalism of the SIGWs generated during early matter domination to the scalar-tensor mixing in the presence of a sudden eMD-RD transition, also working out the linear-order counterpart. Thereby, we aim to isolate the contribution of these components to the observed nHz signal, as measured in the \texttt{NANOGrav 15-year} dataset~\cite{NANOGrav:2023gor}, and to forecast the sensitivity of the future SKA survey~\cite{Janssen:2014dka,Lazio:2013mea}. We do so for STGWs generated during both early matter domination and radiation domination, performing a Bayesian inference on the parameters of the primordial linear fluctuations that source the radiation. We adopt peaked power spectra, modeled as lognormal scalar and tensor seeds, since flat input spectra lead to unphysical enhancements in the integral of the scalar-tensor energy density, as already noted in the literature~\cite{Bari:2023rcw,Picard:2023sbz}. Parameter estimation is carried out under the assumption that the scalar-tensor signal is the dominant component of the SGWB, both with and without an accompanying PGW contribution. We then include the purely scalar-induced contribution in the same analyses with NG15 and in the forecasts with SKA10, for both the eMD and RD eras.

The article is organized as follows. In~\cref{subsec:ev_eqs} we provide an overview of the formalism of the STGWs in the Poisson gauge. We then review the signal produced during a RD epoch in~\cref{subsec:stgw_rd} and compute the energy density generated in a generic MD era in~\cref{subsec:stgw_md}. Next, in~\cref{subsec:stgw_emd_to_rd} we explore the existence of such a signal during an eMD epoch with a sudden eMD-RD transition, studying the kernel analytically and analyzing the associated energy density in case of primordial peaked spectra.~\Cref{sec:stgw_nhz} focuses on the current and future prospects for detecting STGWs in the nHz frequency band. \Cref{subsec:pbh} is dedicated to the constraints from PBH overproduction, which we use to further restrict the parameter space. In~\cref{subsec:ng15_inference} we perform a Bayesian analysis of the \texttt{NANOGrav 15-year} dataset~\cite{NANOGrav:2023hvm} using the software package \href{https://andrea-mitridate.github.io/PTArcade/}{\texttt{PTArcade}}~\cite{Mitridate:2023oar}. In~\cref{subsec:ska10_emd,subsec:ska10_rd} we then forecast parameter uncertainties for future SKA dataset using \href{https://github.com/Mauropieroni/fastPTA}{\texttt{fastPTA}}~\cite{Babak:2024yhu,Cecchini:2025oks}, allowing for STGWs generated during both eMD and RD epochs, respectively.~\Cref{subsec:sigw} is devoted to the same MCMC samplings of the previous sections, including the scalar-induced contribution. We conclude in~\cref{sec:conclusions}. Further discussions are provided in appendices~\ref{sec:matching},~\ref{sec:varying_shape_pars} and~\ref{sec:sigw}.

\section{Formalism for scalar-tensor-induced gravitational waves}\label{sec:stgw}

We work in natural units and define the reduced Planck mass as $M^2_\text{Pl}=(8\pi G)^{-1}$, where $G$ is Newton's constant. The metric of a perturbed flat FLRW space-time in the Poisson gauge is given by~\cite{Bari:2023rcw}
\begin{equation}\label{eq:poisson_gauge}
    ds^2 = -e^{2\Phi}dt^2+a^2(t) e^{-2\Psi} (e^\gamma)_{ij} dx^i dx^j\,,
\end{equation}
where $t$ is the coordinate time, $a(t)$ the scale factor, $\Phi$ and $\Psi$ are scalar perturbations, and $\gamma_{ij}$ tensor perturbations, which are transverse and traceless (TT): $\partial ^i \gamma_{ij}=\delta^{ij}\gamma_{ij}=0$. Gauge dependence is a subtle issue for STGWs: different gauge choices lead to different observable results~\cite{Bari:2023rcw,Picard:2023sbz}. It has recently been shown that the observable GW strain coincides with the tensor modes in Newtonian gauge in the case of SIGWs, although a residual gauge dependence remains in the GW pseudo-stress-energy tensor, and therefore the same conclusion does not apply to the energy density~\cite{Domenech:2025ccu}.

We consider modes produced and stretched during inflation. Vector perturbations are neglected, as they are diluted away during the inflationary stage. The tensor-induced contribution, arising from the coupling of tensor perturbations, is also neglected under the assumption of primordial peaked sources~\cite{Picard:2024ekd}. For more general source spectra, such as power laws, the tensor–tensor coupling can in principle generate a non-negligible contribution~\cite{Chang:2022vlv,Picard:2024ekd,Wu:2025gwt,Chen:2025fcd}. However, our goal is to isolate and quantify the impact of the scalar-tensor component of the signal. Including the tensor-induced component would introduce additional model dependence without affecting the specific effect under investigation. For this reason, we do not take the tensor–tensor term into account in what follows.

The matter content of the universe is modeled as an adiabatic perfect fluid with energy-momentum tensor
\begin{equation}
        T_{\mu\nu} = (\rho + P) u_\mu u_\nu + P g_{\mu\nu}\,,
\end{equation}
where $\rho$ and $P$ are the energy density and pressure, respectively, and $u_\mu$ the 4-velocity. Assuming GR, the anisotropic stress gives a negligible contribution to the signal~\cite{Baumann:2007zm}, so we impose isotropy by setting $\Phi=\Psi$. Deviations from this relation may arise when induced GWs are studied within modified-gravity frameworks~\cite{Kugarajh:2025rbt}. We further assume that the primordial perturbations are adiabatic and Gaussian (see e.g., Refs.~\cite{Perna:2024ehx,LISACosmologyWorkingGroup:2025vdz,Liu:2023ymk} for deviations from this assumption). In addition, we do not consider any primordial parity violation in the tensor sector, as done in Ref.~\cite{Bari:2023rcw}, where scalar-tensor induced GWs were first analyzed in the Poisson gauge.

\subsection{From the evolution equations to the energy density}\label{subsec:ev_eqs}

Ref.~\cite{Bari:2023rcw} shows the derivation of the scalar-tensor signal generated in radiation domination. Here we report only the most important steps, leaving the reader to the reference for a complete derivation; we will further extend to MD epoch, and a sudden eMD-RD transition. Using the ADM formalism they compute the full non-linear Einstein equations (EEs) and, from the traceless part of the ij-th component, the equations of motion for the scalar-tensor induced tensor modes read
\begin{equation}\label{eq:ev_eq}
    \gamma_{ij}^{\prime\prime}+2\mathcal{H}\gamma_{ij}^{\prime}-\nabla^2\gamma_{ij}=4\Phi\nabla^2\gamma_{ij}+4\Phi^\prime\gamma^\prime_{ij}\,,
\end{equation}
where isotropy has already been imposed, and a prime denotes derivatives wrt conformal time, ${}^\prime\equiv d/d\eta$, with $d\eta \equiv dt/a$.~\Cref{eq:ev_eq} is independent of the background equation of state (EoS): it is derived without specifying the dominant energy component and therefore does not depend explicitly on the EoS parameter $w\equiv P/\rho$ or on the sound speed $c_s^2\equiv\delta P/\delta\rho$. Consequently, the formalism applies equally well to radiation domination, matter domination, and even to more exotic phases, such as a stiff EoS with $w=1$ arising, for instance, in kination-dominated scenarios. The dependence on the specific EoS will manifest itself in the transfer functions.

The tensor modes can be decomposed into the homogeneous solution and the scalar-tensor induced contribution,
\begin{equation}
    \gamma_{ij}=\gamma_{ij}^{(0)}+\gamma_{ij}^{(1)}+\dots\,,
\end{equation}
where the first addend is obtained by setting the right-hand side (RHS) of~\cref{eq:ev_eq} to zero, while the second term is sourced by the coupling of linear scalar and tensor fluctuations. The perturbations can be expressed in Fourier space as
\begin{align}
    \Phi(\mathbf{x},\eta)&=\frac{1}{(2\pi)^3}\int d ^3\mathbf{k}\, \Phi_\mathbf{k}(\eta)\, e^{i\mathbf{k}\cdot\mathbf{x}}\,,\\
    \gamma_{ij}(\mathbf{x},\eta)&=\frac{1}{(2\pi)^3}\int d^3 \mathbf{k} \,\gamma_{\mathbf{k},\lambda}(\eta)\epsilon_{ij}^\lambda(\mathbf{\hat{k}})\,e^{i\mathbf{k}\cdot\mathbf{x}}\,,
\end{align}
where the TT polarization tensors $\epsilon_{ij}^\lambda(\mathbf{\hat{k}})$ satisfy the normalization condition
\begin{equation}
    \epsilon_{ij}^{\lambda\ast}(\mathbf{\hat{k}})\epsilon^{\lambda^\prime ij}(\mathbf{\hat{k}})=2\delta^{\lambda\lambda^\prime}\,.
\end{equation}
The evolution equation in Fourier space reads
\begin{equation}\label{eq:evol_fourier_gener}
    \gamma_{\mathbf{k},\lambda}^{(1)\prime\prime}+2\mathcal{H}\gamma^{(1)\prime}_{\mathbf{k},\lambda}+k^2\gamma^{(1)}_{\mathbf{k},\lambda}= \mathcal{S}_{\text{st},\lambda}(\mathbf{k},\eta)\,,
\end{equation}
with the source term given by 
\begin{equation}\label{eq:source_gen}
    \begin{split}
        \mathcal{S}_{\text{st},\lambda}(\mathbf{k},\eta)&=-2 \int \frac{d^3 \textbf{p}}{(2\pi)^3}\Phi^\text{pr}_{\mathbf{k}-\mathbf{p}}\gamma_{\mathbf{p},\sigma}^\text{pr}\epsilon_{ij}^\sigma(\mathbf{\hat{p}})\epsilon_\lambda^{ij\ast}(\mathbf{\hat{k}})\Big[p^2T_\gamma(p\eta)T_\Phi(c_s|\mathbf{k}-\mathbf{p}|\eta)\\
        &
        -T_\gamma^\prime(p\eta)T_\Phi^\prime(c_s|\mathbf{k}-\mathbf{p}|\eta)\Big]\,.
    \end{split}
\end{equation}
We have decomposed the first-order perturbations into a primordial amplitude multiplied by a transfer function, which accounts for their evolution across different epochs~\cite{Ananda:2006af,Watanabe:2006qe}:
\begin{equation}
    \begin{split}
        \Phi_\mathbf{k}(\eta)&=\Phi_\mathbf{k}^\text{pr}T_\Phi(c_sk\eta)\,,\\
        \gamma_{\mathbf{k},\sigma}^{(0)}(\eta)&=\gamma_{\mathbf{k},\sigma}^\text{pr}T_\gamma(k\eta)\,.
    \end{split}
\end{equation}
For a background with conformal Hubble parameter $\mathcal{H}(\eta)=2/((1+3w)\eta)$ and a constant EoS parameter such that $w=c_s^2$, the linear perturbations satisfy~\cite{Domenech:2019quo}
\begin{equation}\label{eq:linear_eqs}
    \begin{split}
        \Phi_\mathbf{k}^{\prime\prime}+6\frac{1+w}{1+3w}\frac{1}{\eta}\Phi_\mathbf{k}^\prime+wk^2\Phi_\mathbf{k}&=0\,,\\
        \gamma_{\mathbf{k},\sigma}^{(0)\prime\prime}+\frac{4}{1+3w}\frac{1}{\eta}\gamma_{\mathbf{k},\sigma}^{(0)\prime}+k^2\gamma^{(0)}_{\mathbf{k},\sigma}&=0\,.
    \end{split}
\end{equation}
From these equations we obtain the transfer functions. In radiation domination, with $c_s^2=1/3$, they read
\begin{subequations}\label{eq:t_func_rd}
    \begin{align}
        \begin{split}
            T^\RD_\Phi(c_sk\eta)&=\frac{3\sqrt{3}}{k\eta}j_1(k\eta)=
            \frac{9}{(k\eta)^2}\left[\frac{\sqrt{3}}{k\eta}\sin{\left(\frac{k\eta}{\sqrt{3}}\right)}-\cos{\left(\frac{k\eta}{\sqrt{3}}\right)}\right]\,,
        \end{split}\\
        \begin{split}
            T^\RD_\gamma(k\eta)&=j_0(k\eta)=\frac{\sin{(k\eta)}}{k\eta}\,.
        \end{split}
    \end{align}
\end{subequations}
In matter domination, where $c_s^2=0$, the transfer functions become
\begin{subequations}\label{eq:t_func_md}
    \begin{align}
        \begin{split}
            T^\MD_\Phi&=1\,,
        \end{split}\\
        \begin{split}
            T^\MD_\gamma(k\eta)&=3\frac{j_1(k\eta)}{k\eta}=\frac{3}{(k\eta)^2}\left[\frac{\sin{(k\eta)}}{k\eta}-\cos{(k\eta)}\right]\,.
        \end{split}
    \end{align}
\end{subequations}
We make use of the spherical Bessel function of the first kind, $j_\nu(x)$, and of the second kind, $y_\nu(x)$; the latter will appear below. For $\nu=0,1$ they read
\begin{equation}
    \begin{split}
        j_0(x)&=\frac{\sin{x}}{x}\,,\quad j_1(x)=\frac{\sin{x}}{x^2}-\frac{\cos{x}}{x}\,,\\
        y_0(x)&=-\frac{\cos{x}}{x}\,,\quad y_1(x)=-\frac{\cos{x}}{x^2}-\frac{\sin{x}}{x}\,.
    \end{split}
\end{equation}
For linear perturbations that cross the horizon during matter domination, the gravitational potential $\Phi$ remains constant on all the scales~\cite{Dodelson:2020bqr}. Consequently, in~\cref{eq:source_gen} the first term retains an explicit dependence only on the tensor transfer function $T_\gamma$, while the second term, proportional to $T_\Phi^\prime$, vanishes. The source function therefore reduces to
\begin{equation}\label{eq:source_md}
    \begin{split}
        \mathcal{S}_{\text{st},\lambda}^\MD(\mathbf{k},\eta)&=-2 \int \frac{d^3 p}{(2\pi)^3}\Phi^\text{pr}_{\mathbf{k}-\mathbf{p}}\gamma_{\mathbf{p},\sigma}^\text{pr}\epsilon_{ij}^\sigma(\mathbf{\hat{p}})\epsilon_\lambda^{ij\ast}(\mathbf{\hat{k}})\left[p^2T_\gamma(p\eta)\right]\,.
    \end{split}
\end{equation}
The general first-order solution can be written as
\begin{equation}\label{eq:sol_gen}
    \gamma_{\mathbf{k},\lambda}^{(1)}(\eta)=\int_0^\eta d \tilde{\eta} \frac{a(\tilde{\eta})}{a(\eta)}\mathcal{S}_{\text{st},\lambda}(\mathbf{k},\tilde{\eta})G_\mathbf{k}(\eta,\tilde{\eta})\,,
\end{equation}
where the Green function obeys to
\begin{equation}\label{eq:gen_green}
    G_\mathbf{k}^{\prime\prime}(\eta,\tilde{\eta})+\left(k^2-\frac{a^{\prime\prime}}{a}\right)G_\mathbf{k}(\eta,\tilde{\eta})=\delta(\eta-\tilde{\eta})\,.
\end{equation}
We are interested in the two-point correlation function of the tensor modes,
\begin{equation}\label{eq:2pf_all}
    \braket{\gamma_{\mathbf{k},\lambda}\gamma_{\mathbf{k^\prime},\lambda^\prime}}=\braket{\gamma^{(0)}_{\mathbf{k},\lambda}\gamma^{(0)}_{\mathbf{k^\prime},\lambda^\prime}}+\braket{\gamma^{(1)}_{\mathbf{k},\lambda}\gamma^{(1)}_{\mathbf{k^\prime},\lambda^\prime}}+\braket{\gamma^{(0)}_{\mathbf{k},\lambda}\gamma^{(2)}_{\mathbf{k^\prime},\lambda^\prime}}+\dots\,.
\end{equation}
Unlike Refs.~\cite{Chen:2022dah,Picard:2025bwq}, we neglect the last term in~\cref{eq:2pf_all}, even though it can in principle affect the total scalar-tensor signal as shown in Ref.~\cite{Picard:2025bwq} in the Newtonian gauge. The dimensionless, isotropic, stationary, and unpolarised power spectrum of the scalar-tensor induced GWB is expressed as the Fourier transform of the two-point correlation function,
\begin{equation}\label{eq:power_spectrum}
    \braket{ \gamma_{\mathbf{k}, \lambda}^{(1)}(\eta) \gamma _{\mathbf{k}^\prime,\lambda^\prime}^{(1)}(\eta)} = (2\pi)^3\delta^{(3)}(\mathbf{k}+\mathbf{k}^\prime)\delta_{\lambda\lambda^\prime}\frac{2\pi^2}{k^3}\Delta^{2,\text{st}}_{\gamma_1,\lambda}(k)\,.
\end{equation}
The two-point function on the LHS of~\cref{eq:power_spectrum} is proportional to the four-point function of the primordial fluctuations. Here, $k$ denotes the comoving wavenumber and $f$ the corresponding frequency, related by
\begin{equation}
    f \simeq 1.6\times10^{-9}\,\rm{Hz} \left(\frac{k}{10^6\,\rm{Mpc}^{\color{black}-1}}\right)\,,
\end{equation}
with $a_0=1$. Given the primordial power spectra
\begin{subequations}
    \begin{align}
        \begin{split}\label{eq:2pts_primordial_phi}
                \braket{\Phi^\text{pr}_\mathbf{k}\Phi^\text{pr}_{\mathbf{k}^\prime}} & = (2\pi)^3\delta^3(\mathbf{k}+\mathbf{k}^\prime) \frac{2\pi^2}{k^3} \Delta^2_\Phi(k),
        \end{split}\\
        \begin{split}\label{eq:2pts_primordial_gamma}
        \braket{\gamma^\text{pr}_{\mathbf{k},\lambda}\gamma^\text{pr}_{\mathbf{k}^\prime,\lambda^\prime}} & = (2\pi)^3\delta^3(\mathbf{k}+\mathbf{k}^\prime) \delta_{\lambda\lambda^\prime}\frac{2\pi^2}{k^3} \Delta^2_{\gamma_0,\lambda}(k)\,,
        \end{split}
    \end{align}
\end{subequations}
after some algebra we obtain, similarly to eq.~(3.5) of~\cite{Bari:2023rcw},
\begin{equation}\label{eq:ps_tbc}
    \begin{split}
        \Delta^2_{\gamma_1,\lambda}&=\frac{k^3}{\pi}\int{d}^3\textbf{p}\frac{\Delta^2_\Phi(|\mathbf{k}-\mathbf{p}|)\Delta^2_{\gamma_0,\sigma}(\mathbf{p})}{p^3|\mathbf{k}-\mathbf{p}|^3}\epsilon^{ij,\sigma}(\mathbf{\hat{p}})\epsilon_{ij}^{\lambda\ast}(\mathbf{\hat{k}})\epsilon^{mn,\sigma}(-\mathbf{\hat{p}})\epsilon_{mn}^{\lambda\ast}(-\mathbf{k})
        \times I^2\,,
    \end{split}
\end{equation}
where the squared kernel $I^2$ encodes the dependence on the background EoS at horizon crossing, as well as the convolution structure of the oscillating primordial modes. Introducing the dimensionless variables
\begin{equation}\label{eq:uv_vars}
v=\frac{p}{k}\,,\quad u=\frac{|\mathbf{k}-\mathbf{p}|}{k}\,,
\end{equation}
the averaged scalar-tensor power spectrum is given by
\begin{equation}\label{eq:gen_ps}
    \begin{split}    
        \overline{\Delta^2_{\gamma_1}(k)} &\equiv\sum_\lambda\overline{\Delta^2_{\gamma_1,\lambda}(k)}\\
        &=  \frac{1}{16}\int_0^\infty d v\int_{|1-v|}^{1+v} \frac{du}{v^6u^2}\left(\left[(v+1)^2-u^2\right]^4+\left[(v-1)^2-u^2\right]^4\right)\\
        &\times\Delta^2_\Phi(uk)\Delta^2_{\gamma_0}(vk)\cdot\overline{I^2}\,,
    \end{split}
\end{equation}
where we have summed over the two equivalent tensor polarizations, and the overline denotes an average over many oscillation periods. This averaging is appropriate because our goal is to relate the spectrum to the GW energy density per logarithmic frequency (or wavenumber) interval,
\begin{equation}\label{eq:Omega}
    \Omega^\st_{\GW}(k)\equiv\frac{\rho_\GW(\eta,k)}{\rho_c}=\frac{1}{12}\left(\frac{k}{\mathcal{H}(\eta)}\right)^2\overline{\Delta_{\gamma_1}^2(k)}\,.
\end{equation}
Using~\cref{eq:gen_ps}, the scalar-tensor induced GW energy density can be written as
\begin{equation}\label{eq:general_Omega}
    \begin{split}
        \Omega_\GW^\st(k)&=\frac{1}{6}\int_0^\infty d v \int_{|1-v|}^{1+v} \frac{du}{v^2u^2}\left[\frac{(1+v^2-u^2)^2}{v^2}+\left(1+\left(\frac{1+v^2-u^2}{2v}\right)^2\right)^2\right]\\
        &\times\overline{\mathcal{I}^2}\cdot \Delta^2_\Phi(uk)\Delta^2_{\gamma_0}(vk)\,,
    \end{split}
\end{equation}
where
\begin{equation}
    \overline{\mathcal{I}^2}\equiv x^2\cdot\overline{I^2}\,,
\end{equation}
and we defined the dimensionless time variable $x\equiv k\eta$.

\subsection{STGWs from a radiation-dominated universe}\label{subsec:stgw_rd}

In the RD epoch, and in the absence of primordial parity violation, the scalar-tensor energy density reads~\cite{Bari:2023rcw}
\begin{equation}
    \begin{split}
        \Omega^\st_\GW(k)\big|_\RD&=\frac{1}{6}\int_0^\infty d v\int_{|1-v|}^{1+v}d u\frac{\Upsilon^\RD(u)}{u^2v^2}\left[\frac{\left(1+v^2-u^2\right)^2}{v^2}+\left(1+\left(\frac{1+v^2-u^2}{2v}\right)^2\right)^2\right]\\
        &\times\overline{\mathcal{I}^2_\RD(u,v)} \Delta^2_\Phi(uk)\Delta^2_{\gamma_0}(vk)\,,
    \end{split}
\end{equation}
where
\begin{align}
    \Upsilon^\RD(u)&=\frac{u^4}{1+u^4}\,,\\\label{eq:I_RD}
    \overline{\mathcal{I}^2_\RD(u,v)}&=\frac{27}{2^7}\left(\frac{v}{u}\right)^2\left[\pi^2\left(1-z^2\right)^2\Theta\left(1-z^2\right)+\left(2z+\left(1-z^2\right)\log{\left|\frac{1+z}{1-z}\right|}\right)^2\right]\,,\\
    z(u,v)&=\frac{\sqrt{3}}{2}\frac{v^2+u^2/3-1}{uv}\,.
\end{align}
The damping function $\Upsilon^\RD(u)$ regularizes an UV divergence of the integrand, which for large scalar wavelengths behaves as $\sim u^{-4}$. We will return to this divergence issue below and address it in an analogous way for the eMD epoch.

\subsection{STGWs from a matter-dominated universe}\label{subsec:stgw_md}

In this section, we derive the kernel in a MD epoch using the Poisson gauge. GWs generated by the scalar-tensor mixing that re-enter during MD period inherit the transfer functions in~\cref{eq:t_func_md} and the source term in~\cref{eq:source_md}. Inserting these into the evolution equation in~\cref{eq:evol_fourier_gener}, the solution for each polarization $\lambda$ reads
\begin{equation}
    \gamma_{\mathbf{k},\lambda}^{(1)}(\eta)=\int_0^\eta d \tilde{\eta} \frac{a(\tilde{\eta})}{a(\eta)}\mathcal{S}_{\text{st},\lambda}^\MD(\mathbf{k},\tilde{\eta})G_\mathbf{k}^\MD(\eta,\tilde{\eta})\,,
\end{equation}
where the Green function,~\cref{eq:gen_green}, for the tensor modes in matter domination satisfies (cf. e.g., Refs.~\cite{Kohri:2018awv,Gong:2019mui})
\begin{equation}
    \begin{split}
        G_\mathbf{k}^\MD(\eta,\tilde{\eta})&=\Theta(\eta-\tilde{\eta})\frac{x\tilde{x}}{k}\left[j_1(\tilde{x})y_1(x)-j_1(x)y_1(\tilde{x})\right]\\
        &= \Theta(\eta-\tilde{\eta})\frac{1}{kx\tilde{x}}\left[(1+x\tilde{x})\sin{(x-\tilde{x})}-(x-\tilde{x})\cos{(x-\tilde{x})}\right]\,.
    \end{split}
\end{equation}
Here, $\Theta(\eta)$ is the Heaviside step function, equal to 1 if $\eta>0$ and zero otherwise. In an MD era one has $a(\tilde{\eta})/a(\eta)=(\tilde{\eta}/\eta)^2$. Using~\cref{eq:source_md}, the first-order solution becomes
\begin{equation}\label{eq:first_ord_sol}
    \gamma_{\mathbf{k},\lambda}^{(1)}(\eta)=-2\sum_\sigma\int\frac{d^3p}{(2\pi)^3}\Phi_{\mathbf{k}-\mathbf{p}}^\text{pr}\gamma_{\mathbf{p},\sigma}^\text{pr}q_\lambda^\sigma(\mathbf{\hat{p}},\mathbf{\hat{k}})I_\MD(x,k,p)\,,
\end{equation}
where the polarization contraction is defined as
\begin{equation}
    q_\lambda^\sigma(\mathbf{\hat{p}},\mathbf{\hat{k}})\equiv\epsilon_{ij}^\sigma(\mathbf{\hat{p}})\epsilon_\lambda^{ij\ast}(\mathbf{\hat{k}})\,.
\end{equation}
Given the MD source function
\begin{equation}\label{eq:f_MD}
    f^\MD(p,\tilde{\eta})\equiv p^2T_\gamma^\MD(p\tilde{\eta})\,,
\end{equation}
we can define the kernel in~\cref{eq:first_ord_sol} as
\begin{equation}\label{eq:I_MD_tbc}
    \begin{split}
        I_\MD(x,k,p)&\equiv\int_0^\eta d\tilde{\eta}\frac{a(\tilde{\eta})}{a(\eta)}f^\MD(p\tilde{\eta})G_\mathbf{k}^\MD(\eta,\tilde{\eta})\\\
        &=3\int_0^x\frac{d \tilde{x}}{k}\,\frac{\tilde{x}^2}{x^2}\cdot p^2 \cdot \frac{j_1(p\tilde{x}/k)}{p\tilde{x}/k}\cdot \frac{x\tilde{x}}{k}\left[j_1(\tilde{x})y_1(x)-j_1(x)y_1(\tilde{x})\right]\\
        &=\frac{3}{x}\frac{p}{k}\int_0^x d\tilde{x}\tilde{x}^2j_1(p\tilde{x}/k)\left[j_1(\tilde{x})y_1(x)-j_1(x)y_1(\tilde{x})\right]\,.
    \end{split}
\end{equation}
We set the lower integration bound to $x=0$, corresponding to the regime in which the relevant fluctuations are frozen on super-horizon scales. We then take the limit $x\gg1$, square the resulting solution, and average over several oscillation periods in order to extract the spectral energy density. To evaluate~\cref{eq:I_MD_tbc}, we use the tabulated integrals of the spherical Bessel functions
\begin{equation}
    \int_0^x d\tilde{x}\tilde{x}^2j_1(a\tilde{x})\left[j_1(\tilde{x})y_1(x)-j_1(x)y_1(\tilde{x})\right]=\frac{1}{a^2-1}\left[aj_1(x)-j_1(ax)\right]\,,
\end{equation}
from which we obtain
\begin{equation}
    \begin{split}
        I_\MD(x,k,p)&=\frac{3p^2}{p^2-k^2}\left[\frac{j_1(x)}{x}-\frac{j_1(px/k)}{px/k}\right]\,.
    \end{split}
\end{equation}
In terms of $(u,v)$ variables, defined in~\cref{eq:uv_vars}, we get
\begin{equation}
    I_\MD(x,k,v)=\frac{3v^2}{v^2-1}\left[\frac{j_1(x)}{x}-\frac{j_1(vx)}{vx}\right]\,.
\end{equation}
In the limit $x\to\infty$, the integral vanishes at leading order. The next-to-leading behavior for $x\gg1$ is
\begin{equation}\label{eq:kernel_MD_asymp}
    I_\MD(x\gg1,k,v)\simeq\frac{3}{x^2}\frac{\cos{vx}-v^2\cos{x}}{v^2-1}+\mathcal{O}\left(\frac{1}{x^3}\right)\,,
\end{equation}
and, after squaring and averaging over many oscillation periods we obtain
\begin{equation}\label{eq:avsq_ker_pois_md}
    \overline{I_\MD^2(x\gg1,k,v)}=\frac{9}{2}\frac{1}{x^4}\frac{v^4+1}{\left(v^2-1\right)^2}\,,
\end{equation}
which is the result we were seeking in this subsection.

When computing the energy density in radiation domination, the factor $(k/\mathcal{H})^2\sim x^2$ renders the kernel time-independent in~\cref{eq:I_RD}, as expected since GWs freely propagate as radiation. In the MD case, instead,
\begin{equation}
    \Omega^\st_\GW(k)\big|_\MD\sim \overline{\mathcal{I}^2_\MD}\cdot \Delta^2_\Phi\Delta^2_{\gamma_0}\sim x^2\cdot\overline{I_\MD^2(x\gg1,k,v)}\cdot\Delta^2_\Phi\Delta^2_{\gamma_0}\sim\frac{1}{x^2}\Delta^2_\Phi\Delta^2_{\gamma_0}\,,
\end{equation}
so the residual $\sim1/x^2$ dependence makes the spectral energy density vanishing at large $x$, in principle independently of the shape of the primordial spectra. This is a distinctive feature of the scalar-tensor induced signal in a purely MD epoch. The suppression applies to modes that re-enter during the standard late-time MD era, $0.2\,\rm{kpc}^{-1}\lesssim k\lesssim0.01\,\rm{Mpc}^{-1}$. On the other hand, as we will discuss below, the behavior in an eMD epoch is different, and STGWs must be treated with a dedicated approach. 

We remark, however, that including the \secrev{last} term in~\cref{eq:2pf_all} could in principle modify this conclusion. A detailed analysis of this contribution is left for future work.

\subsection{STGWs from an early matter-dominated universe}\label{subsec:stgw_emd_to_rd}

We adopt the formalism recently developed in the context of the ``poltergeist mechanism''~\cite{Inomata:2020lmk,Inomata:2025wiv} to describe the matching between an eMD epoch and the subsequent RD era, assuming an instantaneous reheating transition at conformal time $\eta_\R$. The term ``poltergeist'' refers to the behavior of the scalar potential: after horizon crossing during early matter domination it remains constant, then undergoes rapid oscillations immediately after the transition to radiation domination, and decays. Although the scalar perturbation itself quickly diminishes, the gravitational radiation sourced by the coupling of the linear modes during the RD age leaves a non-negligible imprint. Our goal in this section is to extend this framework to the case of STGWs.

An eMD era can generate two distinct GW contributions induced by the scalar perturbations. The first arises from GWs produced during the eMD phase, which subsequently propagate into the RD stage. The second is sourced at the onset of the RD period, immediately after reheating, by fluctuations that re-entered during eMD epoch and then experienced the transition. Since we are interested in the scalar-tensor signal, we must include a third contribution: first-order tensor modes generate PGWs already during eMD age, that then propagate into RD era together with their second-order, scalar-tensor-induced counterpart. This feature is absent in the purely scalar-induced case, where scalar fluctuations cannot produce gravitational radiation at linear order.

All these contributions must be correctly propagated through reheating. For the PGWs, the energy density on sub-horizon scales is given by~\cite{Watanabe:2006qe}
\begin{equation}
    \Omega_\GW(k)\big|_\eMD=\frac{1}{6}\left(\frac{k}{\mathcal{H}}\right)^2\overline{\left[T_\gamma^\MD\right]^2}\Delta^2_{\gamma_0}(k)=\frac{3}{4}\frac{x^2+1}{x^4}\Delta^2_{\gamma_0}(k)\,,
\end{equation}
where we used~\cref{eq:t_func_md}. We account for their propagation through the eMD stage by evaluating the energy density at $x=x_\R\equiv k\eta_\R$. For $x>x_\R$, we then include the standard dilution factor, which encodes the change in relativistic degrees of freedom as the universe expands and cools during radiation domination~\cite{Watanabe:2006qe,Kugarajh:2025pjl}
\begin{equation}
    h^2\Omega_{\GW,0}(k)=h^2\Omega_r\left(\frac{g_{\ast s}(T_\hc)}{g_{\star s0}}\right)^{-4/3}\left(\frac{g_\ast(T_\hc)}{g_{\star 0}}\right)\Omega_{\GW,\hc}(k)\,;
\end{equation}
here, “hc” denotes evaluation at horizon crossing for GWs induced by modes that re-enter during radiation domination.\footnote{The combination $h^2\Omega$, where $H_0\equiv h\times 100\,\rm{km}\,\rm{s}^{-1}\,\rm{Mpc^{-1}}$ is independent of $H_0$~\cite{Allen:1997ad}.} For PGWs originating from early matter domination, we take this reference to be the reheating time $\eta_\R$.

For the second-order contributions, and in close analogy with the scalar-induced case (with a slightly different notation), the kernel to be computed reads~\cite{Kohri:2018awv,Inomata:2020lmk,Inomata:2019ivs,Inomata:2025wiv}
\begin{equation}\label{eq:I_sudden_tbc}
    \begin{split}
        I(x,u,v,x_\R)&=\int_0^{x_\R}d \tilde{x}\left(\frac{\tilde{x}^2/x_\R^2}{2(x/x_\R)-1}\right)G_\mathbf{k}^{\eMD\to\RD}(\eta,\tilde{\eta})f(\tilde{x},u,v,x_\text{R})\\
        &+\int_{x_\R}^x d\tilde{x}\left(\frac{2(\tilde{x}/x_\R)-1}{2(x/x_\R)-1}\right)G_\mathbf{k}^\RD(\eta,\tilde{\eta})f(\tilde{x},u,v,x_\R)\\
        &\equiv I_\eMD(x,u,v,x_\R)+I_\RD(x,u,v,x_\R)\,.
    \end{split}
\end{equation}
Here, $f(\tilde{x},u,v,x_\text{R})$ is the source function that experiences the eMD-RD transition. Before reheating, $f$ reduces to the source function of the pure MD phase in~\cref{eq:source_md}; after reheating, it follows the RD dynamics, with initial conditions fixed by the eMD equations evaluated at the transition. $G_\mathbf{k}^\RD$ denotes the Green function in radiation domination, while $G_\mathbf{k}^{\eMD\to\RD}$ is the matching Green function across the two phases, which we will discuss below.

Starting from the scale factor and conformal Hubble rate, and imposing continuity of $a$ and $\mathcal{H}$ (together with their first derivatives) at $\eta_\R$, we obtain
\begin{equation}\label{eq:a_Hubble_sudden_tr}
    \frac{a(\eta)}{a(\eta_\R)}=\begin{cases}
        \left(\cfrac{\eta}{\eta_\R}\right)^2&(\eta<\eta_\R)\,,\\
        2\cfrac{\eta}{\eta_\R}-1&(\eta\ge \eta_\R)\,,
    \end{cases}\quad \mathcal{H}(\eta)=\begin{cases}
        \cfrac{2}{\eta}&(\eta<\eta_\R)\,,\\
        \cfrac{1}{\eta-\eta_\R/2}&(\eta\ge\eta_\R)\,.
    \end{cases}
\end{equation}
Tensor modes induced during the eMD epoch and subsequently undergoing the sudden transition to radiation domination correspond to the first contribution in~\cref{eq:I_sudden_tbc}, and are described by $G_\mathbf{k}^{\eMD\to\RD}$; its analytical expression is given by the requirement of continuity of the function itself and its derivative at $\eta_\R$, and it reads~\cite{Kohri:2018awv,Inomata:2019ivs,Inomata:2025wiv}
\begin{equation}\label{eq:green_sudden}
    G_\mathbf{k}^{\eMD\to\RD}(\eta;\tilde{\eta})=\frac{C(x,x_\R)}{k}\tilde{x}j_1(\tilde{x})+\frac{D(x,x_\R)}{k}\tilde{x}y_1(\tilde{x})\,,
\end{equation}
where $\tilde{\eta}<\eta_\R<\eta$ and
\begin{align}
    C(x,x_\R)&=\frac{1}{2x_\R^2}\left[\sin x-2x_\R(\cos x+x_\R\sin x)+\sin(x-2x_\R)\right]\,,\\
    D(x,x_\R)&=\frac{1}{2x_\R^2}\left[(2x_\R^2-1)\cos x -2x_\R\sin x +\cos(x-2x_\R)\right]\,.
\end{align}
As shown in~\cite{Pearce:2023kxp}, this analytical matching procedure can induce spurious oscillations in the ultraviolet (UV) tail of the SIGW spectrum within the eMD-to-RD framework. In our setup, the scalar-tensor energy density is not affected by this artefact at large wavenumbers.

We focus only on the GW spectrum induced during the early MD phase, rather than on the signal sourced by the perturbations that re-enter just before reheating, since --- as we show in appendix~\ref{sec:matching} --- the matching of tensor modes across epochs should be treated differently from the analytical prescription typically used for the scalar perturbations. We therefore work out an explicit expression for the first term in~\cref{eq:I_sudden_tbc}. Using the Green function in~\cref{eq:green_sudden}, the corresponding kernel becomes
\begin{equation}
    \begin{split}
        I_\eMD(x,v,x_\R)&=\frac{1}{x_\R(2x-x_\R)}\int_0^{x_\R}\frac{d\tilde{x}}{k}\tilde{x}^3\left[\frac{C(x,x_\R)}{k}j_1(\tilde{x})+\frac{D(x,x_\R)}{k}y_1(\tilde{x})\right]\cdot 3v^2k^2\frac{j_1(v\tilde{x})}{v\tilde{x}}\\
        &=\frac{3v}{x_\R(2x-x_\R)}\\
        &\times\left\{C(x,x_\R)\underbrace{\int_0^{x_\R}d\tilde{x}\tilde{x}^2j_1(\tilde{x})j_1(v\tilde{x})}_{\iota_1}+D(x,x_\R)\underbrace{\int_0^{x_\R}d\tilde{x}\tilde{x}^2y_1(\tilde{x})j_1(v\tilde{x})}_{\iota_2}\right\}\,,
    \end{split}
\end{equation}
where the two integrals over spherical Bessel functions are
\begin{align}
    \iota_1&=\frac{\sin{(vx_\R)}\left[(v^2\left(x_\R\cos{x_\R}-\sin{x_\R}\right)+\sin{x_\R}\right]-vx_\R\cos{(vx_\R)\sin{x_\R}}}{v^2x_\R(v^2-1)}\,,\\
    \iota_2&=\frac{\sin{(vx_\R)}\left[v^2x_\R\sin{x_\R}+(v^2-1)\cos{x_\R}\right]-vx_\R\cos{x_\R}\cos{vx_\R}-v^3x_\R}{v^2x_\R(v^2-1)}\,.
\end{align}
We thus obtain
\begin{equation}
    \begin{split}
        I_\eMD(x,v,x_\R)&=\frac{3}{2vx_\R^3(v^2-1)(2x-x_\R)}\\
        &\times \Big\{v^3[\cos{x}-2x_\R^2\cos{x}-\cos{(x-2x_\R)}+2x_\R\sin{x}]\\
        &+2vx_\R\cos{(vx_\R)}[x_\R\cos{(x-x_\R)}-\sin{(x-x_\R)}]\\
        &-2\sin{(vx_\R)}[x_\R\cos{(x-x_\R)}+(v^2x_\R^2-1)\sin{(x-x_\R)}]\Big\}\,. 
    \end{split}
\end{equation}
After squaring this expression and averaging over oscillations (the intermediate steps are lengthy and not particularly illuminating), we arrive at
\begin{equation}\label{eq:temp_I_eMD}
    \overline{I^2_\eMD(x,v,x_\R)}=\frac{9\left[\alpha_1(v,x_\R)^2+4\alpha_2(v,x_\R)^2\right]}{8v^2x_\R^6(v-1)^2(v+1)^2(2x-x_\R)^2}\,,
\end{equation}
where
\begin{align}
    \alpha_1(v,x_\R)&=2vx_\R\cos(vx_\R)(\cos{x_\R}-x_\R\sin{x_\R})+v^3(\sin{(2x_\R)}-2x_\R)\nonumber\\
    &-2\sin{(vx_\R)}(\cos{x_\R}(v^2x_\R^2-1)+x_\R\sin{x_\R})\,,\\
    \alpha_2(v,x_\R)&=v^3(\sin{x_\R}-x_\R)(\sin{x_\R}+x_\R)+vx_\R\cos{(vx_\R)}(x_\R\cos{x_\R}+\sin{x_\R})\nonumber\\
    &+\sin{(vx_\R)}(\sin{x_\R}(v^2x_\R^2-1)-x_\R\cos{x_\R})\,.
\end{align}
We can factor out a global $1/x^2$ in~\cref{eq:temp_I_eMD}, which cancels against the $(k/\mathcal{H})^2\propto x^2$ factor in the GW energy density. In the limit $x\to\infty$, we can then neglect the \secrev{subleading} term $x_\R/x$ in the last round brackets of the denominator\secrev{: $(2-x_\R/x)^2\sim 2^2$}. This yields
\begin{equation}\label{eq:I_eMD}
    \begin{split}
        \overline{\mathcal{I}^2_\eMD(v,x_\R)}&\equiv x^2\times\overline{I^2_\eMD(x,v,x_\R)}\\
        &=\frac{9}{2^5}\frac{\alpha_1(v,x_\R)^2+4\alpha_2(v,x_\R)^2}{v^2x_\R^6(v-1)^2(v+1)^2}\,,
    \end{split}
\end{equation}
which represents the kernel for GWs induced by scalar-tensor mixing during an eMD phase, and constitutes the main result of this section.
\begin{figure}[t!]
    \centering
    \includegraphics[width=0.8\textwidth]{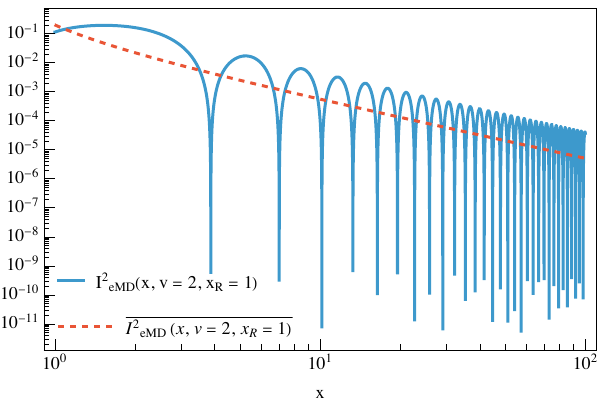}
    \caption{Squared kernel before and after the oscillation average in~\cref{eq:temp_I_eMD} as a function of $x=k\eta$.}
    \label{fig:IeMD}
\end{figure}

~\Cref{eq:I_eMD} is valid as long as the perturbations remain in the linear regime, allowing for a perturbative treatment of the density contrast $\delta\equiv \delta\rho/\overline{\rho}$. The linear approximation breaks down once matter fluctuations become of order unity, $\delta\gtrsim 1$. On sub-horizon scales during matter domination, perturbations grow proportionally to the scale factor $a$. Therefore, if horizon crossing occurs sufficiently before $\eta_\R$, the corresponding modes may enter the nonlinear regime prior to reheating. We define the critical wavenumber $k_\NL$ as the mode that becomes non-linear at the transition, i.e., $\delta_{k_\NL}(\eta_\R)=1$. This scale sets the cutoff of the perturbative treatment. \secrev{We follow the derivation in Ref.~\cite{Inomata:2025wiv}; the general 00-th component of the linearized Einstein equations for a single fluid can be written as~\cite{Mukhanov:2005sc}}
\begin{equation}
    \Delta\Psi -3\mathcal{H}\left(\Psi^\prime+\mathcal{H}\Phi\right)=\frac{a^2}{M_\text{Pl}}\delta\rho\,,
\end{equation}
\secrev{that for sub-horizon perturbations in the MD epoch produces}
\begin{equation}
    \delta_\mathbf{k}\simeq -\frac{2}{3}k^2\mathcal{H}^{-2}\Phi_\mathbf{k}\,,
\end{equation}
\secrev{where $\Phi_\mathbf{k}$ is constant. Therefore, using the matter-domination relations $\mathcal{H}=2/\eta$ and $\Delta^2_\Phi=\frac{9}{25}\Delta^2_\zeta$ (which implies $\Phi_\mathbf{k}\sim-\frac{3}{5}\zeta_\mathbf{k}$) we obtain}
\begin{equation}\label{eq:NL}
    |\delta|\sim\frac{1}{10}(k\eta)^2\sqrt{\Delta^2_\zeta}\Longrightarrow k_\NL\sim\sqrt{10}\eta_\R^{-1}\left(\Delta^2_\zeta\right)^{-1/4}\sim 470/\eta_\R\,.
\end{equation}
In the following analysis, we therefore adopt a conservative upper limit $k_\text{max}=450/\eta_\R$, in line with the existing literature.

\subsubsection{Analytical behavior of the kernel}\label{subsubsec:kernel_eMD_behav}

We now study analytically the behavior of~\cref{eq:I_eMD}. In the IR limit $k\to 0$, one has $u\sim v\sim1/k\gg1$. For small $k$ this yields
\begin{equation}
    \overline{\mathcal{I}^2_\eMD(v=1/k,k\eta_\R)}\Bigg|_{k\to0}\sim\frac{\left(3\sin{\eta_\R}-3\eta_\R\cos{\eta_\R}-3\eta_\R^2\sin{\eta_\R}+2\eta_\R^3\right)^2}{8\eta_\R^6}+\mathcal{O}(k^2)\,.
\end{equation}
In the UV limit we must distinguish two cases. The first corresponds to $v\to0,u\to1$ which describes the regime of large tensor wavelengths. Expanding the kernel for $v\to0$, one finds that it scales as $\sim \mathcal{O}(v^4)$. As a consequence,~\cref{eq:general_Omega} behaves as $\sim1/v^2$, leading to a divergence that requires regularization. The second case corresponds to $v\to1,u\to0$, i.e., the regime of large scalar wavelengths, which also belongs to the UV domain. In this limit, the kernel remains finite as $v\to1$, while the associated energy density in~\cref{eq:general_Omega} diverges as $\sim1/u^2$.

Similar UV divergences arise in the RD epoch~\cite{Bari:2023rcw,Picard:2024ekd,Wu:2025gwt}. In those works, a regularization prescription is adopted in which the power spectrum is multiplied by a damping function to suppress the UV behavior~\cite{Bari:2023rcw}. We follow a similar approach here and introduce a regularizing function
\begin{equation}
    \Upsilon^\eMD(u,v)\equiv\frac{u^2}{u^2+d_1^2}\frac{v^2}{v^2+d_2^2}\,,\quad d_1,d_2\sim\mathcal{O}(0.1)-\mathcal{O}(1)\,,
\end{equation}
which cures the divergences because
\begin{equation}
    \Upsilon^\eMD(u\to0,v\to1)\sim\frac{u^2}{d_1^2(1+d_2^2)}\,,\quad \Upsilon^\eMD(u\to1,v\to0)\sim\frac{v^2}{d_2^2(1+d_1^2)}\,.
\end{equation}
The regularized energy density then reads
\begin{equation}\label{eq:general_Omega_reg}
    \begin{split}
        \Omega^\st_\GW(k)\big|_\eMD&=\frac{1}{12}\int_0^\infty d v \int_{|1-v|}^{1+v} d u\frac{\Upsilon^\eMD(u,v)}{v^2u^2}\left[\frac{(1+v^2-u^2)^2}{v^2}+\left(1+\left(\frac{1+v^2-u^2}{2v}\right)^2\right)^2\right]\\
        &\times \Delta^2_\Phi(uk)\Delta^2_{\gamma_0}(vk)\cdot\overline{\mathcal{I}^2}\,.
    \end{split}
\end{equation}
IR and UV limits are shown for completeness, since the reheating time sets a lower and upper bound on the minimum and maximum frequencies that can contribute to the spectrum. Consequently, the regimes of very small and very large $k$ are not actually reached in this work, as we will discuss below. Nevertheless, the UV regime is explored sufficiently to require the use of the previously defined regularization.

\subsubsection{STGWs with peaked primordial power spectra}\label{subsubsec:omega_emd}

%
\begin{figure}[t!]
    \centering
    \includegraphics[width=0.9\textwidth]{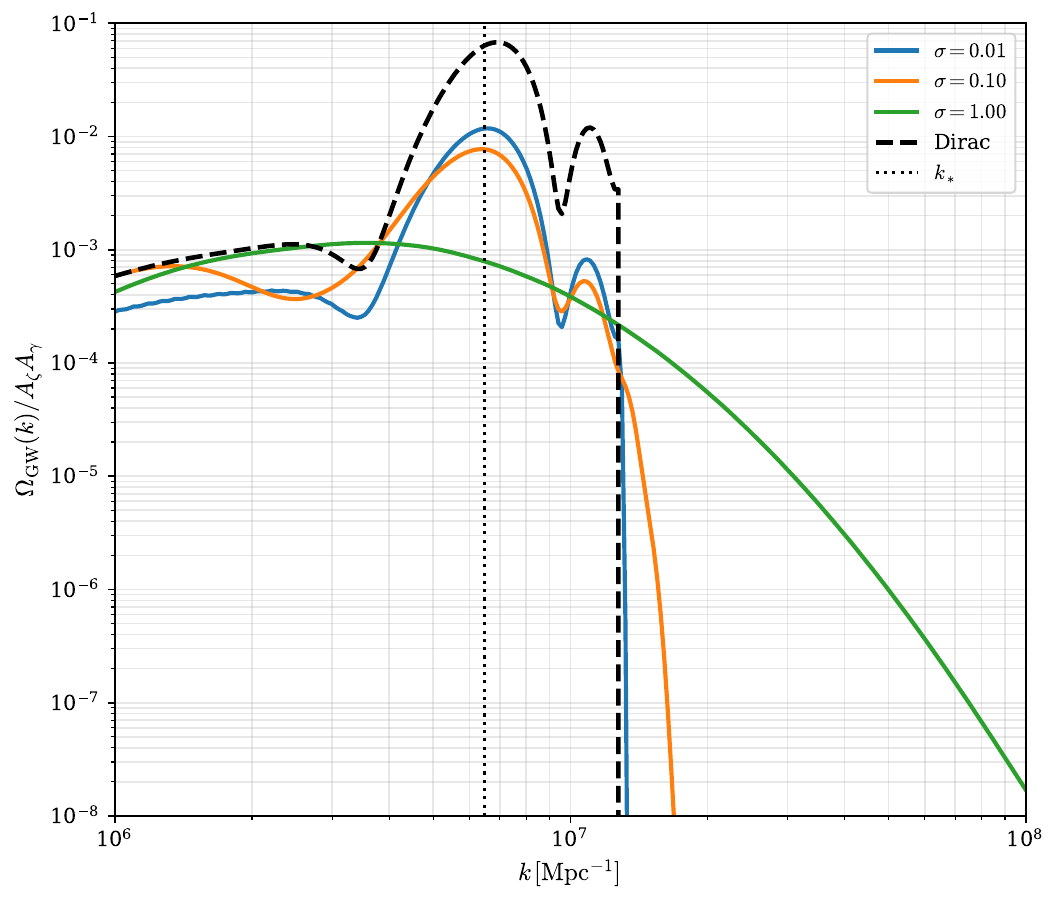}
    \caption{Energy density in~\cref{eq:general_Omega_reg}, normalized by the primordial amplitudes, from an eMD epoch and sudden transition to RD at $\eta_\R=2\times 10^{-6}\,\rm{Mpc}$, as a function of $k$. The log-normal primordial power spectra have same peak $k_\ast=6.47\times 10^{6}\,\rm{Mpc}^{-1}$ (corresponding to a physical frequency $f_\ast \simeq 10^{-8}\,\rm{Hz}$), and various values of $\sigma$, including $\sigma\to0$, i.e., the monochromatic limit.}
    \label{fig:omega_emd_norm}
\end{figure}
We now evaluate the STGW spectral energy density in~\cref{eq:general_Omega_reg} assuming peaked input power spectra, as can arise from many inflationary models that enhance primordial amplitudes, such as axion inflation scenarios~\cite{Garcia-Bellido:2016dkw,Namba:2015gja}. The corresponding results are shown in~\cref{fig:omega_emd_norm}. As a first case, we consider a monochromatic power spectrum
\begin{equation}\label{eq:dirac_input}
    \Delta_\Phi^2(k)=A_\zeta\delta\left(\ln{\frac{k}{k_{s,\ast}}}\right)\,,\quad \Delta_{\gamma_0}^2=A_t\delta\left(\ln{\frac{k}{k_{t,\ast}}}\right)\,,
\end{equation}
with the same peak scale for scalar and tensor source, $k_{s,\ast}=k_{t,\ast}\equiv k_\ast$.\footnote{We label the scalar amplitude with the subscript ``$\zeta$'' to highlight the connection with the comoving curvature perturbation, which is conserved on superhorizon scales.} Multiplying eq.~(4.4) of~\cite{Bari:2023rcw} by 2 to account for the two polarizations contributing equally in the absence of primordial parity violation, we obtain
\begin{equation}\label{eq:omega_st_emd_dirac}
    \Omega^\st_\GW(k)\big|_\eMD=\frac{A_{\zeta}A_t}{384}\left(\frac{k}{k_\ast}\right)^6\left[1+32\left(\frac{k_\ast}{k}\right)^4+48\left(\frac{k_\ast}{k}\right)^2\right]\overline{I^2_\eMD(v=k_\ast/k,x_\R)}\Theta(2k_\ast-k)\,.
\end{equation}
The second case is more physical: we consider two log-normal power spectra with the same peak scale and width,
\begin{align}\label{eq:input_spectra}
        \Delta^2_\Phi(k)  &= \frac{A_\zeta}{\sqrt{2\pi\sigma^2}}\exp{\left[-\frac{\log^2{(k/k_\ast)}}{2\sigma^2}\right]}\,,\\
        \Delta^2_{\gamma_0}(k)  &= \frac{A_t}{\sqrt{2\pi\sigma^2}}\exp{\left[-\frac{\log^2{(k/k_\ast)}}{2\sigma^2}\right]}\,.
\end{align}
Results for various values of $\sigma$, together with the Dirac limit, are shown in~\cref{fig:omega_emd_norm}. In the monochromatic case, the first crest in the spectrum corresponds to the peak of the Dirac delta sources, while the second originates from the specific structure of the kernel in~\cref{eq:I_eMD}, where different trigonometric combinations generate an oscillatory pattern. The curve is finally truncated by the cutoff at $k=2k_*$. Looking at the log-normal-induced signals, as the width decreases, the curves progressively approach the Dirac-induced behavior, as expected. For larger $\sigma$, the spectrum is increasingly flattened, as the broad log-normal profiles smooth out the oscillatory features. A further discussion of the dependence of the energy density on the shape parameters is presented in in appendix~\ref{sec:varying_shape_pars}.

\section{STGWs in the nano-Hertz frequency band: NANOGrav and SKA}\label{sec:stgw_nhz}

In this section, we revisit the state of the art of the STGWs in the 1-100 nHz range, corresponding to the PTA observational window, together with the associated PGWs. Using the log-normal scalar and tensor primordial power spectra with identical shape parameters given in~\cref{eq:input_spectra}, we perform a parameter estimation with the current \texttt{NANOGrav 15-year} dataset and subsequently provide forecasts based on a mock SKA dataset after 10.33 years of observation. The adopted priors are summarized in~\cref{tab:priors}, and the resulting posterior medians and 68\% credible regions are reported in~\cref{tab:values}. The choice of using only log-normal seeds is justified by the fact that, in the limit where the width approaches zero, the distribution becomes monochromatic. In this limit, the Dirac delta case is automatically included and does not require a separate analysis, provided that the parameter estimation favors sufficiently narrow widths.

\subsection{Bounds from PBH overproduction}\label{subsec:pbh}

%
\begin{table}[t!]
    \centering
    \begin{tabular}{Sc|Sc Sc Sc Sc}
            Epoch & $\log_{10}{(A_\zeta)}$ & $\log_{10}{(A_t)}$ & $\log_{10}{(\sigma)}$ & $\log_{10}{(f_\ast/\rm{Hz})}$ \\
          \hline
         eMD& $\mathcal{U}(-3.5,0)$ & $\mathcal{U}(-3.5,0)$ & $\mathcal{U}(-1.5,0.9)$& $\mathcal{U}(-8.5,-6.5)$ \\
         RD & $\mathcal{U}(-3.5,0)$& $\mathcal{U}(-3.5,0)$&$\mathcal{U}(-1.5,0.9)$&  $\mathcal{U}(-10,-6)$
    \end{tabular}
    \caption{Priors for each parameter of the input lognormal power spectra, according to the specific epoch of horizon crossing.}
    \label{tab:priors}
\end{table}
Large density fluctuations that crossed the horizon in the early universe could have collapsed to form compact objects known as PBHs~\cite{Hawking:1971ei,Carr:1974nx}. PBHs have been extensively studied as they may contribute to the dark matter (DM) abundance, explain some of the observed GW events, or seed the formations of supermassive black holes (SMBHs); see~\cite{Carr:2026hot} for a recent review. The scenario in which all DM consists of PBHs sets a stringent upper bound on their abundance. In our analyses, we impose the requirement of avoiding a PBH overproduction, using the infrastructure provided by \texttt{fastPTA}~\cite{Cecchini:2025oks}, and apply these constraints to both NG15 amd SKA10 results.

In~\cref{subsubsec:pbh_rd}, we briefly review the PBH formation mechanism during the RD era and outline how to derive the resulting abundance. Then, in~\cref{subsubsec:pbh_emd}, we describe PBH production in an early matter-dominated era, which differs in several ways from the RD case. In the following, we do not employ peak theory: the derivation in radiation domination relies on threshold statistics, whereas formation during early matter domination is based on the hoop conjecture.

\subsubsection{PBH formation during a radiation-dominated universe}\label{subsubsec:pbh_rd}

Most of the literature has addressed the problem in the canonical RD epoch. In particular,~Ref.~\cite{Ferrante:2022mui} developed an exact formalism for computing the PBH abundance in the presence of completely generic local non-Gaussianity (nG), using threshold statistics on the compaction function.~Ref.~\cite{Cecchini:2025oks} applied this formalism to the much simpler case of Gaussian fluctuations (as in our setup), and implemented the corresponding abundance computation in \texttt{fastPTA}. Similarly, we summarize the main steps of the derivation in Ref.~\cite{Ferrante:2022mui} (where further details can be found) under a general nG assumption, and then specialize to the Gaussian case.

During the RD stage, when the compaction function exceeds a certain threshold value, the corresponding density fluctuation collapses to a PBH. The compaction function is defined as twice the local Misner-Sharp mass excess over the areal radius
\begin{equation}\label{eq:comp_fun}
    \mathcal{C}(r,t) = \frac{2\delta M(r,t)}{R(r,t)}=\frac{2}{R(r,t)}\int_{S_R^2}d^3\mathbf{x}\rho_b(t)\delta(\mathbf{x},t)\,,
\end{equation}
where $\rho_b(t)$ is the background density, $R(r,t)\equiv a(t)re^{\zeta(r)}$ and $\delta(\mathbf{x},t)$ is the density contrast.

Adopting the gradient-expansion approximation and assuming spherical symmetry, the density contrast can be expressed in terms of the curvature perturbation $\zeta$ as
\begin{equation}
    \delta(r,t)=-\frac{2}{3}\Phi\left(\frac{1}{aH}\right)^2e^{-2\zeta(r)}\left[\zeta^{\prime\prime}(r)+\frac{2}{r}\zeta^\prime(r)+\frac{1}{2}\zeta^\prime(r)^2\right]\,;
\end{equation}
$\Phi$ contains the dependence on the EoS parameter, and hence on the thermal history. Integrating the compaction function over the radial coordinate,~\cref{eq:comp_fun} becomes
\begin{equation}
    \mathcal{C}(r)=C_\G(r)\frac{dF}{d\zeta_\G}-\frac{1}{4\Phi}\mathcal{C}^2_\G(r)\left(\frac{dF}{d\zeta_\G}\right)^2\,,
\end{equation}
where $F=F(\zeta_\G)$ is a generic function encoding the presence of primordial nGs. The two-dimensional joint PDF of the two Gaussian variables $\zeta_\G$ and $\mathcal{C}_\G$ is therefore
\begin{equation}\label{eq:joint_prob}
    P_\G(\mathcal{C}_\G,\zeta_\G)=\frac{1}{2\pi\sigma_c\sigma_r\sqrt{1-\gamma_{cr}^2}}\exp\left(-\frac{\zeta_\G^2}{2\sigma_r^2}\right)\exp\left[-\frac{1}{2\left(1-\gamma_{cr}^2\right)}\left(\frac{\mathcal{C}_\G}{\sigma_c}-\frac{\gamma_{cr}\zeta_\G}{\sigma_r}\right)^2\right]\,,
\end{equation}
with $\gamma_{cr}\equiv \sigma^2_{cr}/\sigma_c\sigma_r$ and the two-point functions given by
\begin{align}
    \sigma^2_c&=\braket{\mathcal{C}_\G\mathcal{C}_\G}=\frac{4\Phi^2}{9}\int_0^\infty\frac{dk}{k}\left(kr_m\right)^4W^2(k,r_m)T^2(k,r_m)\Delta^2_\zeta(k)\,,\\
    \sigma^2_{cr}&=\braket{\mathcal{C}_\G\zeta_\G}=\frac{2\Phi}{3}\int_0^\infty\frac{dk}{k}(kr_m)^2W(k,r_m)W_s(k,r_m)T^2(k,r_m)\Delta^2_\zeta(k)\,,\\
    \sigma^2_r&=\braket{\zeta_\G\zeta_\G}=\int_0^\infty\frac{dk}{k}W_s^2(k,r_m)T^2(k,r_m)\Delta^2_\zeta(k)\,;
\end{align}
here, it enters the dependence on the parameters of the chosen primordial power spectrum template. Under the assumption of a perfect fluid of radiation, we have $T(k,\eta)$ as defined in~\cref{eq:t_func_rd} for the scalar perturbation and
\begin{align}
    W_s(k,r)&\,= \frac{\sin(kr)}{kr}\,,\\
    W(k,r)&\, =3 \left[\frac{\sin(kr)-kr\cos(kr)}{(kr)^3}\right]\,.
\end{align}
The Hubble crossing condition is $aHr_m=1$. We can now introduce the PBH mass fraction derived from the collapse of a single mode $M_\HH$
\begin{equation}
    \beta(M_\HH)=\int_\mathcal{D}\mathcal{K}\left(\mathcal{C}-\mathcal{C}_\thh\right)^\gamma{P}_\G(\mathcal{C}_\G,\zeta_\G)d\mathcal{C}_\G d\zeta_\G\,,
\end{equation}
where the integration domain is given such that the threshold condition is imposed (first inequality) and ``type-I'' PBH are selected (second inequality):
\begin{equation}
    \mathcal{D}=\{\mathcal{C}_\G,\zeta_\G\in\mathbb{R}:\,\mathcal{C}(\mathcal{C}_\G,\zeta_\G)>\mathcal{C}_\thh\wedge\mathcal{C}_1(\mathcal{C}_\G,\zeta_\G)<2\Phi\}\,;
\end{equation}
the parameters in Choptuik’s scaling law for critical collapse, $\mathcal{K}\left(\mathcal{C}-\mathcal{C}_\mathrm{th}\right)^\gamma$, depend on $w$, $c_s$, and the shape of the initial configuration~\cite{Choptuik:1992jv,Musco:2012au}. The law is derived under the assumption of spherical symmetry, which is no longer valid for an eMD epoch.

Finally, the present-day fraction of DM in PBHs is given by
\begin{equation}
    f_\PBH\equiv \frac{\Omega_\PBH}{\Omega_\DM}=\frac{1}{\Omega_\DM}\int_{M_{\HH,min}}^{M_{\HH,max}}\frac{dM_\HH}{M_\HH}\left(\frac{1}{M_\HH}\right)^{1/2}\left(\frac{g_*}{106.75}\right)^{3/4}\left(\frac{g_{*s}}{106.75}\right)^{-1}\left(\frac{\beta(M_\HH)}{7.9\times10^{-10}}\right)\,,
\end{equation}
where the horizon mass $M_\HH$ is related to the temperature by
\begin{equation}
    M_\HH\equiv 4.8\times10^{-2}M_\odot\left(\frac{g_*}{106.75}\right)^{-1/2}\left(\frac{T_k}{\rm{GeV}}\right)^{-2}\,.
\end{equation}
We restrict to the case of primordial Gaussian fluctuations: $F(\zeta_\G)=1$ and~\cref{eq:joint_prob} becomes a Gaussian distribution after marginalizing over $\zeta_\G$. In our corner plots, the contour $f_\PBH=1$ is shown as a dashed line, while the ``forbidden'' region $f_\PBH>1$ is shaded. Each two-dimensional parameter subspace is constrained by fixing the remaining parameters to the median values of their posterior distributions.

\subsubsection{PBH formation during an early matter-dominated universe}\label{subsubsec:pbh_emd}

We review the main formulas of the PBH production during an eMD epoch, which is more subtle than in the RD case. On the one hand, because the fluid is pressureless, a large enhancement of PBH formation is expected: in general, $\delta_{\mathrm{th}} \sim w$, so when $w=0$ even a slight overdensity can collapse into a PBH. On the other hand, several important additional effects must be taken into account. The most important is anisotropic collapse: deviations from a perfectly spherical shape grow in time due to gravitational pull, and the resulting ellipsoids undergo one-dimensional squeezing, known as ``pancake'' collapse. Ref.~\cite{Harada:2016mhb} derived the PBH abundance using the hoop conjecture and the Zel’dovich approximation~\cite{Zeldovich:1969sb}, which is particularly well suited for describing the evolution of non-linear density perturbations. In the Zel'dovich approximation the local gradient of the gravitational potential wrt the Lagrangian coordinates determines the motion of each fluid element, with the time dependence encoded in the scale factor $a(t)$.

Under these assumptions, they found that the abundance can be approximated as $\beta\simeq 0.05\sigma_\HH^5$ for $\sigma_\HH\ll1$, where $\sigma_\HH^2=(4/25)\Delta^2_\zeta$ during matter domination (be careful not to confuse $\sigma_\HH$ with the log-normal power spectrum width $\sigma$).

Later, Ref.~\cite{Harada:2017fjm} showed that tidal torques can induce large angular momenta $L$, causing most collapsing masses to bounce back due to centrifugal forces. Thus, the angular momentum provides an additional suppression factor for the production rate; in particular, the abundance can be approximated as
\begin{equation}
    \beta_0 \simeq
    \begin{cases}
        1.984\times10^{-6}f_q(q_c)\mathcal{I}^6\sigma_\HH^2\exp{\left[-0.1474\frac{\mathcal{I}^{4/3}}{\sigma_\HH^{2/3}}\right]} & \sigma_\HH\lesssim 0.005\\
        0.05556\sigma_\HH^5 & 0.005\lesssim \sigma_\HH \lesssim 0.2\,,
    \end{cases}
\end{equation}
where $\mathcal{I}$ is an order-unity parameter, $\mathcal{I}\sim\mathcal{O}(1)$, and $f_q(q_c)$ is the mass fraction with $q<q_c\simeq 2.4\mathcal{I}^{1/3}\sigma_\HH^{1/3}$, where $q$ is a dimensionless initial quadrupole-moment parameter. For $\sigma_\HH\gtrsim0.2$, one should perform the complete numerical integral:
\begin{equation}
    \beta_0\simeq\int_0^\infty d\alpha \int_{-\infty}^{\alpha}d\beta\int_{-\infty}^{\beta}d\gamma \Theta\left[\delta_\HH(\alpha,\beta,\gamma)-\delta_{\thh}\right]\Theta\left[1-h(\alpha,\beta,\gamma)\right]w(\alpha,\beta,\gamma)\,,
\end{equation}
with
\begin{align}
    h(\alpha,\beta,\gamma)&= \frac{2}{\pi}\frac{\alpha-\gamma}{\alpha^2}E\left(\sqrt{1-\left(\frac{\alpha-\beta}{\alpha-\gamma}\right)^2}\right)\,,\\
    w(\alpha,\beta,\gamma)&=-\frac{27}{8\sqrt{5}\pi\sigma_3^6}\exp\left[-\frac{3}{5\sigma_3^2}\left\{(\alpha^2+\beta^2+\gamma^2)-\frac{1}{2}(\alpha\beta+\beta\gamma+\gamma\alpha)\right\}\right]\nonumber\\
    &\times(\alpha-\beta)(\beta-\gamma)(\gamma-\alpha)d\alpha d\beta d\gamma\,;
\end{align}
here, $E(e)$ is the complete elliptic integral of the second kind, $\sigma_3\equiv\sigma_\HH/\sqrt{5}$ and $w(\alpha,\beta,\gamma)$ is the Doroshkevich distribution~\cite{doroshkevich1970spatial}.

Even with these additional suppression mechanisms, the abundance during early matter domination is enhanced relative to the RD case, computed with the same primordial spectrum under similar conditions, as expected from an epoch dominated by a non-relativistic fluid.

Furthermore, they found that most forming PBHs should have large angular momentum, close to the extremal value $a_* = 1$, where $a_* \equiv L/M^2$ is the dimensionless Kerr spin parameter.  Because the majority of PBHs requires large spin, they showed that when the early matter-dominated phase is sufficiently short, or equivalently when the reheating temperature is sufficiently high, the formation of rapidly rotating PBHs is significantly suppressed, and thus the total PBH production vanishes, as the spin distribution is pushed toward high values. This is encoded in their eq.~(B2), which gives an upper bound on the reheating temperature,
\begin{equation}
    T_\mathrm{R} \lesssim 0.1\,\text{GeV} \left(\frac{k}{10^6\,\text{Mpc}^{-1}}\right)\left(\frac{2}{5}\mathcal{I}\sigma_\HH\right)^{1/3}\,.
\end{equation}
We computed our reheating temperature through~\cite{Inomata:2019ivs}
\begin{equation}
    \frac{\eta_\R}{10^{-14}\,\text{Mpc}}=\left(\frac{g_s}{106.75}\right)^{1/3}\left(\frac{g}{106.75}\right)^{-1/2}\left(\frac{T_\R}{1.2\times10^{7}\,\text{GeV}}\right)^{-1}\,,
\end{equation}
finding $T_\R\sim0.1\,\text{GeV}$, that lies below the upper bound within our priors and the PTA bandwidth.

\subsection{STGWs from an early-matter dominated universe with NG15}\label{subsec:ng15_inference}

%
\begin{figure}[t!]
    \centering
    \includegraphics[width=0.8\textwidth]{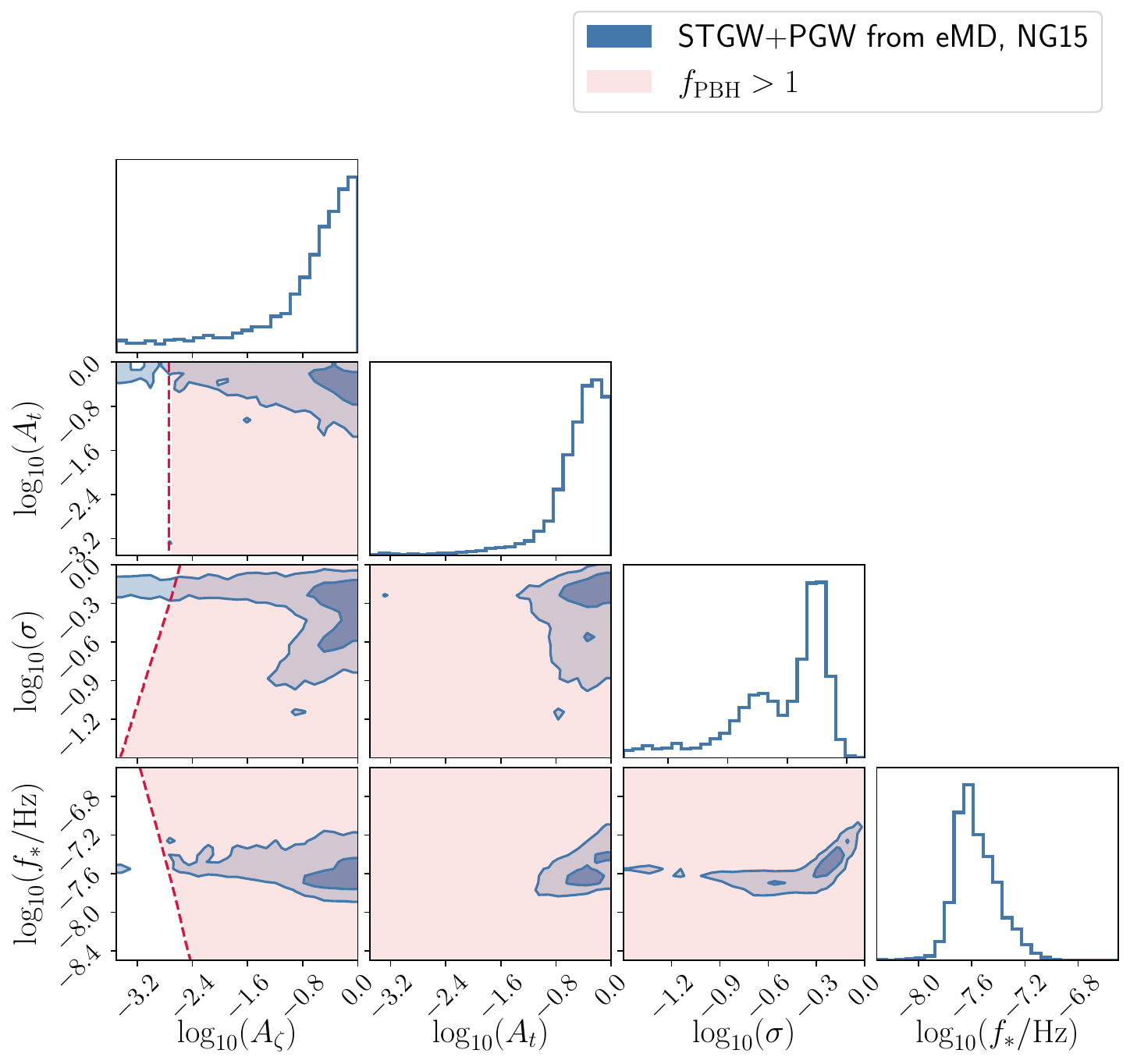}
    \caption{Corner plots for STGWs+PGWs generated during an early matter domination that suddenly interrupts transiting to the standard RD epoch. We used lognormal primordial power spectra with same peak and width, and NG15 data, fixing $\eta_\R=2\times10^{-6}\,\rm{Mpc}$. The plots show the 68\% and 95\% credible regions, whereas the red areas are the subspaces cutted out by the PBH overproduction.}
    \label{fig:stgw_emd_ln_ln_ng15_posteriors}
\end{figure}
We perform a Bayesian analysis using \texttt{PTArcade} in ceffyl mode~\cite{Lamb:2023jls} over the first 14 frequency bins of the \texttt{NANOGrav 15-year} data set~\cite{NANOGrav:2023gor}, injecting a signal given by the sum of STGWs and PGWs. We adopt uninformative flat priors for all parameters, as listed in~\cref{tab:priors}, with one exception: the peak frequency of the GWs generated during the eMD epoch requires informative bounds. The conformal reheating time $\eta_\R$ constrains the range of modes that re-enter during eMD epoch, setting both the smallest and largest scale that cross the horizon before $\eta_\R$. The upper bound is already fixed by the linearity requirement in~\cref{eq:NL}, while the lower bound follows from $\mathcal{H}=2/\eta$ during a matter domination. Since horizon crossing occurs at $k=\mathcal{H}$, the last mode to enter the horizon has comoving wavenumber $k_{\min}=2/\eta_\R$. This corresponds numerically to
\begin{equation}
    \log_{10}{(f_{\min}/\rm{Hz})}=-8.81\,,\quad \log_{10}{(f_{\max}/\rm{Hz})}=-6.45\,.
\end{equation}
In practice, we choose slightly tighter prior limits than these to avoid boundary effects. Consistently, in the numerical integration of both the linear and the second-order contributions, we impose IR and UV cutoffs on the primordial power spectra.

The posteriors obtained from this analysis are shown in~\cref{fig:stgw_emd_ln_ln_ng15_posteriors}. The log-amplitudes of the scalar and tensor power spectra exhibit non-Gaussian marginal distributions that pile up toward large values, indicating that they are only weakly constrained and partially prior-limited. Their pronounced negative correlation in the corner plot reflects the degeneracy inherent in the model, since the STGW energy density depends on the product $A_\zeta \cdot A_t$ and the two amplitudes cannot be disentangled by the data alone.

By contrast, the shape parameters $\log_{10}(f_*/\rm{Hz})$ and $\log_{10}(\sigma)$ are more informative: the posterior for $\sigma$ is bimodal, while $\log_{10}(f_*/\rm{Hz})$ is tightly constrained within a narrow interval. The corresponding corner plot displays a characteristic ``banana-shaped'' contour, suggesting that the data confine the allowed $(f_*, \sigma)$ subspace to a relatively thin, curved region rather than a broad area.

The PBH overproduction constraint excludes all regions of parameter space in which the scalar amplitude is fixed to its median, as its posterior is driven toward very large value (although this may appear as a negative result, we stress that each 2D bounds is obtained by fixing all other parameters to their median values indeed, which is an arbitrary choice). Lower values of the amplitude, still allowed within 2$\sigma$ credible region, can lead to more relaxed PBH constraints, as we explicitly find in the SKA10 case.

\subsection{STGWs from an early-matter dominated universe with SKA10}
\label{subsec:ska10_emd}

%
\begin{figure}[t!]
    \centering
    \includegraphics[width=0.45\textwidth]{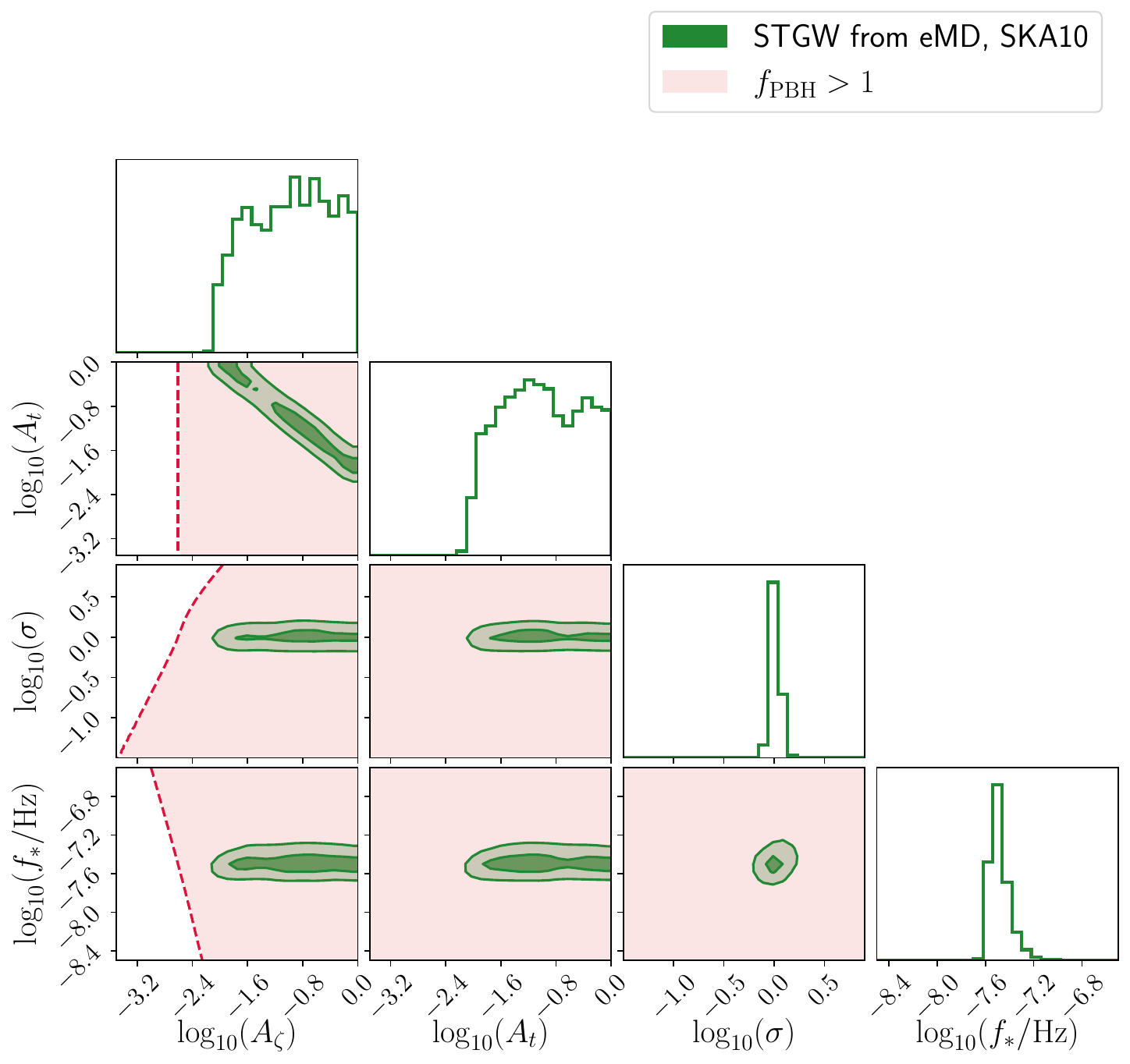}
    \includegraphics[width=0.45\textwidth]{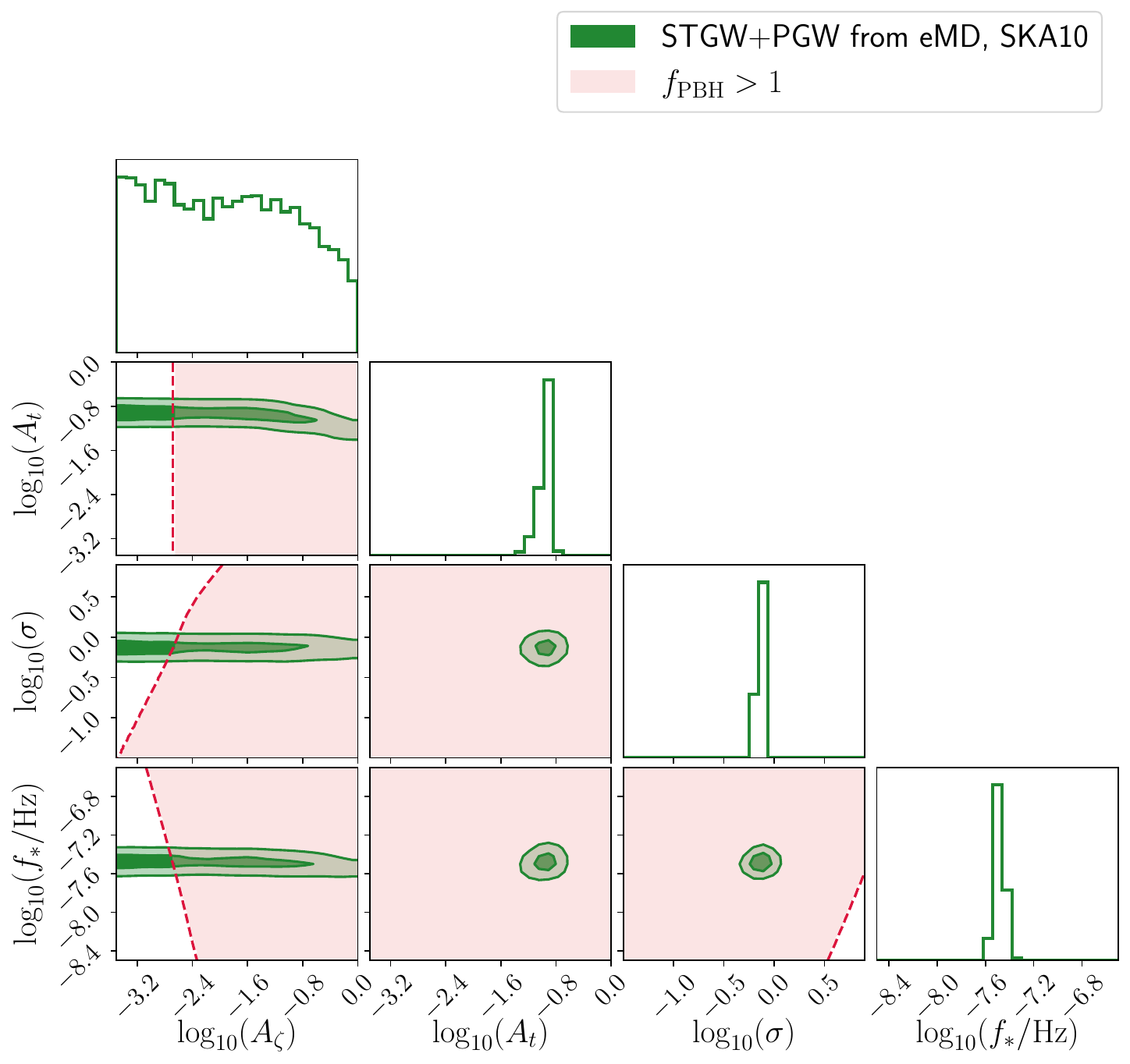}
    \caption{Left panel: Bayesian MCMC parameter estimation obtained from SKA-like configuration with $T_\text{obs}=10.33$ yr and $N_\text{p}=200$, using STGWs from an eMD epoch, fixing $\eta_\R=2\times10^{-6}\,\rm{Mpc}$. Right panel: the same Bayesian estimation, adding the contribution of the linear-order PGWs. Log-normal primordial power spectra with same shape parameters are used as scalar and tensor seeds.}
    \label{fig:ska_stgw_emd}
\end{figure}
We present forecasts for STGWs+PGWs generated during the eMD epoch, assuming SKA-like noise with $N_p=200$ pulsars observed for $T_{\rm obs} = 10.33$ yr. Unlike in the NG15 case, we also show forecasts excluding PGWs, in order to disentangle the contribution of first- and second-order GWs. The adopted priors are provided in~\cref{tab:priors}.

In the left panel of~\cref{fig:ska_stgw_emd}, where only the second-order contribution is included, the scalar and tensor log-amplitudes show posterior distributions pushed toward large values with a clear negative correlation, whereas the posteriors for $\log_{10}(f_*/\rm{Hz})$ and $\log_{10}(\sigma)$ are sharply localized. PBH overproduction excludes the purely second-order signal because the high median value of $\log_{10}(A_\zeta)$ implies $f_\PBH>1$ almost everywhere. A similar behavior will also appear for STGWs generated during RD epoch in the next subsection.

The right panel of~\cref{fig:ska_stgw_emd}, which includes PGWs, shows the expected tightly localized posteriors for the shape parameters and for the tensor log-amplitude, as appropriate for a high signal-to-noise mock dataset centered on fiducial values. The marginalized distributions are nearly symmetric, with no sign of prior-volume effects. On the other hand, the posterior for the scalar log-amplitude is broader and anisotropic, with open contours. This indicates that, once the peaked feature is tightly constrained in frequency, width, and tensor amplitude by the linear PGW signal, the scalar-tensor induced contribution can vary along an overall normalization direction that only weakly affects the total spectrum within the SKA sensitivity band. In other words, after fixing the PGW component, the STGW normalization can be substantially reduced, via a smaller $A_\zeta$, without spoiling the fit.

The absence of strong degeneracies among the shape parameters shows that SKA could effectively resolve the spectral structure of the signal; any residual correlations are weak and approximately linear, visible only as a mild tilt of the contour. Overall, the posterior geometry suggests that the likelihood dominates over prior effects.

Imposing the PBH overproduction bound leads again to a noticeable tightening of the constraints, ruling out almost all the $2\sigma$ credible regions in both the samplings.

\subsection{STGWs from a radiation-dominated universe with SKA10}\label{subsec:ska10_rd}

%
\begin{figure}[t!]
    \centering
    \includegraphics[width=0.45\textwidth]{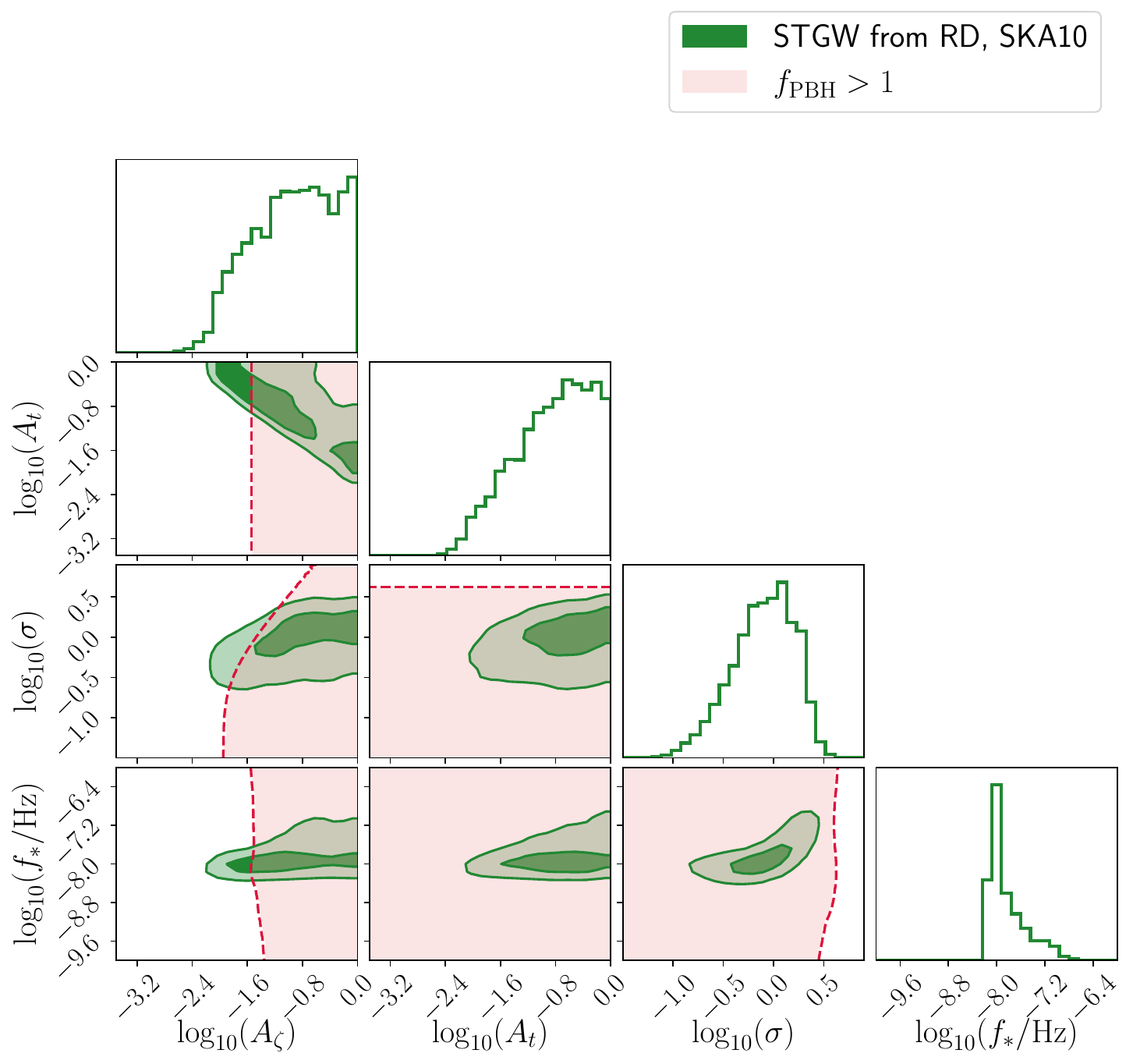}
    \includegraphics[width=0.45\textwidth]{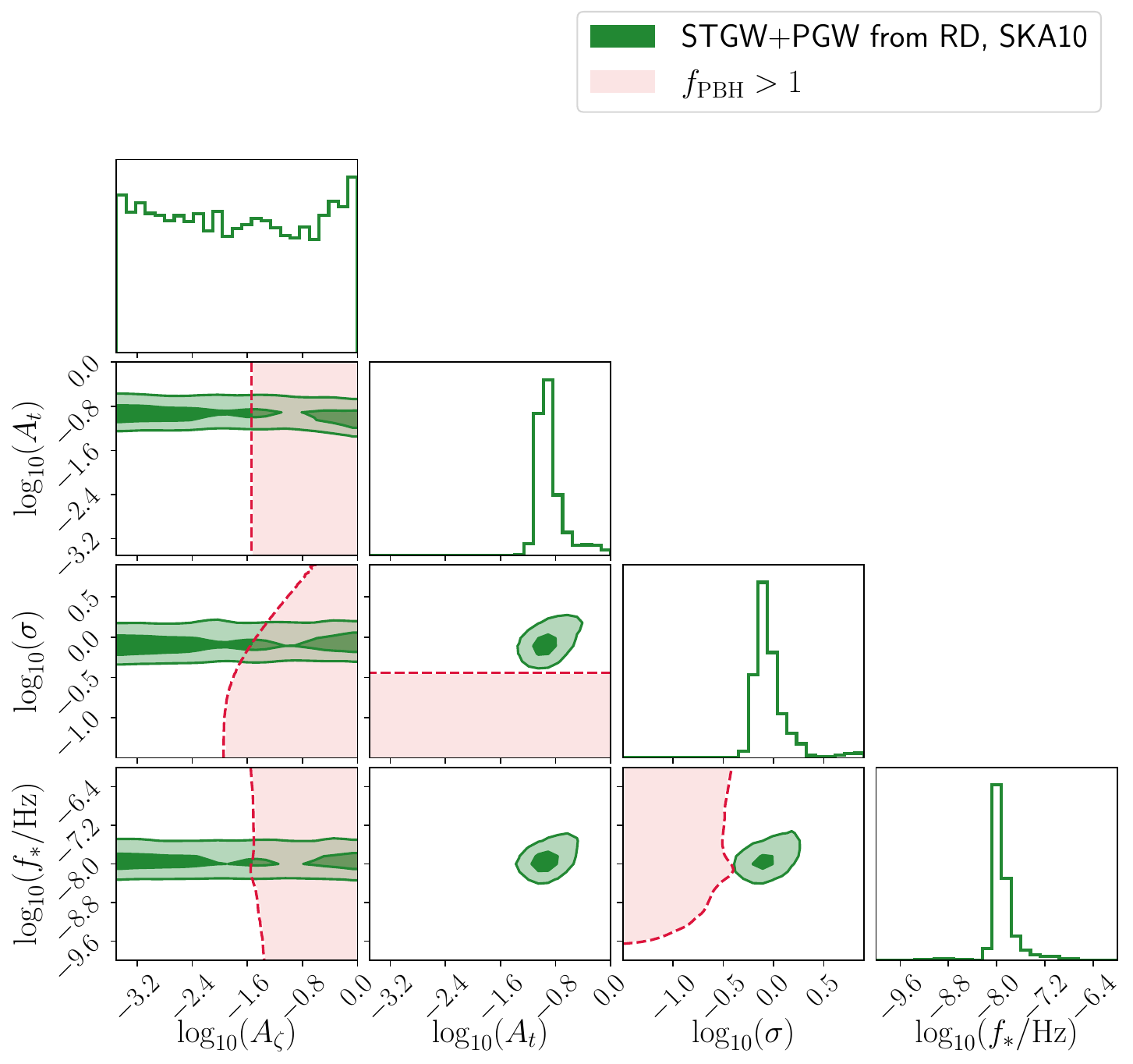}
    \caption{Left panel: Bayesian MCMC parameter estimation obtained from SKA-like configuration with $T_\text{obs}=10.33$ yr and $N_\text{p}=200$, using STGWs generated during the RD epoch. Right panel: same Bayesian estimation adding the contribution of the PGWs. The shaded regions are forbidden by PBH overproduction. Scalar and tensor seeds are both log-normal power spectra with same peak frequency and width.}
    \label{fig:ska_stgw_rd}
\end{figure}
We now present forecasts for STGWs and PGWs generated during the RD epoch with SKA, using the corresponding prior set in~\cref{tab:priors}. In this case, the bounds on $f_*$ are looser, since horizon crossing during RD era is not constrained by the reheating time as the early matter domination case. As before, we compare results with and without the linear PGW contribution. The resulting posteriors are shown in~\cref{fig:ska_stgw_rd}.

In the left panel, where only the second-order STGWs are injected, the log-amplitude posteriors resemble those obtained for the eMD case, while the posteriors for $\log_{10}(f_*/\rm{Hz})$ and $\log_{10}(\sigma)$ are broader. This reflects the different spectral imprints in radiation domination: $\Omega_\GW^\st$ is nearly invariant under shifts of the peak position, and in the monochromatic limit it becomes effectively unimodal. By contrast, in the eMD case fixing the reheating time induces a strong dependence on the shape parameters: the relative weight of the oscillatory kernel in~\cref{eq:kernel_MD_asymp} changes significantly as $f_*$ and $\sigma$ vary. Additional illustrations of this behaviour are provided in appendix~\ref{sec:varying_shape_pars}.

In the right panel, where PGWs are included, both $\log_{10}(f_*/\rm{Hz})$ and $\log_{10}(\sigma)$ are again measured at the percent level. The contours show a slight tilt, signalling a mild positive linear correlation between $\sigma$ and $f_*$, but no strong degeneracy. As in the eMD-SKA case, the scalar amplitude posterior is broader and somewhat anisotropic, while the tensor amplitude is tightly constrained by the injected linear signal.

Qualitatively, the RD-SKA posteriors are very similar to those for eMD–SKA: in both cases the strong curvature degeneracy present in current data is lifted, and the joint posterior is close to a multivariate Gaussian. The main difference is a somewhat more pronounced tilt of the $(f_*,\sigma)$ in the RD scenario, indicating a slightly stronger residual linear correlation between peak width and peak frequency. Nonetheless, in both setups SKA constrains the spectral-shape sector with high precision and reduces parameter degeneracies to very small levels.

The PBH overproduction bound removes roughly half of the posterior support in $\log_{10}(A_\zeta)$ in all RD-SKA corner plots, while leaving the other parameters almost unaffected; the allowed 2D subspaces are, however, more restricted than the prior volumes.

Finally, we evaluate the signal-to-noise ratio (SNR) for the SKA forecasts. For the eMD case, the SNR increases from $\sim$53 to $\sim$60 when PGWs are added, and for the RD epoch it rises from $\sim$27 to $\sim$37, as expected from the additional linear contribution.

\subsection{Adding the contribution of the SIGWs}\label{subsec:sigw}

%
\begin{figure}[t!]
    \centering
    \includegraphics[width=0.45\textwidth]{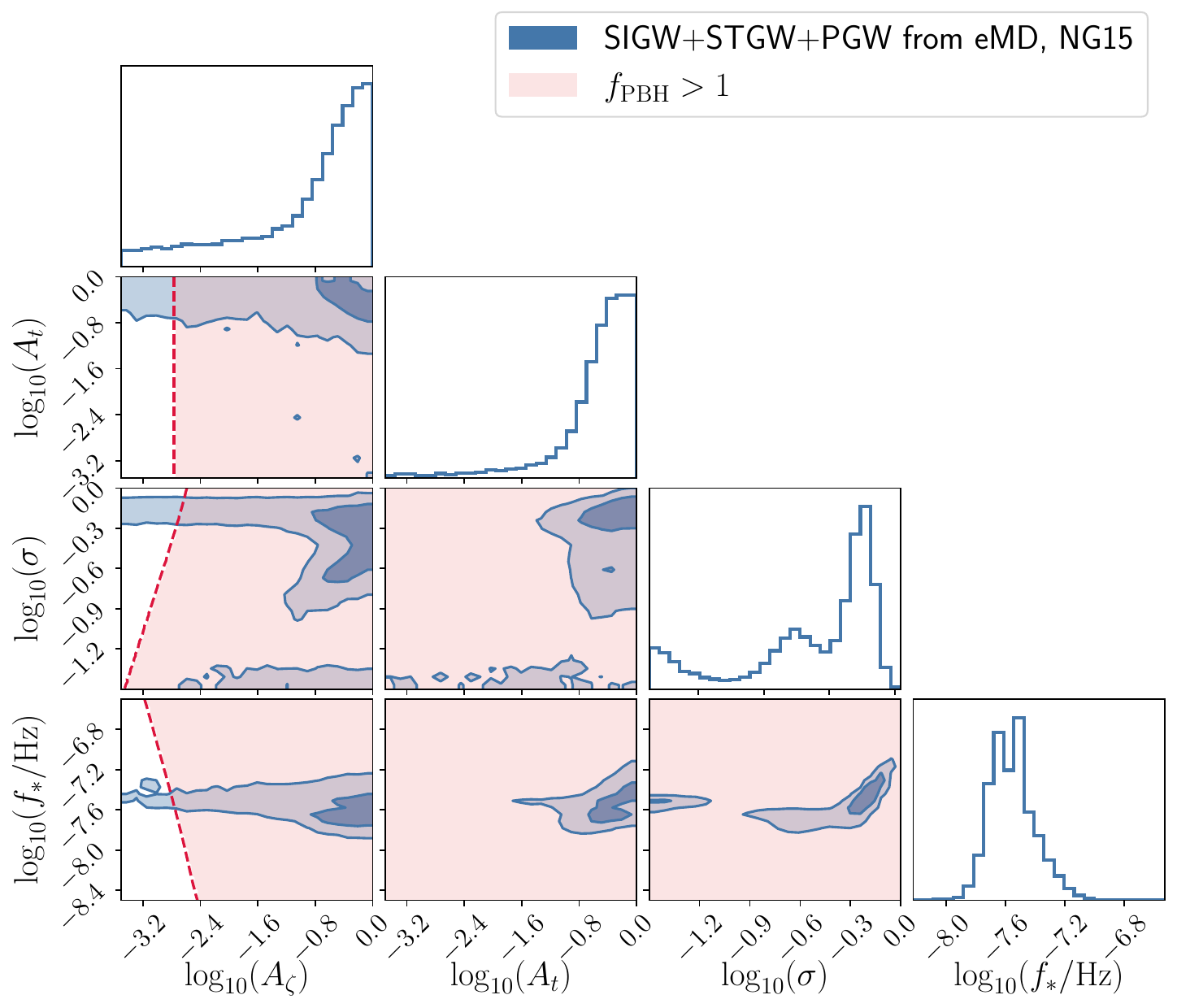}
    \includegraphics[width=0.45\textwidth]{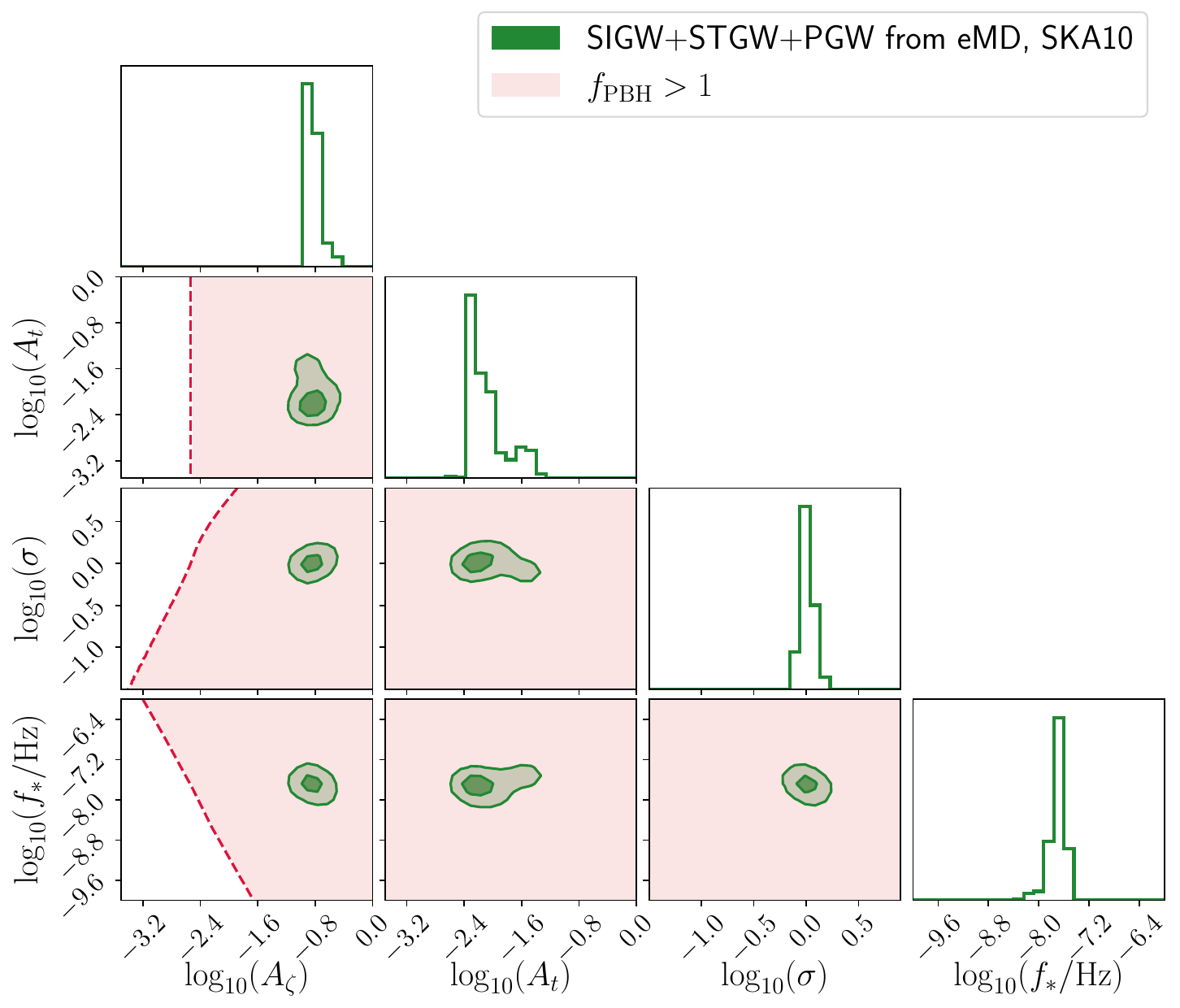}
    \includegraphics[width=0.45\textwidth]{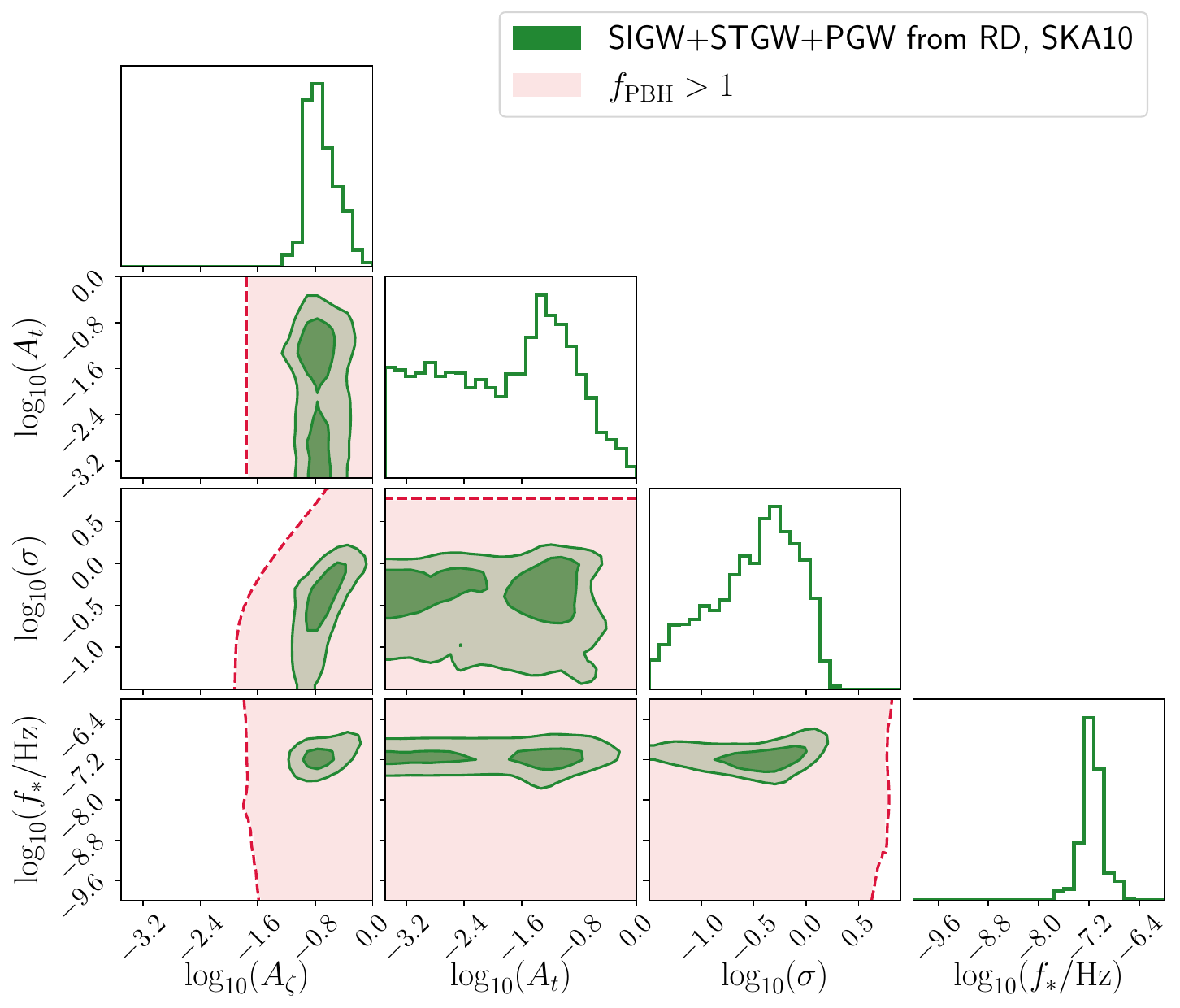}
    \caption{\secrev{Upper left/right} panel: Bayesian MCMC parameter estimation obtained from the NG15/mock SKA10 dataset with $T_\text{obs} = 10.33,\text{yr}$ and $N_\text{p} = 200$, using GWs generated during the eMD epoch with $\eta_\R = 2 \times 10^{-6}\,\text{Mpc}$ from lognormal power spectra with identical shape parameters, including both the linear-order and scalar-induced contributions. \secrev{Lower} panel: parameter estimation for GWs generated during the RD epoch from lognormal power spectra with the same shape parameters, contrasted with the mock SKA10 dataset.}
    \label{fig:pta_gw}
\end{figure}

The main goal of this work is to isolate the effect of the novel scalar–tensor contribution in the nHz frequency band and, in particular, to assess the capability of present and future PTA data to constrain the tensor power spectrum on scales smaller than those probed by the CMB. Nevertheless, the enhancement of the scalar fluctuations also leads to the production of SIGWs, which cannot be neglected a priori. These are expected to be the dominant signal in the RD epoch --- except for very sharp primordial power spectra in the high-frequency range~\cite{Chang:2022vlv} --- and even in the sudden eMD–RD scenario, where we likewise expect them to dominate at high frequencies (STGWs do not experience an enhancement as strong as the SIGW poltergeist mechanism).~\footnote{There is no resonant peak for the scalar–tensor signal in radiation domination, unlike the case of SIGWs. Furthermore, in a generic MD era, GWs sourced by non-linear structures will dominate in the high-frequency regime\cite{Fernandez:2023ddy,Escriva:2026blk}.}

For the reasons discussed above, in this subsection we include the contribution of SIGWs in the total signal and compare it once again with the NG15 and mock SKA10 datasets. We refer to appendix~\ref{sec:sigw} for a review of the main formulas for SIGWs generated during both the RD and eMD epochs. In that appendix, we also present the results of the parameter estimation performed using only the scalar-scalar seed, and the corresponding SNRs.

In~\cref{fig:pta_gw} we show the results of these new MCMC samplings. While the posteriors obtained with the NG15 dataset do not exhibit any additional features compared to the STGW+PGW results shown in~\cref{fig:stgw_emd_ln_ln_ng15_posteriors}, the SKA10 contour plots break the degeneracy in the scalar amplitude that was present in~\cref{fig:ska_stgw_emd,fig:ska_stgw_rd}. As a consequence, the analysis of the signal generated during the eMD epoch yields tightly constrained parameter posteriors, whereas in the RD case new degeneracies between the tensor amplitude and the width emerge.

The inclusion of the SIGW contribution increases the SNR substantially from $\sim 60$ to $\sim 100$ in the eMD case, and more moderately from $\sim 37$ to $\sim 38$ for the RD results. This can be interpreted as additional evidence of the strength of the poltergeist enhancement in the sudden eMD–RD scenario. By contrast, no such enhancement appears in the pure RD era, and consequently the increase in SNR is much more moderate.

Finally, for all the sampling we note again that almost the entire $2\sigma$ parameter regions are forbidden due to PBH overproduction.

\section{Conclusions}
\label{sec:conclusions}

%
\begin{table}[t!]
    \centering
    \begin{tabular}{Sc Sc Sc Sc|Sc Sc Sc Sc}
            Dataset & Epoch & +PGW & +SIGW & $\log_{10}{(A_\zeta)}$ & $\log_{10}{(A_t)}$ & $\log_{10}{(\sigma)}$ & $\log_{10}{(f_\ast/\rm{Hz})}$\\
          \hline
          NG15 & eMD & yes & no &$-0.55^{+0.39}_{-1.14}$& $-0.40^{+0.27}_{-0.41}$ & $-0.30^{+0.14}_{-0.42}$ &  $-7.59^{+0.18}_{-0.12}$ \\
          NG15 & eMD & yes & yes & $-0.56^{+0.40}_{-1.24}$ & $-0.38^{+0.26}_{-0.38}$ & $-0.30^{+0.13}_{-0.41}$ &  $-7.58^{+0.20}_{-0.12}$ \\
         SKA10 & eMD & no & no & $-0.94^{+0.63}_{-0.71}$ & $-1.0^{+0.7}_{-0.6}$ & $0.002^{+0.054}_{-0.041}$ & $-7.50^{+0.09}_{-0.06}$\\
         SKA10 & eMD & yes & no & $-1.92^{+1.15}_{-1.11}$ & $-0.95^{+0.05}_{-0.09}$ & $-0.145^{+0.018}_{-0.017}$ & $-7.49^{+0.04}_{-0.04}$\\
         SKA10 & eMD & yes & yes & $-0.85^{+0.10}_{-0.08}$ & $-2.19^{+0.40}_{-0.13}$ & $-0.0013^{+0.0669}_{-0.0424}$ & $-7.70^{+0.10}_{-0.09}$\\
         SKA10 & RD & no & no &$-0.90^{+0.62}_{-0.72}$ & $-0.77^{+0.52}_{-0.70}$ & $-0.08^{+0.29}_{-0.36}$ & $-7.96^{+0.45}_{-0.13}$\\
         SKA10 & RD & yes & no & $-1.8^{+1.3}_{-1.2}$ & $-0.94^{+0.19}_{-0.09}$ & $-0.08^{+0.15}_{-0.09}$ & $-7.95^{+0.19}_{-0.09}$\\
         SKA10 & RD & yes & yes & $-0.74^{+0.27}_{-0.17}$ & $-1.77^{+0.84}_{-1.19}$ & $-0.44^{+0.36}_{-0.56}$ & $-7.16^{+0.15}_{-0.13}$
    \end{tabular}
    \caption{Medians and 68\% credible regions for each Bayesian inference with the STGW signal, adding the linear order and scalar-induced contributions.}
    \label{tab:values}
\end{table}
GWs induced at second order by primordial perturbations, enhanced at small scales that are not constrained by current observations, represent one of the most compelling cosmological candidates for dominating the SGWB in the frequency range probed by PTA. Such a signal provides a powerful probe of the early universe thermal history, since its energy density can retain distinctive features associated with the epoch of the GW production. Motivated by these considerations, it is natural to explore different scenarios that could give rise to an induced GWB observable in the nHz frequency band.

In this work we have investigated the prospects for probing the STGW signal at nHz scales. Our goal was to forecast constraints on the signal generated during the standard RD epoch, while also exploring new scenarios: the contribution arising from scalar-tensor mixing in a generic MD epoch, as well as a sudden transition from an eMD phase to RD, which has not been studied prior to this work.

We have shown that, unlike the purely scalar-induced case, STGWs are not amplified during matter domination: scalar perturbations remain constant on all scales, while the tensor modes decay, so the scalar-tensor signal becomes negligible on sub-horizon scales in a pure MD epoch. We then demonstrated that the formalism developed by Inomata, Kohri and Terada for early matter domination with a sudden reheating can be extended beyond the purely quadratic scalar-scalar source to include scalar-tensor mixing. However, the exact analogue of the ``poltergeist'' contribution cannot be obtained in the same way, due to the specific behavior of tensor perturbations across the eMD-RD transition. Additional fine-tuning is also required, such as the use of peaked input spectra to avoid unphysical enhancements of the energy density and the choice of fixing the conformal reheating time, which sets the duration of the eMD era.

We found that the linear PGWs and second-order STGWs generated during an eMD epoch can produce a distinctive observable signal which, for certain combinations of the input parameters, is non-negligible in the PTA band. The resulting spectral shape is strongly controlled by the shape parameters: the kernel induces visible oscillations for narrow widths and smoothly approaches the monochromatic limit as the width decreases.

We performed MCMC parameter estimation using the \texttt{NANOGrav 15-year} dataset and a mock SKA dataset, injecting log-normal scalar and tensor seeds with same shape parameters.

Both NANOGrav results and SKA-like forecasts favour scenarios where STGWs provide the dominant contribution to the GWB and yield well-constrained posteriors, especially for the mock SKA10 dataset. When the second-order signal is accompanied by the linear PGW component, the shape parameters and tensor amplitude are sharply localized, while the scalar amplitude can span a much broader range. Similar conclusions hold in a RD epoch for SKA, although the shape parameters are slightly less constrained because the RD spectrum is less sensitive to changes in $f_*$ and $\sigma$ than in the eMD scenario. The median scalar log-amplitude allows, within its 2$\sigma$ region, parameter combinations in $\log_{10}(A_t),\log_{10}(\sigma),\log_{10}(f_*/\rm{Hz})$ that satisfy PBH constraints, when the linear-order contribution is included.

The results change once the scalar-induced contribution is added to the injected signal. For the SKA forecasts, the degeneracy in the scalar amplitude is lifted, while new degeneracies in the tensor amplitude and the width appear for production during radiation domination. The doubling of the SNR for the signal from the eMD epoch indicates that, when included, SIGWs provide the dominant contribution; the same does not hold in the radiation-dominated case. Moreover, the samplings performed with only SIGWs in appendix~\ref{sec:sigw} yield SNR values comparable to those obtained for STGW+PGW and for the combination of all three signals. These values highlight the prominence of SIGWs within the PTA band, even in the absence of any enhancement of the primordial tensor power spectrum, whose investigation was, in any case, the primary aim of this work. PBH overproduction excludes almost all the $2\sigma$ regions of the parameter subspaces, especially in those related to an early-matter domination (the only signal whose cut is weaker is the purely scalar-induced one in~\cref{fig:sigw_posteriors}). However, we emphasize that these constraints are obtained by fixing all other parameters to their median values; therefore, this result does not imply that the signal cannot dominate the SGWB component in future detections with SKA.

One might wonder whether all SKA10 forecasts could instead be obtained using the Fisher Information Matrix (FIM), which is implemented in \texttt{fastPTA} and would be computationally cheaper. However, FIM method requires high SNR and approximately Gaussian posteriors. While the first condition is roughly satisfied, our posteriors are evidently non-Gaussian, so we have conservatively relied on a full MCMC analysis.

There are several directions in which our work can be improved. First, the inclusion of third-order contributions could impact the induced signal, and as a consequence our forecasts. Second, the issue of unphysical enhancements in the energy-density integral, which we cured adopting peaked spectra and a damping regularization function, still lacks an appropriate well-motivated theoretical explanation. Third, fully numerical Green functions that interpolate between eMD and RD epochs, as in SIGW studies with gradual transitions, could modify the detailed shape of the spectrum compared to the sudden-transition analytical matching. Finally, a more rigorous strategy for matching tensor fluctuations across different eras, in order to obtain the exact analogue of the ``poltergeist mechanism'' for scalar-tensor induced GWs, still requires dedicated investigation. We leave these developments for future works.

\acknowledgments

This study was financed in part by the Coordenação de Aperfeiçoamento de Pessoal de Nível Superior - Brasil (CAPES) – Finance Code  001. The authors thank Ilaria Caporali, Sasha Allegrini and Antonio Junior Iovino for valuable discussions and feedback on the draft.

\appendix

\section{Matching perturbations at reheating}\label{sec:matching}

We review here the analytical matching of scalar perturbations~\cite{Kohri:2018awv,Inomata:2019ivs,Inomata:2025wiv} and show that the same procedure does not apply to tensor modes.

The gravitational potential, which is constant during eMD epoch and then oscillates and decays in RD age, can be written by matching the transfer functions across the transition as
\begin{equation}
    \Phi_\mathbf{k}(x,x_\R)=\Phi_\mathbf{k}^\text{pr}\times\begin{cases}
        1&(x\le x_\R),\,\\
        A(x_\R)\mathcal{J}(x,x_\R)+B(x_\R)\mathcal{Y}(x,x_\R)&(x\ge x_\R)\,,
    \end{cases}
\end{equation}
with
\begin{equation}
    \mathcal{J}(x,x_\R)\equiv3\sqrt{3}\cdot \frac{j_1\left(\frac{x-x_\R/2}{\sqrt{3}}\right)}{x-x_\R/2}\,,\quad
    \mathcal{Y}(x,x_\R)\equiv 3\sqrt{3}\cdot \frac{y_1\left(\frac{x-x_\R/2}{\sqrt{3}}\right)}{x-x_\R/2}\,,
\end{equation}
and
\begin{equation}
    A(x_\R)=\frac{1}{\mathcal{J}(x,x_\R)-\frac{\mathcal{Y}(x,x_\R)}{\mathcal{Y}^\prime(x,x_\R)}\mathcal{J}^\prime(x,x_\R)}\Bigg\vert_{x\to x_\R}\,,\quad 
    B(x_\R)=-\frac{\mathcal{J}^\prime(x,x_\R)}{\mathcal{Y}^\prime(x,x_\R)}A(x_\R)\Bigg\vert_{x\to x_\R}\,.
\end{equation}
For tensor modes, we must instead solve the equation of motion in RD epoch starting at $\eta_\R$
\begin{equation}\label{eq:gamma_rd}
    \gamma^{(0)\prime\prime}_{\mathbf{k},\lambda}+2\mathcal{H}\gamma^{(0)\prime}_{\mathbf{k},\lambda}+k^2\gamma^{(0)}_{\mathbf{k},\lambda}=0\,,\quad (\eta>\eta_\R)\,,
\end{equation}
with initial conditions inherited from the preceding eMD phase
\begin{equation}
     \gamma^{(0)}_{\mathbf{k},\lambda}(\eta_\R)=3\frac{j_1(x_\R)}{x_\R}\,,\quad  \gamma^{(0)\prime}_{\mathbf{k},\lambda}(\eta_\R)=3\frac{\sin{x_\R}-3j_1(x_\R)}{x_\R^2}\,.
\end{equation}
Using the standard RD solution, we introduce the auxiliary variable $\xi=\eta-\eta_\R/2$. From~\cref{eq:a_Hubble_sudden_tr} we have $\mathcal{H}(\xi)=1/\xi$, so omitting subscripts and superscripts~\cref{eq:gamma_rd} becomes
\begin{equation}
    \gamma^{\prime\prime}+\frac{2}{\xi}\gamma^{\prime}+k^2\gamma=0\,.
\end{equation}
Defining $\chi\equiv k\xi=x-x_\R/2$, we can rewrite this as
\begin{equation}
    \chi^2\frac{\partial^2\gamma}{\partial \chi^2}+2\chi\frac{\partial\gamma}{\partial \chi}+\chi^2\gamma=0\,,
\end{equation}
whose general solution is
\begin{equation}
    \gamma = c_1(x_\R)j_0\left(x-\frac{x_\R}{2}\right)+c_2(x_\R)y_0\left(x-\frac{x_\R}{2}\right)\,.
\end{equation}
Imposing continuity of $\gamma$ at reheating yields
\begin{equation}
    c_1(x_\R)=3\frac{j_1(x_\R)}{x_\R}\,,\quad c_2(x_\R)=0\,.
\end{equation}
The second coefficient must be identically null, because $y_0(z)$ diverges at the origin, which prevents us from using it to enforce the continuity of the derivative. Therefore, the analytical matching procedure that works for scalar perturbations cannot be straightforwardly extended to tensor seeds.

To further confirm this, we can instead work with the auxiliary variable $h\equiv a\gamma$ (see e.g. sec.~(3.3) of~\cite{Gong:2019mui}). In a pure MD phase, the general solution for $h$ is a linear combination of $xj_1(x)$ and $xy_1(x)$. As $x\to0$, one has $xj_1(x)\to0$ but $xy_1(x)\to-\infty$, making it impossible to connect the perturbations across the two phases analytically without encountering divergences.

The above considerations are not sufficient, however, to claim that the signal generated immediately after reheating by the coupling of scalar and tensor perturbations is negligible. In the purely scalar-induced case, Kohri and Terada initially neglected this contribution~\cite{Kohri:2018awv}, but later work showed that the associated spectral density can exceed the originally considered term by several orders of magnitude~\cite{Inomata:2019ivs}. 

For scalar-tensor mixing, the situation is different: tensor modes begin to decay after horizon crossing in eMD epoch, while the gravitational potential remains constant. We therefore do not expect an enhancement as strong as in the scalar ``poltergeist mechanism''. Nevertheless, this expectation must be checked quantitatively, and modes re-entering just before the transition may provide an exception, since they have not decayed enough to be safely neglected. Moreover, a more realistic gradual reheating could also modify the result.

In light of this caveat, in this paper we have focused on the kernel $I_\eMD(x,u,v,x_\R)$ in~\cref{eq:I_sudden_tbc}. A proper treatment of the sudden eMD-RD transition for primordial tensor fluctuations --- and thus a fully consistent analogue of the ``poltergeist mechanism'' in the scalar-tensor case --- is left for future work.

\section{Varying the shape parameters of the input spectra}\label{sec:varying_shape_pars}

%
\begin{figure}[t!]
    \centering
    \includegraphics[width=0.45\textwidth]{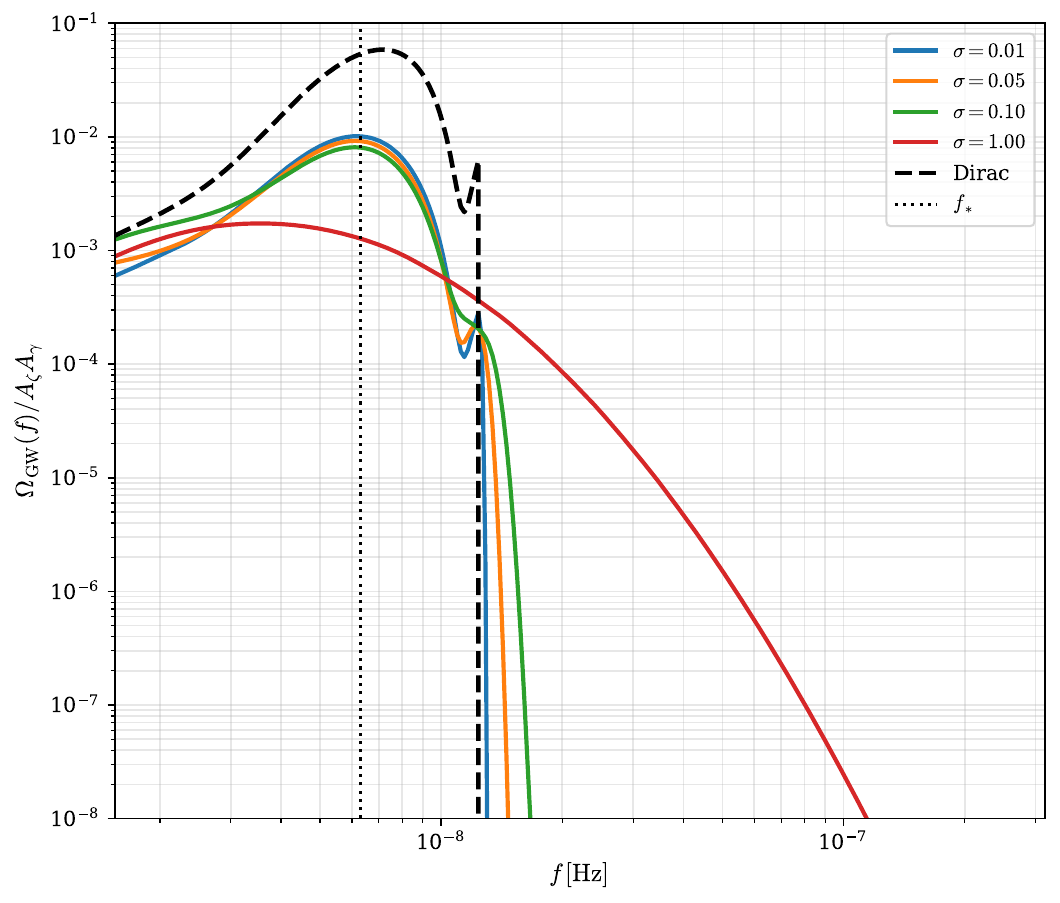}
    \includegraphics[width=0.45\textwidth]{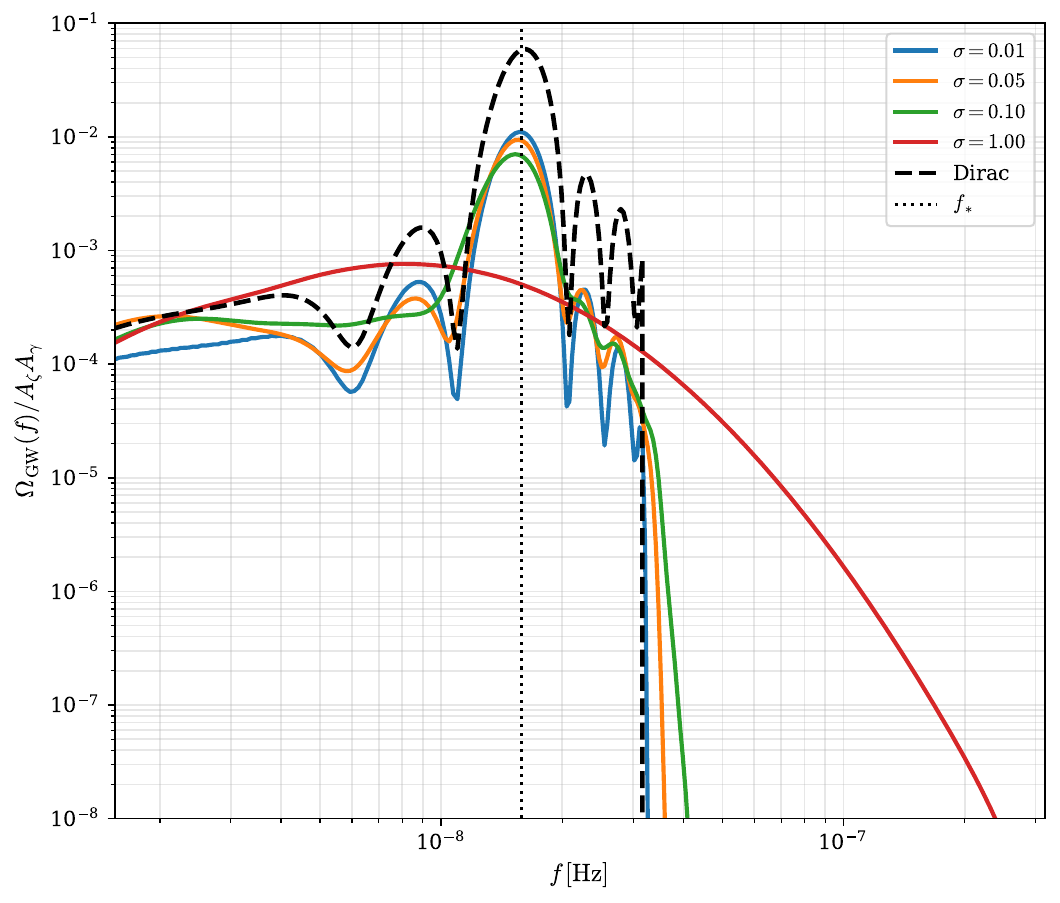}
    \includegraphics[width=0.45\textwidth]{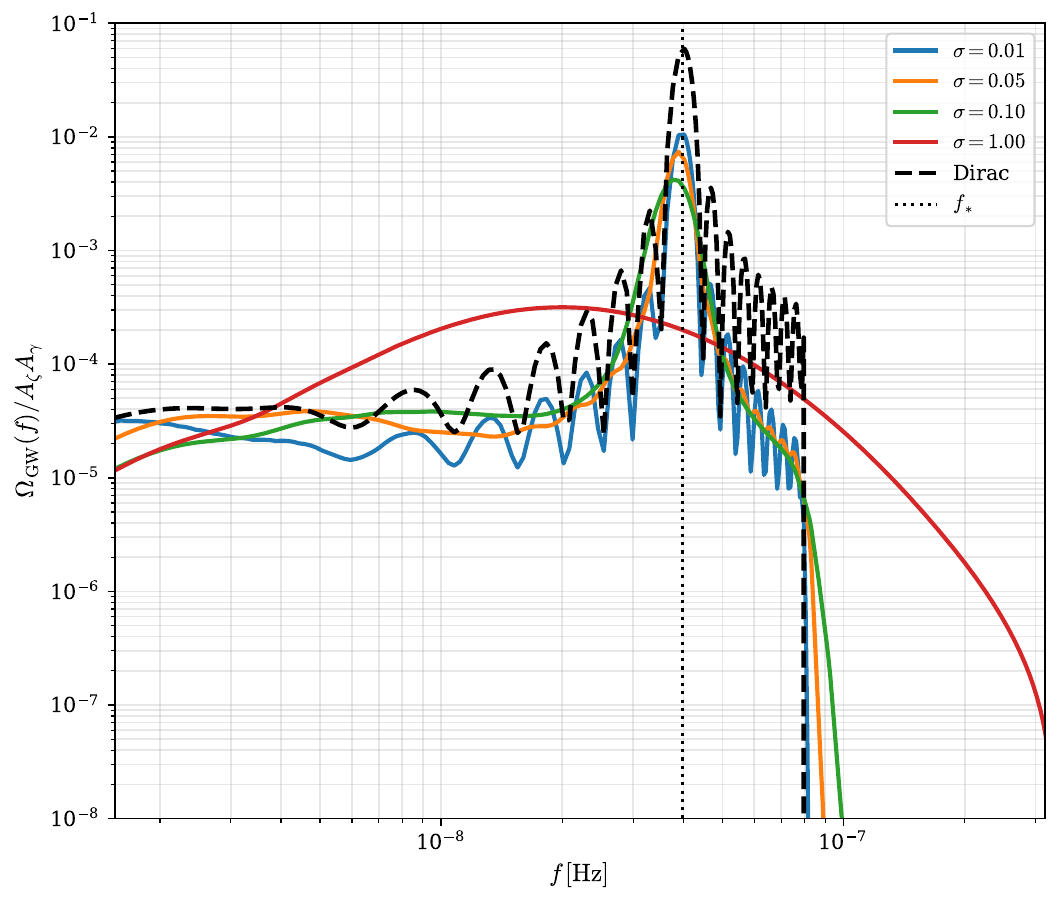}
    \includegraphics[width=0.45\textwidth]{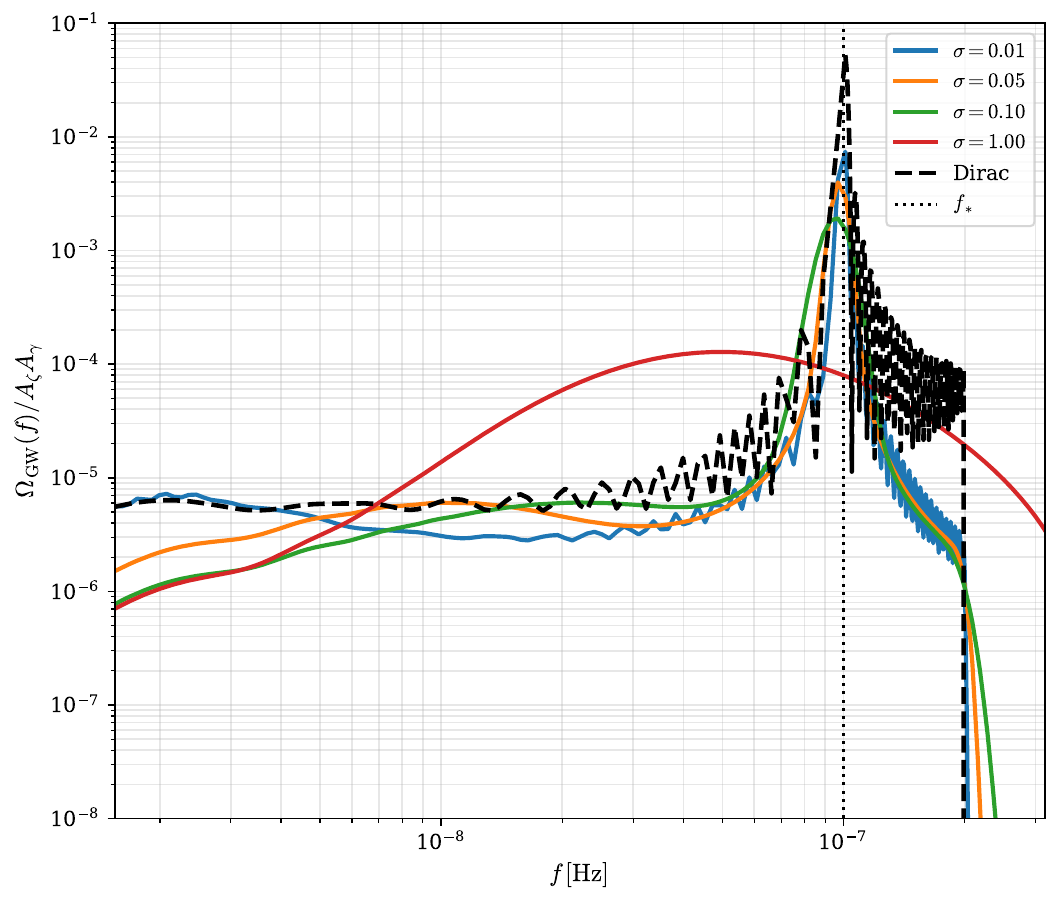}
    \caption{STGWs generated during early matter domination by log-normal and monochromatic input power spectra, as a function of $f$, for some values of peak frequency and width.}
    \label{fig:shape_pars}
\end{figure}
The scalar-tensor signal generated during early matter domination, from peaked primordial power spectra depends sensitively on the values of the shape parameters.~\Cref{fig:shape_pars} illustrates how this dependence can drastically modify the profile of the (amplitude-normalized) GW energy density. Throughout, we adopt the monochromatic and log-normal input templates of~\cref{eq:dirac_input,eq:input_spectra}, consistent with the rest of the paper.

In many induced-GW scenarios, the energy density is invariant under a shift of the peak frequency, so that plots are typically shown as $\Omega_\GW$ versus $k/k_*$ or $f/f_*$. In our case this invariance is broken. The reheating time $\eta_\R$ introduces an explicit numerical scale that strongly affects the kernel in~\cref{eq:input_spectra}; even after oscillation averaging, sines and cosines of $x_\R$ and $vx_\R$ remain so their relative weight in the numerical integration of~\cref{eq:general_Omega_reg} is directly influenced by the choice of $f_*$.

In particular, shifting the peak frequency to higher values increases the number of visible oscillations sourced by the kernel in the monochromatic limit, where $v=k_*/k$. For the Dirac-delta power spectra in~\cref{eq:omega_st_emd_dirac}, the square-bracket term (a purely kinematical factor from polarization contractions) grows with $k$ up to the peak and is then abruptly cut at $k=2k_*$. The kernel superimposes its characteristic oscillatory features on this envelope, as seen in~\cref{fig:shape_pars}. For narrow widths, the fully numerical grid integration reproduces the analytical monochromatic behavior. These oscillations are strongly washed out when broader widths are considered, because the log-normal profiles smooth the kernel-induced structure - precisely the regime preferred by the Bayesian analyses performed in this work.

\section{SIGWs from a radiation- and early matter-dominated universe}\label{sec:sigw}

%
\begin{figure}
    \centering
    \includegraphics[width=0.32\textwidth]{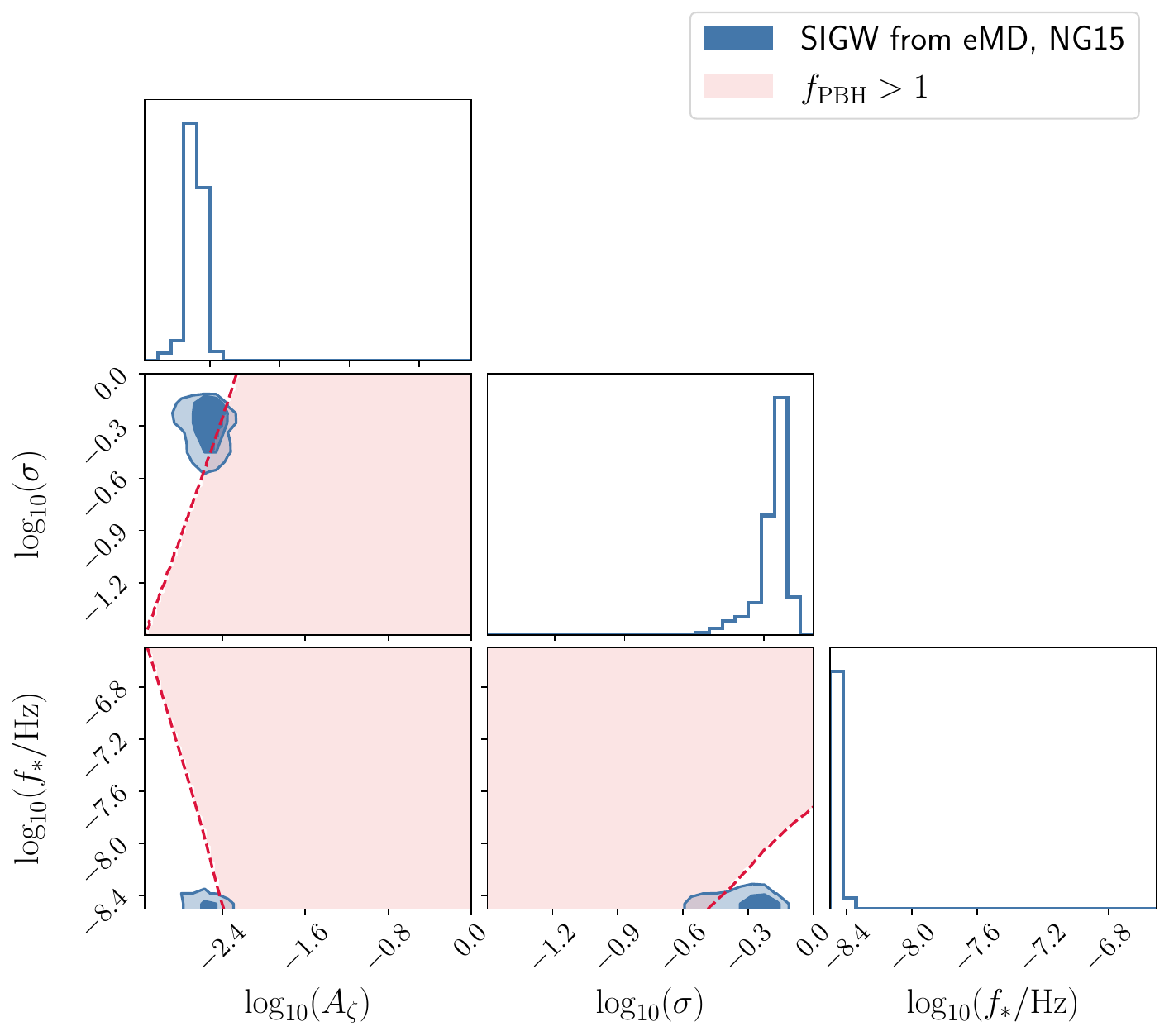}
    \includegraphics[width=0.32\textwidth]{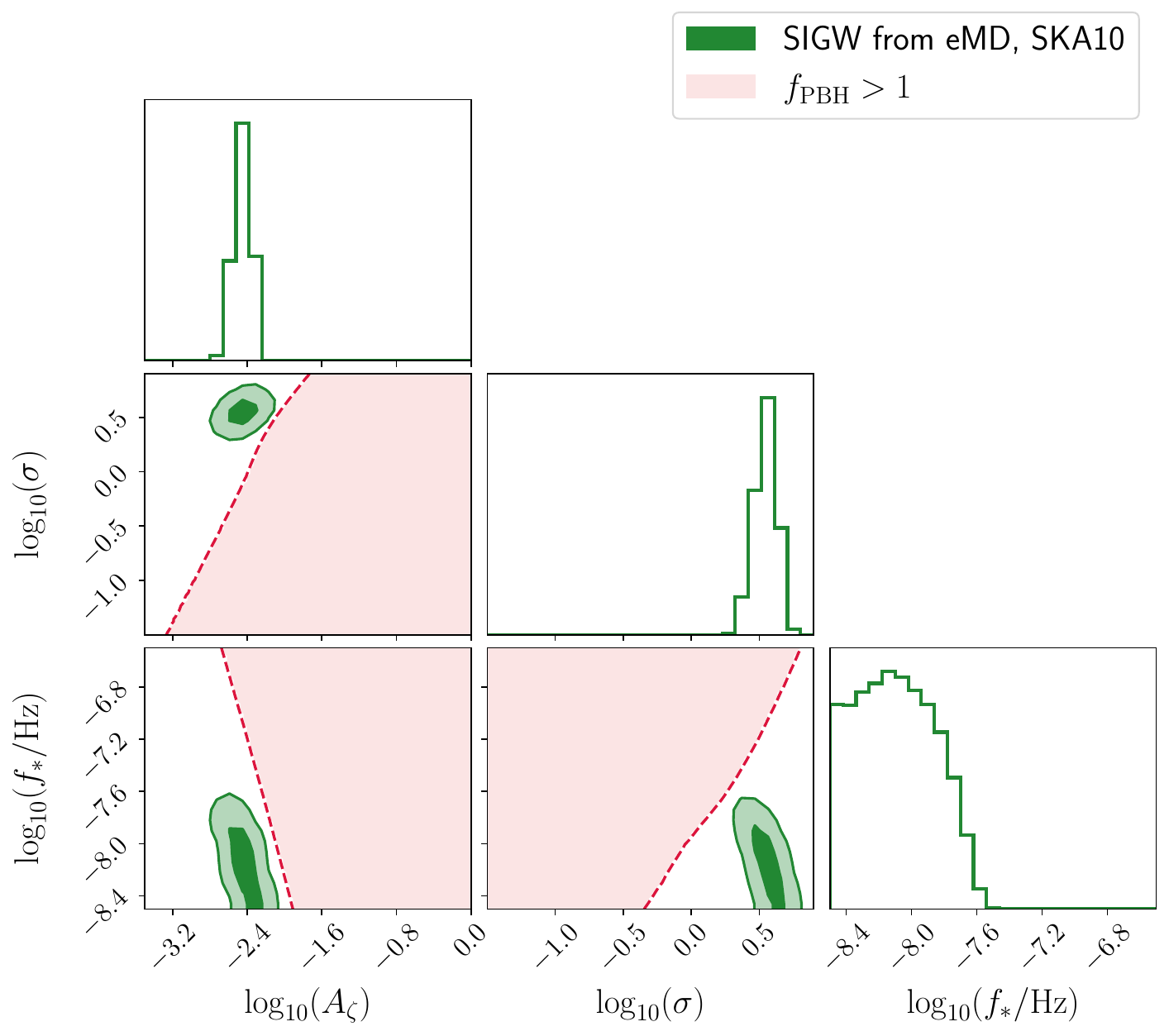}
    \includegraphics[width=0.32\textwidth]{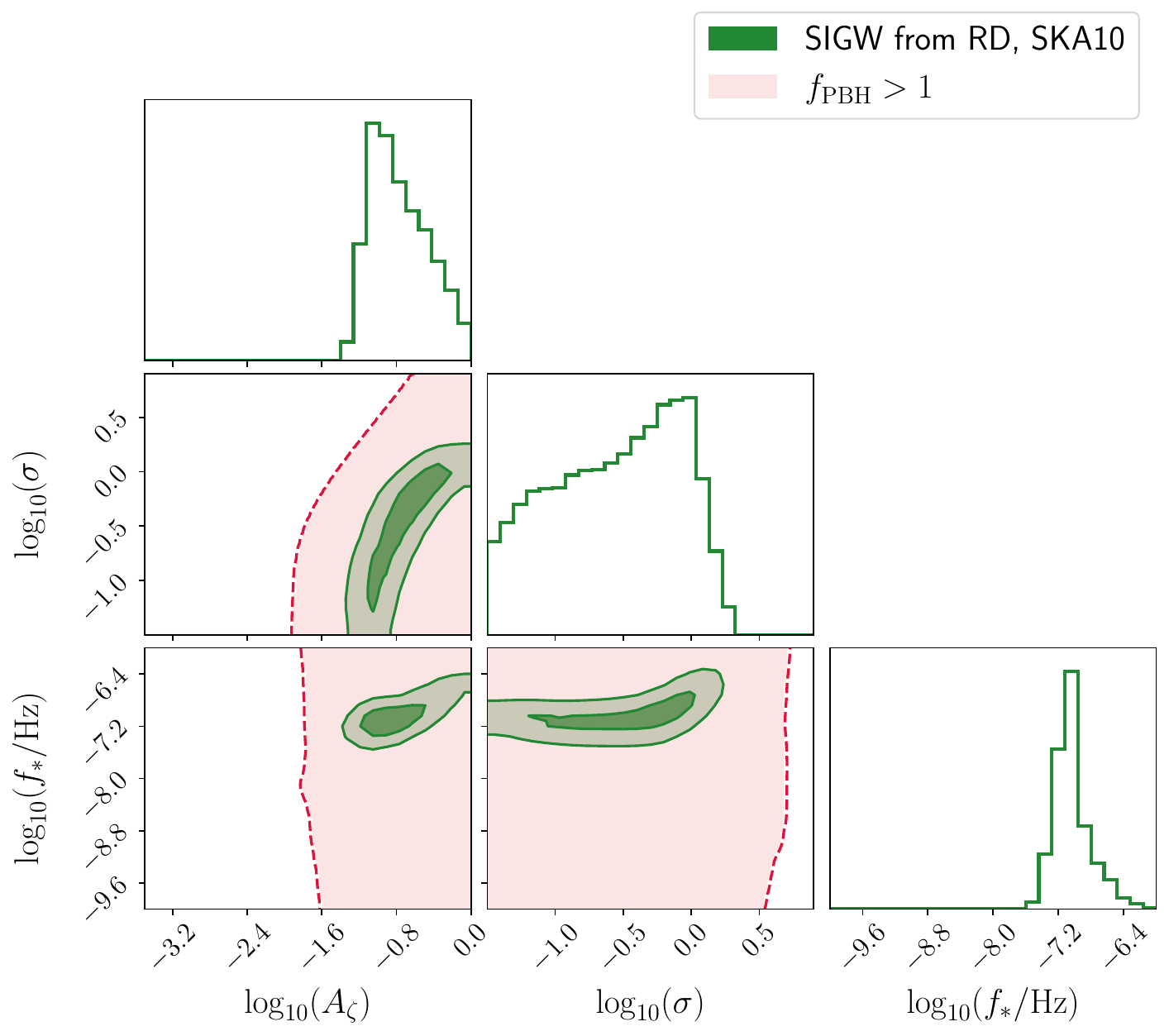}
    \caption{Left panel: Bayesian MCMC parameter estimation obtained from NG15 dataset, using SIGWs generated during the eMD epoch ended at $\eta_\R=2\times10^{-6}\,\rm{Mpc}$. Central/right panel: Bayesian estimation using SKA-like configuration with $T_\text{obs}=10.33\,\rm{yr}$ and $N_p=200$, injecting SIGWs generated during eMD (central) and RD (right) eras, from primordial lognormal power spectra.}
    \label{fig:sigw_posteriors}
\end{figure}
In this appendix, we briefly provide the analytical formulas used for the computation of the SIGWs coming from RD epoch and from the sudden eMD-RD transition, as in Refs.~\cite{Inomata:2019ivs,Inomata:2025wiv}. The numerical integration is performed using the software package~\href{https://github.com/jonaselgammal/SIGWAY/tree/main}{\texttt{SIGWAY}}~\cite{LISACosmologyWorkingGroup:2025vdz}. The energy density in~\cref{eq:Omega} can be conveniently reshaped for the scalar-scalar case as
\begin{equation}\label{eq:Omega_ss}
    \begin{split}
        \Omega_\GW^{\scsc}(k)&=\frac{1}{12}\left(\frac{3(1+w)}{5+3w}\right)^4\left(\frac{k}{\mathcal{H}}\right)^2\int_0^\infty dt\int_{-1}^1 ds\left[\frac{t(2+t)(s^2-1)}{(1-s+t)(1+s+t)}\right]^2\\
        &\times \overline{I^2(t,s,x)}\Delta^2_\Phi\left(k\frac{t+s+1}{2}\right)\Delta^2_\Phi\left(k\frac{t-s+1}{2}\right)\,,
    \end{split}
\end{equation}
where the $(t,s)$ coordinates are given by the following rotation:
\begin{equation}
    t\equiv u+v-1\,,\quad s\equiv u-v\,.
\end{equation}
The RD case reads
\begin{equation}
    \overline{\mathcal{I}^2_\RD(t,s,x\to\infty)}\equiv x^2\cdot \overline{{I}^2_\RD(t,s,x\to\infty)}=\frac{I_A^2(u,v)}{2}\left[I^2_B(u,v)+I_C^2(u,v)\right]\,,
\end{equation}
where
\begin{equation}
    \begin{split}
        I_A(u,v)&= \frac{3\left(u^2+v^2-3\right)}{4u^3v^3}\,,\\
        I_B(u,v) &= -4uv+\left(u^2+v^2-3\right)\log\left|\frac{3-(u+v)^2}{3-(u-v)^2}\right|\,,\\
        I_C(u,v) &= \pi\left(u^2+v^2-3\right)\Theta\left(u+v-\sqrt{3}\right)\,.
    \end{split}
\end{equation}
Instead, under the assumption of a sudden transition, the main contributions to the kernel come from the resonance and the large scale components (from here on, $\mathcal{I}\equiv(x-x_\R/2)I$)
\begin{equation}
    \overline{\mathcal{I}^2(t,s,x_\R)}\simeq\overline{\mathcal{I}^2_\res(t\approx t_0,s,x_\R)}+\overline{\mathcal{I}^2_\LS(t\gg1,s\simeq0,x_\R)}\,,
\end{equation}
The resonance component corresponds to $k\approx k_{\max}$, so $t\approx t_0\equiv\sqrt{3}-1$. It reads
\begin{equation}
    \overline{\mathcal{I}^2_\res(t\approx t_0,s,x_\R)}\simeq  \frac{9\left(-5+s^2+2t+t^2\right)^4x_\R^8}{2^{17}5^4(1-s+t)^2(1+s+t)^2}\Ci^2(y)\,,
\end{equation}
where $y\equiv \left|t+1-\sqrt{3}\right|x_\R/2\sqrt{3}$. A fudge factor, $Y \simeq 2.3$, sometimes appears in the literature and should be included in the calculation of the energy density to absorb the uncertainty in the integration boundary over the $y$ (or, equivalently, $t$) coordinate.

The large-scale component corresponds to $u\sim v\sim t \gg1$ and thus $s\simeq0$. Its expression is given by
\begin{equation}
    \overline{\mathcal{I}^2_\LS(t\gg1,s\simeq0,x_\R)}\simeq\frac{9t^4x_\R^8}{2^{17}5^4}\left[4\Ci^2(x_\R/2)+\left(\pi-2\Si(x_\R/2)\right)^2\right]\,.
\end{equation}
Since~\cref{eq:Omega_ss} is an even function of $s$, the integration region can be expressed as
\begin{equation}
    0\leq s\leq 1\,,\quad 0\leq t \leq -s+2\frac{k_{\max}}{k}-1\,,
\end{equation}
for $k\leq k_{\max}$ (when $k>k_{\max}$, it changes to $0\leq 2k/k_{\max}-1$), and the result of the integral is then doubled to keep the parity into account.

\Cref{fig:sigw_posteriors} shows the results of the same MCMC analysis described in~\cref{sec:stgw_nhz}, but with only the scalar-induced signal injected. In the eMD case, the resulting posteriors exhibit similar 1–2$\sigma$ credible regions for NG15 and SKA10. For the RD case, the amplitude is pushed toward high values, as expected given the absence of a poltergeist mechanism in pure radiation domination. Finally, we note that PBH overproduction severely restricts the allowed parameter space in the RD results. Instead, the eMD chains show 2$\sigma$ regions whose PBH restrictions are much weaker.

When sampling SIGWs alone, the SNR for the eMD case against the mock SKA dataset is $\sim97$, while it reaches $\sim 33$ for SIGWs generated during radiation domination.

\bibliographystyle{JHEP}
\bibliography{bibliography.bib}

@article{NANOGrav:2023gor,
    author = "Agazie, Gabriella and others",
    collaboration = "NANOGrav",
    title = "{The NANOGrav 15 yr Data Set: Evidence for a Gravitational-wave Background}",
    eprint = "2306.16213",
    archivePrefix = "arXiv",
    primaryClass = "astro-ph.HE",
    doi = "10.3847/2041-8213/acdac6",
    journal = "Astrophys. J. Lett.",
    volume = "951",
    number = "1",
    pages = "L8",
    year = "2023"
}

@article{Das:2023nmm,
    author = "Das, Barnali and Jaman, Nur and Sami, M.",
    title = "{Gravitational wave background from quintessential inflation and NANOGrav data}",
    eprint = "2307.12913",
    archivePrefix = "arXiv",
    primaryClass = "gr-qc",
    doi = "10.1103/PhysRevD.108.103510",
    journal = "Phys. Rev. D",
    volume = "108",
    number = "10",
    pages = "103510",
    year = "2023"
}

@article{Liu:2023ymk,
    author = "Liu, Lang and Chen, Zu-Cheng and Huang, Qing-Guo",
    title = "{Implications for the non-Gaussianity of curvature perturbation from pulsar timing arrays}",
    eprint = "2307.01102",
    archivePrefix = "arXiv",
    primaryClass = "astro-ph.CO",
    doi = "10.1103/PhysRevD.109.L061301",
    journal = "Phys. Rev. D",
    volume = "109",
    number = "6",
    pages = "L061301",
    year = "2024"
}

@article{Liu:2023pau,
    author = "Liu, Lang and Chen, Zu-Cheng and Huang, Qing-Guo",
    title = "{Probing the equation of state of the early Universe with pulsar timing arrays}",
    eprint = "2307.14911",
    archivePrefix = "arXiv",
    primaryClass = "astro-ph.CO",
    doi = "10.1088/1475-7516/2023/11/071",
    journal = "JCAP",
    volume = "11",
    pages = "071",
    year = "2023"
}

@article{Liu:2023hpw,
    author = "Liu, Lang and Wu, You and Chen, Zu-Cheng",
    title = "{Simultaneously probing the sound speed and equation of state of the early Universe with pulsar timing arrays}",
    eprint = "2310.16500",
    archivePrefix = "arXiv",
    primaryClass = "astro-ph.CO",
    doi = "10.1088/1475-7516/2024/04/011",
    journal = "JCAP",
    volume = "04",
    pages = "011",
    year = "2024"
}

@article{Domenech:2020kqm,
    author = "Dom{\`e}nech, Guillem and Pi, Shi and Sasaki, Misao",
    title = "{Induced gravitational waves as a probe of thermal history of the universe}",
    eprint = "2005.12314",
    archivePrefix = "arXiv",
    primaryClass = "gr-qc",
    reportNumber = "YITP-20-70, IPMU20-0053",
    doi = "10.1088/1475-7516/2020/08/017",
    journal = "JCAP",
    volume = "08",
    pages = "017",
    year = "2020"
}

@article{Balaji:2023ehk,
    author = "Balaji, Shyam and Dom{\`e}nech, Guillem and Franciolini, Gabriele",
    title = "{Scalar-induced gravitational wave interpretation of PTA data: the role of scalar fluctuation propagation speed}",
    eprint = "2307.08552",
    archivePrefix = "arXiv",
    primaryClass = "gr-qc",
    doi = "10.1088/1475-7516/2023/10/041",
    journal = "JCAP",
    volume = "10",
    pages = "041",
    year = "2023"
}

@article{EPTA:2023sfo,
    author = "Antoniadis, J. and others",
    collaboration = "EPTA",
    title = "{The second data release from the European Pulsar Timing Array - I. The dataset and timing analysis}",
    eprint = "2306.16224",
    archivePrefix = "arXiv",
    primaryClass = "astro-ph.HE",
    doi = "10.1051/0004-6361/202346841",
    journal = "Astron. Astrophys.",
    volume = "678",
    pages = "A48",
    year = "2023"
}

@article{EPTA:2023fyk,
    author = "Antoniadis, J. and others",
    collaboration = "EPTA, InPTA:",
    title = "{The second data release from the European Pulsar Timing Array - III. Search for gravitational wave signals}",
    eprint = "2306.16214",
    archivePrefix = "arXiv",
    primaryClass = "astro-ph.HE",
    doi = "10.1051/0004-6361/202346844",
    journal = "Astron. Astrophys.",
    volume = "678",
    pages = "A50",
    year = "2023"
}

@article{EPTA:2023xxk,
    author = "Antoniadis, J. and others",
    collaboration = "EPTA, InPTA",
    title = "{The second data release from the European Pulsar Timing Array - IV. Implications for massive black holes, dark matter, and the early Universe}",
    eprint = "2306.16227",
    archivePrefix = "arXiv",
    primaryClass = "astro-ph.CO",
    doi = "10.1051/0004-6361/202347433",
    journal = "Astron. Astrophys.",
    volume = "685",
    pages = "A94",
    year = "2024"
}

@article{Xu:2023wog,
    author = "Xu, Heng and others",
    title = "{Searching for the Nano-Hertz Stochastic Gravitational Wave Background with the Chinese Pulsar Timing Array Data Release I}",
    eprint = "2306.16216",
    archivePrefix = "arXiv",
    primaryClass = "astro-ph.HE",
    doi = "10.1088/1674-4527/acdfa5",
    journal = "Res. Astron. Astrophys.",
    volume = "23",
    number = "7",
    pages = "075024",
    year = "2023"
}

@article{Reardon:2023gzh,
    author = "Reardon, Daniel J. and others",
    title = "{Search for an Isotropic Gravitational-wave Background with the Parkes Pulsar Timing Array}",
    eprint = "2306.16215",
    archivePrefix = "arXiv",
    primaryClass = "astro-ph.HE",
    doi = "10.3847/2041-8213/acdd02",
    journal = "Astrophys. J. Lett.",
    volume = "951",
    number = "1",
    pages = "L6",
    year = "2023"
}

@article{Zic:2023gta,
    author = "Zic, Andrew and others",
    title = "{The Parkes Pulsar Timing Array third data release}",
    eprint = "2306.16230",
    archivePrefix = "arXiv",
    primaryClass = "astro-ph.HE",
    doi = "10.1017/pasa.2023.36",
    journal = "Publ. Astron. Soc. Austral.",
    volume = "40",
    pages = "e049",
    year = "2023"
}

@article{Agazie:2026tui,
    author = "Agazie, Gabriella and others",
    title = "{The NANOGrav 15 yr Data Set: Piecewise Power-Law Reconstruction of the Gravitational-Wave Background}",
    eprint = "2601.09481",
    archivePrefix = "arXiv",
    primaryClass = "astro-ph.HE",
    month = "1",
    year = "2026"
}

@article{NANOGrav:2023pdq,
    author = "Agazie, Gabriella and others",
    collaboration = "NANOGrav",
    title = "{The NANOGrav 15 yr Data Set: Bayesian Limits on Gravitational Waves from Individual Supermassive Black Hole Binaries}",
    eprint = "2306.16222",
    archivePrefix = "arXiv",
    primaryClass = "astro-ph.HE",
    doi = "10.3847/2041-8213/ace18a",
    journal = "Astrophys. J. Lett.",
    volume = "951",
    number = "2",
    pages = "L50",
    year = "2023"
}

@article{NANOGrav:2025gqp,
    author = "Agarwal, Nikita and others",
    collaboration = "NANOGrav",
    title = "{The NANOGrav 15 yr Dataset: Targeted Searches for Supermassive Black Hole Binaries}",
    eprint = "2508.16534",
    archivePrefix = "arXiv",
    primaryClass = "astro-ph.HE",
    doi = "10.3847/2041-8213/ae3719",
    journal = "Astrophys. J. Lett.",
    volume = "998",
    number = "1",
    pages = "L11",
    year = "2026"
}

@article{NANOGrav:2023hvm,
    author = "Afzal, Adeela and others",
    collaboration = "NANOGrav",
    title = "{The NANOGrav 15 yr Data Set: Search for Signals from New Physics}",
    eprint = "2306.16219",
    archivePrefix = "arXiv",
    primaryClass = "astro-ph.HE",
    reportNumber = "FERMILAB-PUB-23-589-T",
    doi = "10.3847/2041-8213/acdc91",
    journal = "Astrophys. J. Lett.",
    volume = "951",
    number = "1",
    pages = "L11",
    year = "2023",
    note = "[Erratum: Astrophys.J.Lett. 971, L27 (2024), Erratum: Astrophys.J. 971, L27 (2024)]"
}

@article{Caprini:2018mtu,
    author = "Caprini, Chiara and Figueroa, Daniel G.",
    title = "{Cosmological Backgrounds of Gravitational Waves}",
    eprint = "1801.04268",
    archivePrefix = "arXiv",
    primaryClass = "astro-ph.CO",
    doi = "10.1088/1361-6382/aac608",
    journal = "Class. Quant. Grav.",
    volume = "35",
    number = "16",
    pages = "163001",
    year = "2018"
}

@article{Grishchuk:1974ny,
    author = "Grishchuk, L. P.",
    title = "{Amplification of gravitational waves in an isotropic universe}",
    journal = "Sov. Phys. JETP",
    volume = "40",
    number = "3",
    pages = "409--415",
    year = "1975"
}

@article{Starobinsky:1979ty,
    author = "Starobinsky, Alexei A.",
    editor = "Khalatnikov, I. M. and Mineev, V. P.",
    title = "{Spectrum of relict gravitational radiation and the early state of the universe}",
    journal = "JETP Lett.",
    volume = "30",
    pages = "682--685",
    year = "1979"
}

@article{Rubakov:1982df,
    author = "Rubakov, V. A. and Sazhin, M. V. and Veryaskin, A. V.",
    title = "{Graviton Creation in the Inflationary Universe and the Grand Unification Scale}",
    doi = "10.1016/0370-2693(82)90641-4",
    journal = "Phys. Lett. B",
    volume = "115",
    pages = "189--192",
    year = "1982"
}

@article{Fabbri:1983us,
    author = "Fabbri, R. and Pollock, M. d.",
    title = "{The Effect of Primordially Produced Gravitons upon the Anisotropy of the Cosmological Microwave Background Radiation}",
    doi = "10.1016/0370-2693(83)91322-9",
    journal = "Phys. Lett. B",
    volume = "125",
    pages = "445--448",
    year = "1983"
}

@article{Tomita:1967wkp,
    author = "Tomita, Kenji",
    title = "{Non-Linear Theory of Gravitational Instability in the Expanding Universe}",
    doi = "10.1143/PTP.37.831",
    journal = "Prog. Theor. Phys.",
    volume = "37",
    number = "5",
    pages = "831--846",
    year = "1967"
}

@article{Matarrese:1994wa,
    author = "Matarrese, Sabino and Pantano, Ornella and Saez, Diego",
    title = "{A Relativistic approach to gravitational instability in the expanding Universe: Second order Lagrangian solutions}",
    eprint = "astro-ph/9403032",
    archivePrefix = "arXiv",
    reportNumber = "DFPD-94-A-18",
    doi = "10.1093/mnras/271.3.513",
    journal = "Mon. Not. Roy. Astron. Soc.",
    volume = "271",
    pages = "513--522",
    year = "1994"
}

@article{Matarrese:1993zf,
    author = "Matarrese, Sabino and Pantano, Ornella and Saez, Diego",
    title = "{General relativistic dynamics of irrotational dust: Cosmological implications}",
    eprint = "astro-ph/9310036",
    archivePrefix = "arXiv",
    reportNumber = "DFPD-93-A-67",
    doi = "10.1103/PhysRevLett.72.320",
    journal = "Phys. Rev. Lett.",
    volume = "72",
    pages = "320--323",
    year = "1994"
}

@article{Matarrese:1997ay,
    author = "Matarrese, Sabino and Mollerach, Silvia and Bruni, Marco",
    title = "{Second order perturbations of the Einstein-de Sitter universe}",
    eprint = "astro-ph/9707278",
    archivePrefix = "arXiv",
    reportNumber = "SISSA-83-97-A",
    doi = "10.1103/PhysRevD.58.043504",
    journal = "Phys. Rev. D",
    volume = "58",
    pages = "043504",
    year = "1998"
}

@article{Kamionkowski:1993fg,
    author = "Kamionkowski, Marc and Kosowsky, Arthur and Turner, Michael S.",
    title = "{Gravitational radiation from first order phase transitions}",
    eprint = "astro-ph/9310044",
    archivePrefix = "arXiv",
    reportNumber = "IASSNS-HEP-93-44, FERMILAB-PUB-93-235-A",
    doi = "10.1103/PhysRevD.49.2837",
    journal = "Phys. Rev. D",
    volume = "49",
    pages = "2837--2851",
    year = "1994"
}

@article{Caprini:2007xq,
    author = "Caprini, Chiara and Durrer, Ruth and Servant, Geraldine",
    title = "{Gravitational wave generation from bubble collisions in first-order phase transitions: An analytic approach}",
    eprint = "0711.2593",
    archivePrefix = "arXiv",
    primaryClass = "astro-ph",
    reportNumber = "CERN-PH-TH-2007-206, SACLAY-T07-142",
    doi = "10.1103/PhysRevD.77.124015",
    journal = "Phys. Rev. D",
    volume = "77",
    pages = "124015",
    year = "2008"
}

@article{Hindmarsh:2017gnf,
    author = "Hindmarsh, Mark and Huber, Stephan J. and Rummukainen, Kari and Weir, David J.",
    title = "{Shape of the acoustic gravitational wave power spectrum from a first order phase transition}",
    eprint = "1704.05871",
    archivePrefix = "arXiv",
    primaryClass = "astro-ph.CO",
    reportNumber = "HIP-2017-02-TH, HIP-2017-02/TH",
    doi = "10.1103/PhysRevD.96.103520",
    journal = "Phys. Rev. D",
    volume = "96",
    number = "10",
    pages = "103520",
    year = "2017",
    note = "[Erratum: Phys.Rev.D 101, 089902 (2020)]"
}

@article{Cutting:2018tjt,
    author = "Cutting, Daniel and Hindmarsh, Mark and Weir, David J.",
    title = "{Gravitational waves from vacuum first-order phase transitions: from the envelope to the lattice}",
    eprint = "1802.05712",
    archivePrefix = "arXiv",
    primaryClass = "astro-ph.CO",
    reportNumber = "HIP-2018-4-TH",
    doi = "10.1103/PhysRevD.97.123513",
    journal = "Phys. Rev. D",
    volume = "97",
    number = "12",
    pages = "123513",
    year = "2018"
}

@article{Cutting:2019zws,
    author = "Cutting, Daniel and Hindmarsh, Mark and Weir, David J.",
    title = "{Vorticity, kinetic energy, and suppressed gravitational wave production in strong first order phase transitions}",
    eprint = "1906.00480",
    archivePrefix = "arXiv",
    primaryClass = "hep-ph",
    reportNumber = "HIP-2019-15/TH",
    doi = "10.1103/PhysRevLett.125.021302",
    journal = "Phys. Rev. Lett.",
    volume = "125",
    number = "2",
    pages = "021302",
    year = "2020"
}

@article{RoperPol:2019wvy,
    author = "Roper Pol, Alberto and Mandal, Sayan and Brandenburg, Axel and Kahniashvili, Tina and Kosowsky, Arthur",
    title = "{Numerical simulations of gravitational waves from early-universe turbulence}",
    eprint = "1903.08585",
    archivePrefix = "arXiv",
    primaryClass = "astro-ph.CO",
    reportNumber = "NORDITA-2019-024",
    doi = "10.1103/PhysRevD.102.083512",
    journal = "Phys. Rev. D",
    volume = "102",
    number = "8",
    pages = "083512",
    year = "2020"
}

@article{Caprini:2019egz,
    author = "Caprini, Chiara and others",
    title = "{Detecting gravitational waves from cosmological phase transitions with LISA: an update}",
    eprint = "1910.13125",
    archivePrefix = "arXiv",
    primaryClass = "astro-ph.CO",
    reportNumber = "DESY-19-159, IPPP/19/27, HIP-2019-14/TH, MITP/19-066, IFT-UAM/CSIC-19-139",
    doi = "10.1088/1475-7516/2020/03/024",
    journal = "JCAP",
    volume = "03",
    pages = "024",
    year = "2020"
}

@article{Cutting:2020nla,
    author = "Cutting, Daniel and Escartin, Elba Granados and Hindmarsh, Mark and Weir, David J.",
    title = "{Gravitational waves from vacuum first order phase transitions II: from thin to thick walls}",
    eprint = "2005.13537",
    archivePrefix = "arXiv",
    primaryClass = "astro-ph.CO",
    reportNumber = "HIP-2020-13/TH",
    doi = "10.1103/PhysRevD.103.023531",
    journal = "Phys. Rev. D",
    volume = "103",
    number = "2",
    pages = "023531",
    year = "2021"
}

@article{Han:2023olf,
    author = "Han, Chengcheng and Xie, Ke-Pan and Yang, Jin Min and Zhang, Mengchao",
    title = "{Self-interacting dark matter implied by nano-Hertz gravitational waves}",
    eprint = "2306.16966",
    archivePrefix = "arXiv",
    primaryClass = "hep-ph",
    doi = "10.1103/PhysRevD.109.115025",
    journal = "Phys. Rev. D",
    volume = "109",
    number = "11",
    pages = "115025",
    year = "2024"
}

@article{Ashoorioon:2022raz,
    author = "Ashoorioon, Amjad and Rezazadeh, Kazem and Rostami, Abasalt",
    title = "{NANOGrav signal from the end of inflation and the LIGO mass and heavier primordial black holes}",
    eprint = "2202.01131",
    archivePrefix = "arXiv",
    primaryClass = "astro-ph.CO",
    reportNumber = "IPM/P-2022/06",
    doi = "10.1016/j.physletb.2022.137542",
    journal = "Phys. Lett. B",
    volume = "835",
    pages = "137542",
    year = "2022"
}

@article{Athron:2023mer,
    author = "Athron, Peter and Fowlie, Andrew and Lu, Chih-Ting and Morris, Lachlan and Wu, Lei and Wu, Yongcheng and Xu, Zhongxiu",
    title = "{Can Supercooled Phase Transitions Explain the Gravitational Wave Background Observed by Pulsar Timing Arrays?}",
    eprint = "2306.17239",
    archivePrefix = "arXiv",
    primaryClass = "hep-ph",
    doi = "10.1103/PhysRevLett.132.221001",
    journal = "Phys. Rev. Lett.",
    volume = "132",
    number = "22",
    pages = "221001",
    year = "2024"
}

@article{Li:2023yaj,
    author = "Li, Yao-Yu and Zhang, Chi and Wang, Ziwei and Cui, Ming-Yang and Tsai, Yue-Lin Sming and Yuan, Qiang and Fan, Yi-Zhong",
    title = "{Primordial magnetic field as a common solution of nanohertz gravitational waves and the Hubble tension}",
    eprint = "2306.17124",
    archivePrefix = "arXiv",
    primaryClass = "astro-ph.HE",
    doi = "10.1103/PhysRevD.109.043538",
    journal = "Phys. Rev. D",
    volume = "109",
    number = "4",
    pages = "043538",
    year = "2024"
}

@article{Vachaspati:1984gt,
    author = "Vachaspati, Tanmay and Vilenkin, Alexander",
    title = "{Gravitational Radiation from Cosmic Strings}",
    reportNumber = "HUTP-84/A065",
    doi = "10.1103/PhysRevD.31.3052",
    journal = "Phys. Rev. D",
    volume = "31",
    pages = "3052",
    year = "1985"
}

@article{Sakellariadou:1990ne,
    author = "Sakellariadou, M.",
    title = "{Gravitational waves emitted from infinite strings}",
    doi = "10.1103/PhysRevD.42.354",
    journal = "Phys. Rev. D",
    volume = "42",
    pages = "354--360",
    year = "1990",
    note = "[Erratum: Phys.Rev.D 43, 4150 (1991)]"
}

@article{Damour:2000wa,
    author = "Damour, Thibault and Vilenkin, Alexander",
    title = "{Gravitational wave bursts from cosmic strings}",
    eprint = "gr-qc/0004075",
    archivePrefix = "arXiv",
    reportNumber = "IHES-P-00-32",
    doi = "10.1103/PhysRevLett.85.3761",
    journal = "Phys. Rev. Lett.",
    volume = "85",
    pages = "3761--3764",
    year = "2000"
}

@article{Damour:2001bk,
    author = "Damour, Thibault and Vilenkin, Alexander",
    title = "{Gravitational wave bursts from cusps and kinks on cosmic strings}",
    eprint = "gr-qc/0104026",
    archivePrefix = "arXiv",
    reportNumber = "IHES-P-01-15",
    doi = "10.1103/PhysRevD.64.064008",
    journal = "Phys. Rev. D",
    volume = "64",
    pages = "064008",
    year = "2001"
}

@article{Damour:2004kw,
    author = "Damour, Thibault and Vilenkin, Alexander",
    title = "{Gravitational radiation from cosmic (super)strings: Bursts, stochastic background, and observational windows}",
    eprint = "hep-th/0410222",
    archivePrefix = "arXiv",
    doi = "10.1103/PhysRevD.71.063510",
    journal = "Phys. Rev. D",
    volume = "71",
    pages = "063510",
    year = "2005"
}

@article{Figueroa:2012kw,
    author = "Figueroa, Daniel G. and Hindmarsh, Mark and Urrestilla, Jon",
    title = "{Exact Scale-Invariant Background of Gravitational Waves from Cosmic Defects}",
    eprint = "1212.5458",
    archivePrefix = "arXiv",
    primaryClass = "astro-ph.CO",
    doi = "10.1103/PhysRevLett.110.101302",
    journal = "Phys. Rev. Lett.",
    volume = "110",
    number = "10",
    pages = "101302",
    year = "2013"
}

@article{Hiramatsu:2013qaa,
    author = "Hiramatsu, Takashi and Kawasaki, Masahiro and Saikawa, Ken'ichi",
    title = "{On the estimation of gravitational wave spectrum from cosmic domain walls}",
    eprint = "1309.5001",
    archivePrefix = "arXiv",
    primaryClass = "astro-ph.CO",
    reportNumber = "ICRR-REPORT-659-2013-8, IPMU13-0182, YITP-13-87",
    doi = "10.1088/1475-7516/2014/02/031",
    journal = "JCAP",
    volume = "02",
    pages = "031",
    year = "2014"
}

@article{Blanco-Pillado:2017oxo,
    author = "Blanco-Pillado, Jose J. and Olum, Ken D.",
    title = "{Stochastic gravitational wave background from smoothed cosmic string loops}",
    eprint = "1709.02693",
    archivePrefix = "arXiv",
    primaryClass = "astro-ph.CO",
    doi = "10.1103/PhysRevD.96.104046",
    journal = "Phys. Rev. D",
    volume = "96",
    number = "10",
    pages = "104046",
    year = "2017"
}

@article{Auclair:2019wcv,
    author = "Auclair, Pierre and others",
    title = "{Probing the gravitational wave background from cosmic strings with LISA}",
    eprint = "1909.00819",
    archivePrefix = "arXiv",
    primaryClass = "astro-ph.CO",
    doi = "10.1088/1475-7516/2020/04/034",
    journal = "JCAP",
    volume = "04",
    pages = "034",
    year = "2020"
}

@article{Gouttenoire:2019kij,
    author = "Gouttenoire, Yann and Servant, G{\'e}raldine and Simakachorn, Peera",
    title = "{Beyond the Standard Models with Cosmic Strings}",
    eprint = "1912.02569",
    archivePrefix = "arXiv",
    primaryClass = "hep-ph",
    reportNumber = "DESY-19-204",
    doi = "10.1088/1475-7516/2020/07/032",
    journal = "JCAP",
    volume = "07",
    pages = "032",
    year = "2020"
}

@article{Figueroa:2020lvo,
    author = "Figueroa, Daniel G. and Hindmarsh, Mark and Lizarraga, Joanes and Urrestilla, Jon",
    title = "{Irreducible background of gravitational waves from a cosmic defect network: update and comparison of numerical techniques}",
    eprint = "2007.03337",
    archivePrefix = "arXiv",
    primaryClass = "astro-ph.CO",
    doi = "10.1103/PhysRevD.102.103516",
    journal = "Phys. Rev. D",
    volume = "102",
    number = "10",
    pages = "103516",
    year = "2020"
}

@article{Chang:2021afa,
    author = "Chang, Chia-Feng and Cui, Yanou",
    title = "{Gravitational waves from global cosmic strings and cosmic archaeology}",
    eprint = "2106.09746",
    archivePrefix = "arXiv",
    primaryClass = "hep-ph",
    doi = "10.1007/JHEP03(2022)114",
    journal = "JHEP",
    volume = "03",
    pages = "114",
    year = "2022"
}

@article{Yamada:2022aax,
    author = "Yamada, Masaki and Yonekura, Kazuya",
    title = "{Cosmic F- and D-strings from pure Yang{\textendash}Mills theory}",
    eprint = "2204.13125",
    archivePrefix = "arXiv",
    primaryClass = "hep-th",
    reportNumber = "TU-1153",
    doi = "10.1016/j.physletb.2023.137724",
    journal = "Phys. Lett. B",
    volume = "838",
    pages = "137724",
    year = "2023"
}

@article{Yamada:2022imq,
    author = "Yamada, Masaki and Yonekura, Kazuya",
    title = "{Cosmic strings from pure Yang{\textendash}Mills theory}",
    eprint = "2204.13123",
    archivePrefix = "arXiv",
    primaryClass = "hep-th",
    reportNumber = "TU-1152",
    doi = "10.1103/PhysRevD.106.123515",
    journal = "Phys. Rev. D",
    volume = "106",
    number = "12",
    pages = "123515",
    year = "2022"
}

@article{Giovannini:1998bp,
    author = "Giovannini, Massimo",
    title = "{Gravitational waves constraints on postinflationary phases stiffer than radiation}",
    eprint = "hep-ph/9806329",
    archivePrefix = "arXiv",
    doi = "10.1103/PhysRevD.58.083504",
    journal = "Phys. Rev. D",
    volume = "58",
    pages = "083504",
    year = "1998"
}

@article{Giovannini:1999bh,
    author = "Giovannini, Massimo",
    title = "{Production and detection of relic gravitons in quintessential inflationary models}",
    eprint = "astro-ph/9903004",
    archivePrefix = "arXiv",
    reportNumber = "TUPT-01-99",
    doi = "10.1103/PhysRevD.60.123511",
    journal = "Phys. Rev. D",
    volume = "60",
    pages = "123511",
    year = "1999"
}

@article{Boyle:2007zx,
    author = "Boyle, Latham A. and Buonanno, Alessandra",
    title = "{Relating gravitational wave constraints from primordial nucleosynthesis, pulsar timing, laser interferometers, and the CMB: Implications for the early Universe}",
    eprint = "0708.2279",
    archivePrefix = "arXiv",
    primaryClass = "astro-ph",
    doi = "10.1103/PhysRevD.78.043531",
    journal = "Phys. Rev. D",
    volume = "78",
    pages = "043531",
    year = "2008"
}

@article{Li:2016mmc,
    author = "Li, Bohua and Shapiro, Paul R. and Rindler-Daller, Tanja",
    title = "{Bose-Einstein-condensed scalar field dark matter and the gravitational wave background from inflation: new cosmological constraints and its detectability by LIGO}",
    eprint = "1611.07961",
    archivePrefix = "arXiv",
    primaryClass = "astro-ph.CO",
    doi = "10.1103/PhysRevD.96.063505",
    journal = "Phys. Rev. D",
    volume = "96",
    number = "6",
    pages = "063505",
    year = "2017"
}

@article{Li:2021htg,
    author = "Li, Bohua and Shapiro, Paul R.",
    title = "{Precision cosmology and the stiff-amplified gravitational-wave background from inflation: NANOGrav, Advanced LIGO-Virgo and the Hubble tension}",
    eprint = "2107.12229",
    archivePrefix = "arXiv",
    primaryClass = "astro-ph.CO",
    doi = "10.1088/1475-7516/2021/10/024",
    journal = "JCAP",
    volume = "10",
    pages = "024",
    year = "2021"
}

@article{Figueroa:2018twl,
    author = "Figueroa, Daniel G. and Tanin, Erwin H.",
    title = "{Inconsistency of an inflationary sector coupled only to Einstein gravity}",
    eprint = "1811.04093",
    archivePrefix = "arXiv",
    primaryClass = "astro-ph.CO",
    doi = "10.1088/1475-7516/2019/10/050",
    journal = "JCAP",
    volume = "10",
    pages = "050",
    year = "2019"
}

@article{Figueroa:2019paj,
    author = "Figueroa, Daniel G. and Tanin, Erwin H.",
    title = "{Ability of LIGO and LISA to probe the equation of state of the early Universe}",
    eprint = "1905.11960",
    archivePrefix = "arXiv",
    primaryClass = "astro-ph.CO",
    doi = "10.1088/1475-7516/2019/08/011",
    journal = "JCAP",
    volume = "08",
    pages = "011",
    year = "2019"
}

@article{Gouttenoire:2021wzu,
    author = "Gouttenoire, Yann and Servant, G{\'e}raldine and Simakachorn, Peera",
    title = "{Revealing the Primordial Irreducible Inflationary Gravitational-Wave Background with a Spinning Peccei-Quinn Axion}",
    eprint = "2108.10328",
    archivePrefix = "arXiv",
    primaryClass = "hep-ph",
    reportNumber = "DESY 21-126",
    month = "8",
    year = "2021"
}

@article{Co:2021lkc,
    author = "Co, Raymond T. and Dunsky, David and Fernandez, Nicolas and Ghalsasi, Akshay and Hall, Lawrence J. and Harigaya, Keisuke and Shelton, Jessie",
    title = "{Gravitational wave and CMB probes of axion kination}",
    eprint = "2108.09299",
    archivePrefix = "arXiv",
    primaryClass = "hep-ph",
    reportNumber = "UMN-TH-4023/21, FTPI-MINN-21-15, CERN-TH-2021-124",
    doi = "10.1007/JHEP09(2022)116",
    journal = "JHEP",
    volume = "09",
    pages = "116",
    year = "2022"
}

@article{Gouttenoire:2021jhk,
    author = "Gouttenoire, Yann and Servant, Geraldine and Simakachorn, Peera",
    title = "{Kination cosmology from scalar fields and gravitational-wave signatures}",
    eprint = "2111.01150",
    archivePrefix = "arXiv",
    primaryClass = "hep-ph",
    reportNumber = "DESY 21-134",
    month = "11",
    year = "2021"
}

@article{Oikonomou:2023qfz,
    author = "Oikonomou, V. K.",
    title = "{Flat energy spectrum of primordial gravitational waves versus peaks and the NANOGrav 2023 observation}",
    eprint = "2306.17351",
    archivePrefix = "arXiv",
    primaryClass = "astro-ph.CO",
    doi = "10.1103/PhysRevD.108.043516",
    journal = "Phys. Rev. D",
    volume = "108",
    number = "4",
    pages = "043516",
    year = "2023"
}

@article{Anber:2006xt,
    author = "Anber, Mohamed M. and Sorbo, Lorenzo",
    title = "{N-flationary magnetic fields}",
    eprint = "astro-ph/0606534",
    archivePrefix = "arXiv",
    doi = "10.1088/1475-7516/2006/10/018",
    journal = "JCAP",
    volume = "10",
    pages = "018",
    year = "2006"
}

@article{Sorbo:2011rz,
    author = "Sorbo, Lorenzo",
    title = "{Parity violation in the Cosmic Microwave Background from a pseudoscalar inflaton}",
    eprint = "1101.1525",
    archivePrefix = "arXiv",
    primaryClass = "astro-ph.CO",
    doi = "10.1088/1475-7516/2011/06/003",
    journal = "JCAP",
    volume = "06",
    pages = "003",
    year = "2011"
}

@article{Pajer:2013fsa,
    author = "Pajer, Enrico and Peloso, Marco",
    title = "{A review of Axion Inflation in the era of Planck}",
    eprint = "1305.3557",
    archivePrefix = "arXiv",
    primaryClass = "hep-th",
    doi = "10.1088/0264-9381/30/21/214002",
    journal = "Class. Quant. Grav.",
    volume = "30",
    pages = "214002",
    year = "2013"
}

@article{Adshead:2013qp,
    author = "Adshead, Peter and Martinec, Emil and Wyman, Mark",
    title = "{Gauge fields and inflation: Chiral gravitational waves, fluctuations, and the Lyth bound}",
    eprint = "1301.2598",
    archivePrefix = "arXiv",
    primaryClass = "hep-th",
    doi = "10.1103/PhysRevD.88.021302",
    journal = "Phys. Rev. D",
    volume = "88",
    number = "2",
    pages = "021302",
    year = "2013"
}

@article{Adshead:2013nka,
    author = "Adshead, Peter and Martinec, Emil and Wyman, Mark",
    title = "{Perturbations in Chromo-Natural Inflation}",
    eprint = "1305.2930",
    archivePrefix = "arXiv",
    primaryClass = "hep-th",
    doi = "10.1007/JHEP09(2013)087",
    journal = "JHEP",
    volume = "09",
    pages = "087",
    year = "2013"
}

@article{Maleknejad:2016qjz,
    author = "Maleknejad, Azadeh",
    title = "{Axion Inflation with an SU(2) Gauge Field: Detectable Chiral Gravity Waves}",
    eprint = "1604.03327",
    archivePrefix = "arXiv",
    primaryClass = "hep-ph",
    doi = "10.1007/JHEP07(2016)104",
    journal = "JHEP",
    volume = "07",
    pages = "104",
    year = "2016"
}

@article{Namba:2015gja,
    author = "Namba, Ryo and Peloso, Marco and Shiraishi, Maresuke and Sorbo, Lorenzo and Unal, Caner",
    title = "{Scale-dependent gravitational waves from a rolling axion}",
    eprint = "1509.07521",
    archivePrefix = "arXiv",
    primaryClass = "astro-ph.CO",
    doi = "10.1088/1475-7516/2016/01/041",
    journal = "JCAP",
    volume = "01",
    pages = "041",
    year = "2016"
}

@article{Ferreira:2015omg,
    author = "Ferreira, Ricardo Z. and Ganc, Jonathan and Nore{\~n}a, Jorge and Sloth, Martin S.",
    title = "{On the validity of the perturbative description of axions during inflation}",
    eprint = "1512.06116",
    archivePrefix = "arXiv",
    primaryClass = "astro-ph.CO",
    doi = "10.1088/1475-7516/2016/04/039",
    journal = "JCAP",
    volume = "04",
    pages = "039",
    year = "2016",
    note = "[Erratum: JCAP 10, E01 (2016)]"
}

@article{Peloso:2016gqs,
    author = "Peloso, Marco and Sorbo, Lorenzo and Unal, Caner",
    title = "{Rolling axions during inflation: perturbativity and signatures}",
    eprint = "1606.00459",
    archivePrefix = "arXiv",
    primaryClass = "astro-ph.CO",
    reportNumber = "ACFI-T16-16, UMN--TH--3531-16",
    doi = "10.1088/1475-7516/2016/09/001",
    journal = "JCAP",
    volume = "09",
    pages = "001",
    year = "2016"
}

@article{Domcke:2016bkh,
    author = "Domcke, Valerie and Pieroni, Mauro and Bin{\'e}truy, Pierre",
    title = "{Primordial gravitational waves for universality classes of pseudoscalar inflation}",
    eprint = "1603.01287",
    archivePrefix = "arXiv",
    primaryClass = "astro-ph.CO",
    doi = "10.1088/1475-7516/2016/06/031",
    journal = "JCAP",
    volume = "06",
    pages = "031",
    year = "2016"
}

@article{Caldwell:2017chz,
    author = "Caldwell, R. R. and Devulder, C.",
    title = "{Axion Gauge Field Inflation and Gravitational Leptogenesis: A Lower Bound on B Modes from the Matter-Antimatter Asymmetry of the Universe}",
    eprint = "1706.03765",
    archivePrefix = "arXiv",
    primaryClass = "astro-ph.CO",
    doi = "10.1103/PhysRevD.97.023532",
    journal = "Phys. Rev. D",
    volume = "97",
    number = "2",
    pages = "023532",
    year = "2018"
}

@article{Guzzetti:2016mkm,
    author = "Guzzetti, M. C. and Bartolo, N. and Liguori, M. and Matarrese, S.",
    title = "{Gravitational waves from inflation}",
    eprint = "1605.01615",
    archivePrefix = "arXiv",
    primaryClass = "astro-ph.CO",
    doi = "10.1393/ncr/i2016-10127-1",
    journal = "Riv. Nuovo Cim.",
    volume = "39",
    number = "9",
    pages = "399--495",
    year = "2016"
}

@article{Bartolo:2016ami,
    author = "Bartolo, Nicola and others",
    title = "{Science with the space-based interferometer LISA. IV: Probing inflation with gravitational waves}",
    eprint = "1610.06481",
    archivePrefix = "arXiv",
    primaryClass = "astro-ph.CO",
    reportNumber = "ACFI-T16-19, UMN-TH-3608-16, CERN-TH-2016-222, KCL-PH-TH-2016-58, IFT-UAM-CSIC-16-104",
    doi = "10.1088/1475-7516/2016/12/026",
    journal = "JCAP",
    volume = "12",
    pages = "026",
    year = "2016"
}

@article{DAmico:2021fhz,
    author = "D'Amico, Guido and Kaloper, Nemanja and Westphal, Alexander",
    title = "{General double monodromy inflation}",
    eprint = "2112.13861",
    archivePrefix = "arXiv",
    primaryClass = "hep-th",
    doi = "10.1103/PhysRevD.105.103527",
    journal = "Phys. Rev. D",
    volume = "105",
    number = "10",
    pages = "103527",
    year = "2022"
}

@article{DAmico:2021vka,
    author = "D'Amico, Guido and Kaloper, Nemanja and Westphal, Alexander",
    title = "{Double Monodromy Inflation: A Gravity Waves Factory for CMB-S4, LiteBIRD and LISA}",
    eprint = "2101.05861",
    archivePrefix = "arXiv",
    primaryClass = "hep-th",
    reportNumber = "DESY 21-004",
    doi = "10.1103/PhysRevD.104.L081302",
    journal = "Phys. Rev. D",
    volume = "104",
    number = "8",
    pages = "L081302",
    year = "2021"
}

@article{Sesana:2025udx,
    author = "Sesana, Alberto and Figueroa, Daniel G.",
    title = "{Nanohertz Gravitational Waves}",
    eprint = "2512.18822",
    archivePrefix = "arXiv",
    primaryClass = "astro-ph.CO",
    month = "12",
    year = "2025"
}

@article{Ananda:2006af,
    author = "Ananda, Kishore N. and Clarkson, Chris and Wands, David",
    title = "{The Cosmological gravitational wave background from primordial density perturbations}",
    eprint = "gr-qc/0612013",
    archivePrefix = "arXiv",
    doi = "10.1103/PhysRevD.75.123518",
    journal = "Phys. Rev. D",
    volume = "75",
    pages = "123518",
    year = "2007"
}

@article{Baumann:2007zm,
    author = "Baumann, Daniel and Steinhardt, Paul J. and Takahashi, Keitaro and Ichiki, Kiyotomo",
    title = "{Gravitational Wave Spectrum Induced by Primordial Scalar Perturbations}",
    eprint = "hep-th/0703290",
    archivePrefix = "arXiv",
    doi = "10.1103/PhysRevD.76.084019",
    journal = "Phys. Rev. D",
    volume = "76",
    pages = "084019",
    year = "2007"
}

@article{Domenech:2021ztg,
    author = "Dom\`enech, Guillem",
    title = "{Scalar Induced Gravitational Waves Review}",
    eprint = "2109.01398",
    archivePrefix = "arXiv",
    primaryClass = "gr-qc",
    doi = "10.3390/universe7110398",
    journal = "Universe",
    volume = "7",
    number = "11",
    pages = "398",
    year = "2021"
}

@article{Figueroa:2023zhu,
    author = "Figueroa, Daniel G. and Pieroni, Mauro and Ricciardone, Angelo and Simakachorn, Peera",
    title = "{Cosmological Background Interpretation of Pulsar Timing Array Data}",
    eprint = "2307.02399",
    archivePrefix = "arXiv",
    primaryClass = "astro-ph.CO",
    reportNumber = "CERN-TH-2023-132",
    doi = "10.1103/PhysRevLett.132.171002",
    journal = "Phys. Rev. Lett.",
    volume = "132",
    number = "17",
    pages = "171002",
    year = "2024"
}

@article{Saito:2008jc,
    author = "Saito, Ryo and Yokoyama, Jun'ichi",
    title = "{Gravitational wave background as a probe of the primordial black hole abundance}",
    eprint = "0812.4339",
    archivePrefix = "arXiv",
    primaryClass = "astro-ph",
    reportNumber = "RESCEU-63-08",
    doi = "10.1103/PhysRevLett.102.161101",
    journal = "Phys. Rev. Lett.",
    volume = "102",
    pages = "161101",
    year = "2009",
    note = "[Erratum: Phys.Rev.Lett. 107, 069901 (2011)]"
}

@article{saito2010gravitational,
  title={Gravitational-wave constraints on the abundance of primordial black holes},
  author={Saito, Ryo and Yokoyama, Jun'ichi},
  journal={Progress of theoretical physics},
  volume={123},
  number={5},
  pages={867--886},
  year={2010},
  publisher={Oxford University Press}
}

@article{LIGOScientific:2018mvr,
    author = "Abbott, B. P. and others",
    collaboration = "LIGO Scientific, Virgo",
    title = "{GWTC-1: A Gravitational-Wave Transient Catalog of Compact Binary Mergers Observed by LIGO and Virgo during the First and Second Observing Runs}",
    eprint = "1811.12907",
    archivePrefix = "arXiv",
    primaryClass = "astro-ph.HE",
    reportNumber = "LIGO-P1800307",
    doi = "10.1103/PhysRevX.9.031040",
    journal = "Phys. Rev. X",
    volume = "9",
    number = "3",
    pages = "031040",
    year = "2019"
}

@article{LIGOScientific:2020ibl,
    author = "Abbott, R. and others",
    collaboration = "LIGO Scientific, Virgo",
    title = "{GWTC-2: Compact Binary Coalescences Observed by LIGO and Virgo During the First Half of the Third Observing Run}",
    eprint = "2010.14527",
    archivePrefix = "arXiv",
    primaryClass = "gr-qc",
    reportNumber = "P2000061",
    doi = "10.1103/PhysRevX.11.021053",
    journal = "Phys. Rev. X",
    volume = "11",
    pages = "021053",
    year = "2021"
}

@article{KAGRA:2021vkt,
    author = "Abbott, R. and others",
    collaboration = "KAGRA, VIRGO, LIGO Scientific",
    title = "{GWTC-3: Compact Binary Coalescences Observed by LIGO and Virgo during the Second Part of the Third Observing Run}",
    eprint = "2111.03606",
    archivePrefix = "arXiv",
    primaryClass = "gr-qc",
    reportNumber = "LIGO-P2000318",
    doi = "10.1103/PhysRevX.13.041039",
    journal = "Phys. Rev. X",
    volume = "13",
    number = "4",
    pages = "041039",
    year = "2023"
}

@article{LIGOScientific:2025slb,
    author = "Abac, A. G. and others",
    collaboration = "LIGO Scientific, VIRGO, KAGRA",
    title = "{GWTC-4.0: Updating the Gravitational-Wave Transient Catalog with Observations from the First Part of the Fourth LIGO-Virgo-KAGRA Observing Run}",
    eprint = "2508.18082",
    archivePrefix = "arXiv",
    primaryClass = "gr-qc",
    reportNumber = "LIGO-P2400386",
    month = "8",
    year = "2025"
}

@article{Ando:2022tpj,
    author = "Ando, Shin'ichiro and Hiroshima, Nagisa and Ishiwata, Koji",
    title = "{Constraining the primordial curvature perturbation using dark matter substructure}",
    eprint = "2207.05747",
    archivePrefix = "arXiv",
    primaryClass = "astro-ph.CO",
    reportNumber = "RIKEN-iTHEMS-Report-22, UT-HET-138, KANAZAWA-22-03",
    doi = "10.1103/PhysRevD.106.103014",
    journal = "Phys. Rev. D",
    volume = "106",
    number = "10",
    pages = "103014",
    year = "2022"
}

@article{Garcia-Bellido:2017fdg,
    author = "Garc{\'\i}a-Bellido, Juan",
    editor = "Giardini, Domencio and Jetzer, Philippe",
    title = "{Massive Primordial Black Holes as Dark Matter and their detection with Gravitational Waves}",
    eprint = "1702.08275",
    archivePrefix = "arXiv",
    primaryClass = "astro-ph.CO",
    reportNumber = "IFT-UAM-CSIC-17-016",
    doi = "10.1088/1742-6596/840/1/012032",
    journal = "J. Phys. Conf. Ser.",
    volume = "840",
    number = "1",
    pages = "012032",
    year = "2017"
}

@article{LISACosmologyWorkingGroup:2023njw,
    author = "Bagui, Eleni and others",
    collaboration = "LISA Cosmology Working Group",
    title = "{Primordial black holes and their gravitational-wave signatures}",
    eprint = "2310.19857",
    archivePrefix = "arXiv",
    primaryClass = "astro-ph.CO",
    doi = "10.1007/s41114-024-00053-w",
    journal = "Living Rev. Rel.",
    volume = "28",
    number = "1",
    pages = "1",
    year = "2025"
}

@article{Hawking:1971ei,
    author = "Hawking, Stephen",
    title = "{Gravitationally collapsed objects of very low mass}",
    doi = "10.1093/mnras/152.1.75",
    journal = "Mon. Not. Roy. Astron. Soc.",
    volume = "152",
    pages = "75",
    year = "1971"
}

@article{Carr:1974nx,
    author = "Carr, Bernard J. and Hawking, S. W.",
    title = "{Black holes in the early Universe}",
    doi = "10.1093/mnras/168.2.399",
    journal = "Mon. Not. Roy. Astron. Soc.",
    volume = "168",
    pages = "399--415",
    year = "1974"
}

@article{Coleman:1985ki,
    author = "Coleman, Sidney R.",
    title = "{Q-balls}",
    reportNumber = "HUTP-85/A050",
    doi = "10.1016/0550-3213(86)90520-1",
    journal = "Nucl. Phys. B",
    volume = "262",
    number = "2",
    pages = "263",
    year = "1985",
    note = "[Addendum: Nucl.Phys.B 269, 744 (1986)]"
}

@article{Turner:1983he,
    author = "Turner, Michael S.",
    title = "{Coherent Scalar Field Oscillations in an Expanding Universe}",
    reportNumber = "EFI-83-29-CHICAGO",
    doi = "10.1103/PhysRevD.28.1243",
    journal = "Phys. Rev. D",
    volume = "28",
    pages = "1243",
    year = "1983"
}

@article{Bogolyubsky:1976pw,
    author = "Bogolyubsky, I. L.",
    title = "{Oscillating Particle-Like Solutions of Nonlinear Klein-Gordon Equation}",
    reportNumber = "JINR-E2-10129",
    journal = "JETP Lett.",
    volume = "24",
    pages = "535",
    year = "1976"
}

@article{Lozanov:2022yoy,
    author = "Lozanov, Kaloian D. and Takhistov, Volodymyr",
    title = "{Enhanced Gravitational Waves from Inflaton Oscillons}",
    eprint = "2204.07152",
    archivePrefix = "arXiv",
    primaryClass = "astro-ph.CO",
    reportNumber = "IPMU22-0016, KEK-QUP-2023-0008, KEK-TH-2518, KEK-Cosmo-0311",
    doi = "10.1103/PhysRevLett.130.181002",
    journal = "Phys. Rev. Lett.",
    volume = "130",
    number = "18",
    pages = "181002",
    year = "2023"
}

@article{Sui:2024grm,
    author = "Sui, Xiao-Bin and Liu, Jing and Cai, Rong-Gen",
    title = "{Enhancement of small-scale induced gravitational waves from solitons and oscillons}",
    eprint = "2412.08057",
    archivePrefix = "arXiv",
    primaryClass = "astro-ph.CO",
    doi = "10.1103/d2rp-xyzt",
    journal = "Phys. Rev. D",
    volume = "111",
    number = "12",
    pages = "123503",
    year = "2025"
}

@article{Domenech:2025ccu,
    author = "Dom{\`e}nech, Guillem and Pi, Shi and Wang, Ao",
    title = "{Physical Effects of Gravitational Waves at Second Order}",
    eprint = "2512.15704",
    archivePrefix = "arXiv",
    primaryClass = "gr-qc",
    month = "12",
    year = "2025"
}

@article{Co:2019jts,
    author = "Co, Raymond T. and Hall, Lawrence J. and Harigaya, Keisuke",
    title = "{Axion Kinetic Misalignment Mechanism}",
    eprint = "1910.14152",
    archivePrefix = "arXiv",
    primaryClass = "hep-ph",
    reportNumber = "LCTP-19-28",
    doi = "10.1103/PhysRevLett.124.251802",
    journal = "Phys. Rev. Lett.",
    volume = "124",
    number = "25",
    pages = "251802",
    year = "2020"
}

@article{Co:2019wyp,
    author = "Co, Raymond T. and Harigaya, Keisuke",
    title = "{Axiogenesis}",
    eprint = "1910.02080",
    archivePrefix = "arXiv",
    primaryClass = "hep-ph",
    reportNumber = "LCTP-19-27",
    doi = "10.1103/PhysRevLett.124.111602",
    journal = "Phys. Rev. Lett.",
    volume = "124",
    number = "11",
    pages = "111602",
    year = "2020"
}

@article{Co:2022qpr,
    author = "Co, Raymond T. and Harigaya, Keisuke and Pierce, Aaron",
    title = "{Cosmic perturbations from a rotating field}",
    eprint = "2202.01785",
    archivePrefix = "arXiv",
    primaryClass = "hep-ph",
    reportNumber = "UMN-TH-4113/22, FTPI-MINN-22-04, CERN-TH-2022-007, LCTP-22-02",
    doi = "10.1088/1475-7516/2022/10/037",
    journal = "JCAP",
    volume = "10",
    pages = "037",
    year = "2022"
}

@article{Assadullahi:2009nf,
    author = "Assadullahi, Hooshyar and Wands, David",
    title = "{Gravitational waves from an early matter era}",
    eprint = "0901.0989",
    archivePrefix = "arXiv",
    primaryClass = "astro-ph.CO",
    doi = "10.1103/PhysRevD.79.083511",
    journal = "Phys. Rev. D",
    volume = "79",
    pages = "083511",
    year = "2009"
}

@article{Kohri:2018awv,
    author = "Kohri, Kazunori and Terada, Takahiro",
    title = "{Semianalytic calculation of gravitational wave spectrum nonlinearly induced from primordial curvature perturbations}",
    eprint = "1804.08577",
    archivePrefix = "arXiv",
    primaryClass = "gr-qc",
    reportNumber = "KEK-TH-2046, KEK-COSMO-223",
    doi = "10.1103/PhysRevD.97.123532",
    journal = "Phys. Rev. D",
    volume = "97",
    number = "12",
    pages = "123532",
    year = "2018"
}

@article{Inomata:2019ivs,
    author = "Inomata, Keisuke and Kohri, Kazunori and Nakama, Tomohiro and Terada, Takahiro",
    title = "{Enhancement of Gravitational Waves Induced by Scalar Perturbations due to a Sudden Transition from an Early Matter Era to the Radiation Era}",
    eprint = "1904.12879",
    archivePrefix = "arXiv",
    primaryClass = "astro-ph.CO",
    reportNumber = "IPMU 19-0067, KEK-TH-2122, KEK-Cosmo-237",
    doi = "10.1103/PhysRevD.108.049901",
    journal = "Phys. Rev. D",
    volume = "100",
    pages = "043532",
    year = "2019",
    note = "[Erratum: Phys.Rev.D 108, 049901 (2023)]"
}

@article{Inomata:2019zqy,
    author = "Inomata, Keisuke and Kohri, Kazunori and Nakama, Tomohiro and Terada, Takahiro",
    title = "{Gravitational Waves Induced by Scalar Perturbations during a Gradual Transition from an Early Matter Era to the Radiation Era}",
    eprint = "1904.12878",
    archivePrefix = "arXiv",
    primaryClass = "astro-ph.CO",
    reportNumber = "IPMU 19-0066, KEK-TH-2121, KEK-Cosmo-236",
    doi = "10.1088/1475-7516/2019/10/071",
    journal = "JCAP",
    volume = "10",
    pages = "071",
    year = "2019",
    note = "[Erratum: JCAP 08, E01 (2023)]"
}

@article{Inomata:2020lmk,
    author = "Inomata, Keisuke and Kawasaki, Masahiro and Mukaida, Kyohei and Terada, Takahiro and Yanagida, Tsutomu T.",
    title = "{Gravitational Wave Production right after a Primordial Black Hole Evaporation}",
    eprint = "2003.10455",
    archivePrefix = "arXiv",
    primaryClass = "astro-ph.CO",
    reportNumber = "IPMU 20-0029, DESY 20-042, DESY-20-042, CTPU-PTC-20-05",
    doi = "10.1103/PhysRevD.101.123533",
    journal = "Phys. Rev. D",
    volume = "101",
    number = "12",
    pages = "123533",
    year = "2020"
}

@article{Inomata:2025wiv,
    author = "Inomata, Keisuke and Kohri, Kazunori and Terada, Takahiro",
    title = "{The poltergeist mechanism -- Enhancement of scalar-induced gravitational waves with early matter-dominated era}",
    eprint = "2511.07266",
    archivePrefix = "arXiv",
    primaryClass = "astro-ph.CO",
    reportNumber = "KEK-TH-2781, KEK-Cosmo-0400",
    month = "11",
    year = "2025"
}

@article{Pearce:2023kxp,
    author = "Pearce, Matthew and Pearce, Lauren and White, Graham and Balazs, Csaba",
    title = "{Gravitational wave signals from early matter domination: interpolating between fast and slow transitions}",
    eprint = "2311.12340",
    archivePrefix = "arXiv",
    primaryClass = "astro-ph.CO",
    doi = "10.1088/1475-7516/2024/06/021",
    journal = "JCAP",
    volume = "06",
    pages = "021",
    year = "2024"
}

@article{Pearce:2025ywc,
    author = "Pearce, Matthew and Pearce, Lauren and White, Graham and Bal{\'a}zs, Csaba",
    title = "{Using gravitational wave signals to disentangle early matter dominated epochs}",
    eprint = "2503.03101",
    archivePrefix = "arXiv",
    primaryClass = "astro-ph.CO",
    doi = "10.1088/1475-7516/2025/11/004",
    journal = "JCAP",
    volume = "11",
    pages = "004",
    year = "2025"
}

@article{Gong:2019mui,
    author = "Gong, Jinn-Ouk",
    title = "{Analytic Integral Solutions for Induced Gravitational Waves}",
    eprint = "1909.12708",
    archivePrefix = "arXiv",
    primaryClass = "gr-qc",
    doi = "10.3847/1538-4357/ac3a6c",
    journal = "Astrophys. J.",
    volume = "925",
    number = "1",
    pages = "102",
    year = "2022"
}

@article{Chang:2022vlv,
    author = "Chang, Zhe and Zhang, Xukun and Zhou, Jing-Zhi",
    title = "{Gravitational waves from primordial scalar and tensor perturbations}",
    eprint = "2209.07693",
    archivePrefix = "arXiv",
    primaryClass = "astro-ph.CO",
    doi = "10.1103/PhysRevD.107.063510",
    journal = "Phys. Rev. D",
    volume = "107",
    number = "6",
    pages = "063510",
    year = "2023"
}

@article{Chen:2025fcd,
    author = "Chen, Fei-Yu and Zhou, Jing-Zhi and Wu, Di and Li, Zhi-Chao and Wu, Peng-Yu",
    title = "{Tensor induced gravitational waves}",
    eprint = "2507.22688",
    archivePrefix = "arXiv",
    primaryClass = "astro-ph.CO",
    month = "7",
    year = "2025"
}

@article{Bari:2023rcw,
    author = "Bari, Pritha and Bartolo, Nicola and Dom\`enech, Guillem and Matarrese, Sabino",
    title = "{Gravitational waves induced by scalar-tensor mixing}",
    eprint = "2307.05404",
    archivePrefix = "arXiv",
    primaryClass = "astro-ph.CO",
    doi = "10.1103/PhysRevD.109.023509",
    journal = "Phys. Rev. D",
    volume = "109",
    number = "2",
    pages = "023509",
    year = "2024"
}

@article{Picard:2023sbz,
    author = "Picard, Raphael and Malik, Karim A.",
    title = "{Induced gravitational waves: the effect of first order tensor perturbations}",
    eprint = "2311.14513",
    archivePrefix = "arXiv",
    primaryClass = "astro-ph.CO",
    doi = "10.1088/1475-7516/2024/10/010",
    journal = "JCAP",
    volume = "10",
    pages = "010",
    year = "2024"
}

@article{Yu:2023lmo,
    author = "Yu, Yan-Heng and Wang, Sai",
    title = "{Primordial gravitational waves assisted by cosmological scalar perturbations}",
    eprint = "2303.03897",
    archivePrefix = "arXiv",
    primaryClass = "astro-ph.CO",
    doi = "10.1140/epjc/s10052-024-12937-w",
    journal = "Eur. Phys. J. C",
    volume = "84",
    number = "6",
    pages = "555",
    year = "2024"
}

@article{Picard:2024ekd,
    author = {Picard, Rapha{\"e}l and Davies, Matthew W.},
    title = "{Effects of scalar non-Gaussianity on induced scalar-tensor gravitational waves}",
    eprint = "2410.17819",
    archivePrefix = "arXiv",
    primaryClass = "astro-ph.CO",
    doi = "10.1088/1475-7516/2025/02/037",
    journal = "JCAP",
    volume = "02",
    pages = "037",
    year = "2025"
}

@article{Picard:2025bwq,
    author = "Picard, Raphael and Padilla, Luis E. and Malik, Karim A. and Mulryne, David J.",
    title = "{Suppression of the induced gravitational wave background due to third-order perturbations}",
    eprint = "2509.07811",
    archivePrefix = "arXiv",
    primaryClass = "astro-ph.CO",
    doi = "10.1088/1475-7516/2026/01/056",
    journal = "JCAP",
    volume = "01",
    pages = "056",
    year = "2026"
}

@article{Wu:2024qdb,
    author = "Wu, Di and Zhou, Jing-Zhi and Kuang, Yu-Ting and Li, Zhi-Chao and Chang, Zhe and Huang, Qing-Guo",
    title = "{Can tensor-scalar induced GWs dominate PTA observations?}",
    eprint = "2501.00228",
    archivePrefix = "arXiv",
    primaryClass = "astro-ph.CO",
    doi = "10.1088/1475-7516/2025/03/045",
    journal = "JCAP",
    volume = "03",
    pages = "045",
    year = "2025"
}

@article{Wu:2025gwt,
    author = "Wu, Di and Li, Zhi-Chao and Wu, Peng-Yu and Chen, Fei-Yu and Zhou, Jing-Zhi",
    title = "{Probing small-scale primordial power spectra with induced gravitational waves}",
    eprint = "2507.07836",
    archivePrefix = "arXiv",
    primaryClass = "astro-ph.CO",
    doi = "10.1103/qd6t-cxz8",
    journal = "Phys. Rev. D",
    volume = "113",
    number = "6",
    pages = "063521",
    year = "2026"
}

@article{Janssen:2014dka,
    author = "Janssen, Gemma and others",
    editor = "Bourke, Tyler L. and others",
    title = "{Gravitational wave astronomy with the SKA}",
    eprint = "1501.00127",
    archivePrefix = "arXiv",
    primaryClass = "astro-ph.IM",
    doi = "10.22323/1.215.0037",
    journal = "PoS",
    volume = "AASKA14",
    pages = "037",
    year = "2015"
}

@article{Lazio:2013mea,
    author = "Lazio, T. J. W.",
    title = "{The Square Kilometre Array pulsar timing array}",
    doi = "10.1088/0264-9381/30/22/224011",
    journal = "Class. Quant. Grav.",
    volume = "30",
    pages = "224011",
    year = "2013"
}

@article{Mitridate:2023oar,
    author = {Mitridate, Andrea and Wright, David and von Eckardstein, Richard and Schr{\"o}der, Tobias and Nay, Jonathan and Olum, Ken and Schmitz, Kai and Trickle, Tanner},
    title = "{PTArcade}",
    eprint = "2306.16377",
    archivePrefix = "arXiv",
    primaryClass = "hep-ph",
    reportNumber = "FERMILAB-PUB-23-588-T",
    month = "6",
    year = "2023"
}

@article{Babak:2024yhu,
    author = "Babak, Stanislav and Falxa, Mikel and Franciolini, Gabriele and Pieroni, Mauro",
    title = "{Forecasting the sensitivity of pulsar timing arrays to gravitational wave backgrounds}",
    eprint = "2404.02864",
    archivePrefix = "arXiv",
    primaryClass = "astro-ph.CO",
    reportNumber = "CERN-TH-2024-039",
    doi = "10.1103/PhysRevD.110.063022",
    journal = "Phys. Rev. D",
    volume = "110",
    number = "6",
    pages = "063022",
    year = "2024"
}

@article{Cecchini:2025oks,
    author = "Cecchini, Chiara and Franciolini, Gabriele and Pieroni, Mauro",
    title = "{Forecasting constraints on scalar-induced gravitational waves with future pulsar timing array observations}",
    eprint = "2503.10805",
    archivePrefix = "arXiv",
    primaryClass = "astro-ph.CO",
    reportNumber = "CERN-TH-2025-045",
    doi = "10.1103/nxx5-gx7d",
    journal = "Phys. Rev. D",
    volume = "111",
    number = "12",
    pages = "123536",
    year = "2025"
}

@article{Kugarajh:2025rbt,
    author = "Kugarajh, Anjali Abirami and Traforetti, Marisol and Maselli, Andrea and Matarrese, Sabino and Ricciardone, Angelo",
    title = "{Scalar-Induced Gravitational Waves in Modified Gravity}",
    eprint = "2502.20137",
    archivePrefix = "arXiv",
    primaryClass = "gr-qc",
    doi = "10.1088/1475-7516/2025/07/022",
    month = "2",
    year = "2025"
}

@article{Perna:2024ehx,
    author = "Perna, Gabriele and Testini, Chiara and Ricciardone, Angelo and Matarrese, Sabino",
    title = "{Fully non-Gaussian Scalar-Induced Gravitational Waves}",
    eprint = "2403.06962",
    archivePrefix = "arXiv",
    primaryClass = "astro-ph.CO",
    doi = "10.1088/1475-7516/2024/05/086",
    journal = "JCAP",
    volume = "05",
    pages = "086",
    year = "2024"
}

@article{LISACosmologyWorkingGroup:2025vdz,
    author = "Gammal, Jonas El and others",
    collaboration = "LISA Cosmology Working Group",
    title = "{Reconstructing primordial curvature perturbations via scalar-induced gravitational waves with LISA}",
    eprint = "2501.11320",
    archivePrefix = "arXiv",
    primaryClass = "astro-ph.CO",
    reportNumber = "CERN-TH-2024-217",
    doi = "10.1088/1475-7516/2025/05/062",
    journal = "JCAP",
    volume = "05",
    pages = "062",
    year = "2025"
}

@article{Watanabe:2006qe,
    author = "Watanabe, Yuki and Komatsu, Eiichiro",
    title = "{Improved Calculation of the Primordial Gravitational Wave Spectrum in the Standard Model}",
    eprint = "astro-ph/0604176",
    archivePrefix = "arXiv",
    doi = "10.1103/PhysRevD.73.123515",
    journal = "Phys. Rev. D",
    volume = "73",
    pages = "123515",
    year = "2006"
}

@article{Dimastrogiovanni:2016fuu,
    author = "Dimastrogiovanni, Emanuela and Fasiello, Matteo and Fujita, Tomohiro",
    title = "{Primordial Gravitational Waves from Axion-Gauge Fields Dynamics}",
    eprint = "1608.04216",
    archivePrefix = "arXiv",
    primaryClass = "astro-ph.CO",
    doi = "10.1088/1475-7516/2017/01/019",
    journal = "JCAP",
    volume = "01",
    pages = "019",
    year = "2017"
}

@article{Chen:2022dah,
    author = "Chen, Chao and Ota, Atsuhisa and Zhu, Hui-Yu and Zhu, Yuhang",
    title = "{Missing one-loop contributions in secondary gravitational waves}",
    eprint = "2210.17176",
    archivePrefix = "arXiv",
    primaryClass = "astro-ph.CO",
    doi = "10.1103/PhysRevD.107.083518",
    journal = "Phys. Rev. D",
    volume = "107",
    number = "8",
    pages = "083518",
    year = "2023"
}

@article{Kugarajh:2025pjl,
    author = "Kugarajh, Anjali Abirami",
    title = "{Gauge-dependence of Scalar Induced Gravitational Waves}",
    eprint = "2503.00083",
    archivePrefix = "arXiv",
    primaryClass = "gr-qc",
    doi = "10.1088/1361-6382/ade2b3",
    journal = "Class. Quant. Grav.",
    volume = "42",
    number = "12",
    pages = "127001",
    year = "2025"
}

@article{Domenech:2019quo,
    author = "Dom{\`e}nech, Guillem",
    title = "{Induced gravitational waves in a general cosmological background}",
    eprint = "1912.05583",
    archivePrefix = "arXiv",
    primaryClass = "gr-qc",
    doi = "10.1142/S0218271820500285",
    journal = "Int. J. Mod. Phys. D",
    volume = "29",
    number = "03",
    pages = "2050028",
    year = "2020"
}

@book{Dodelson:2020bqr,
    author = "Dodelson, Scott and Schmidt, Fabian",
    title = "{Modern Cosmology}",
    doi = "10.1016/C2017-0-01943-2",
    publisher = "Academic Press",
    year = "2020"
}

@article{Allen:1997ad,
    author = "Allen, Bruce and Romano, Joseph D.",
    title = "{Detecting a stochastic background of gravitational radiation: Signal processing strategies and sensitivities}",
    eprint = "gr-qc/9710117",
    archivePrefix = "arXiv",
    reportNumber = "WISC-MILW-97-TH-14",
    doi = "10.1103/PhysRevD.59.102001",
    journal = "Phys. Rev. D",
    volume = "59",
    pages = "102001",
    year = "1999"
}

@article{Garcia-Bellido:2016dkw,
    author = "Garcia-Bellido, Juan and Peloso, Marco and Unal, Caner",
    title = "{Gravitational waves at interferometer scales and primordial black holes in axion inflation}",
    eprint = "1610.03763",
    archivePrefix = "arXiv",
    primaryClass = "astro-ph.CO",
    reportNumber = "IFT-UAM-CSIC-16-100, UMN-TH-3607-16",
    doi = "10.1088/1475-7516/2016/12/031",
    journal = "JCAP",
    volume = "12",
    pages = "031",
    year = "2016"
}

@article{Carr:2026hot,
    author = {Carr, Bernard and Iovino, Antonio J. and Perna, Gabriele and Vaskonen, Ville and Veerm{\"a}e, Hardi},
    title = "{Primordial black holes: constraints, potential evidence and prospects}",
    eprint = "2601.06024",
    archivePrefix = "arXiv",
    primaryClass = "astro-ph.CO",
    doi = "10.1007/s40766-026-00080-z",
    month = "1",
    year = "2026"
}

@article{Lamb:2023jls,
    author = "Lamb, William G. and Taylor, Stephen R. and van Haasteren, Rutger",
    title = "{Rapid refitting techniques for Bayesian spectral characterization of the gravitational wave background using pulsar timing arrays}",
    eprint = "2303.15442",
    archivePrefix = "arXiv",
    primaryClass = "astro-ph.HE",
    doi = "10.1103/PhysRevD.108.103019",
    journal = "Phys. Rev. D",
    volume = "108",
    number = "10",
    pages = "103019",
    year = "2023"
}

@article{Garoffolo:2022usx,
    author = "Garoffolo, Alice",
    title = "{Wave-optics limit of the stochastic gravitational wave background}",
    eprint = "2210.05718",
    archivePrefix = "arXiv",
    primaryClass = "astro-ph.CO",
    doi = "10.1016/j.dark.2024.101475",
    journal = "Phys. Dark Univ.",
    volume = "44",
    pages = "101475",
    year = "2024"
}

@article{NANOGrav:2023hfp,
    author = "Agazie, Gabriella and others",
    collaboration = "NANOGrav",
    title = "{The NANOGrav 15 yr Data Set: Constraints on Supermassive Black Hole Binaries from the Gravitational-wave Background}",
    eprint = "2306.16220",
    archivePrefix = "arXiv",
    primaryClass = "astro-ph.HE",
    doi = "10.3847/2041-8213/ace18b",
    journal = "Astrophys. J. Lett.",
    volume = "952",
    number = "2",
    pages = "L37",
    year = "2023"
}

@article{Zhu:2023gmx,
    author = "Zhu, Qing-Hua and Zhao, Zhi-Chao and Wang, Sai and Zhang, Xin",
    title = "{Unraveling the early universe{\textquoteright}s equation of state and primordial black hole production with PTA, BBN, and CMB observations*}",
    eprint = "2307.13574",
    archivePrefix = "arXiv",
    primaryClass = "astro-ph.CO",
    doi = "10.1088/1674-1137/ad79d5",
    journal = "Chin. Phys. C",
    volume = "48",
    number = "12",
    pages = "125105",
    year = "2024"
}

@article{Ghoshal:2026ros,
    author = "Ghoshal, Anish and Spalding, Angus and White, Graham",
    title = "{Irreducible Gravitational Wave Background as a Particle Detector}",
    eprint = "2604.20792",
    archivePrefix = "arXiv",
    primaryClass = "hep-ph",
    month = "4",
    year = "2026"
}

@article{Fernandez:2023ddy,
    author = "Fernandez, Nicolas and Foster, Joshua W. and Lillard, Benjamin and Shelton, Jessie",
    title = "{Stochastic Gravitational Waves from Early Structure Formation}",
    eprint = "2312.12499",
    archivePrefix = "arXiv",
    primaryClass = "astro-ph.CO",
    reportNumber = "MIT-CTP/5660",
    doi = "10.1103/PhysRevLett.133.111002",
    journal = "Phys. Rev. Lett.",
    volume = "133",
    number = "11",
    pages = "111002",
    year = "2024"
}

@article{Escriva:2026blk,
    author = "Escriv{\`a}, Albert and Harada, Tomohiro and Kohri, Kazunori and Terada, Takahiro and Yoo, Chul-Moon",
    title = "{Gravitational wave emission from nonspherical collapse in an early matter-dominated era using N-body simulations}",
    eprint = "2605.04487",
    archivePrefix = "arXiv",
    primaryClass = "astro-ph.CO",
    reportNumber = "NU-QG-17, RUP-26-8, KEK-TH-2833, KEK-Cosmo-0420",
    month = "5",
    year = "2026"
}

@article{Ferrante:2022mui,
    author = "Ferrante, Giacomo and Franciolini, Gabriele and Iovino, Junior., Antonio and Urbano, Alfredo",
    title = "{Primordial non-Gaussianity up to all orders: Theoretical aspects and implications for primordial black hole models}",
    eprint = "2211.01728",
    archivePrefix = "arXiv",
    primaryClass = "astro-ph.CO",
    doi = "10.1103/PhysRevD.107.043520",
    journal = "Phys. Rev. D",
    volume = "107",
    number = "4",
    pages = "043520",
    year = "2023"
}

@article{Choptuik:1992jv,
    author = "Choptuik, Matthew W.",
    title = "{Universality and scaling in gravitational collapse of a massless scalar field}",
    reportNumber = "FPRINT-92-33",
    doi = "10.1103/PhysRevLett.70.9",
    journal = "Phys. Rev. Lett.",
    volume = "70",
    pages = "9--12",
    year = "1993"
}

@article{Musco:2012au,
    author = "Musco, Ilia and Miller, John C.",
    title = "{Primordial black hole formation in the early universe: critical behaviour and self-similarity}",
    eprint = "1201.2379",
    archivePrefix = "arXiv",
    primaryClass = "gr-qc",
    doi = "10.1088/0264-9381/30/14/145009",
    journal = "Class. Quant. Grav.",
    volume = "30",
    pages = "145009",
    year = "2013"
}

@article{Zeldovich:1969sb,
    author = "Zeldovich, Ya. B.",
    title = "{Gravitational instability: An Approximate theory for large density perturbations}",
    journal = "Astron. Astrophys.",
    volume = "5",
    pages = "84--89",
    year = "1970"
}

@article{Harada:2016mhb,
    author = "Harada, Tomohiro and Yoo, Chul-Moon and Kohri, Kazunori and Nakao, Ken-ichi and Jhingan, Sanjay",
    title = "{Primordial black hole formation in the matter-dominated phase of the Universe}",
    eprint = "1609.01588",
    archivePrefix = "arXiv",
    primaryClass = "astro-ph.CO",
    reportNumber = "RUP-16-25, OCU-PHYS-454, AP-GR-134, KEK-TH-1929, KEK-COSMO-197",
    doi = "10.3847/1538-4357/833/1/61",
    journal = "Astrophys. J.",
    volume = "833",
    number = "1",
    pages = "61",
    year = "2016"
}

@article{Harada:2017fjm,
    author = "Harada, Tomohiro and Yoo, Chul-Moon and Kohri, Kazunori and Nakao, Ken-Ichi",
    title = "{Spins of primordial black holes formed in the matter-dominated phase of the Universe}",
    eprint = "1707.03595",
    archivePrefix = "arXiv",
    primaryClass = "gr-qc",
    reportNumber = "KEK-COSMO-206, RUP-17-12, KEK-Cosmo-206, KEK-TH-1987, OCU-PHYS-466, AP-GR-139",
    doi = "10.1103/PhysRevD.96.083517",
    journal = "Phys. Rev. D",
    volume = "96",
    number = "8",
    pages = "083517",
    year = "2017",
    note = "[Erratum: Phys.Rev.D 99, 069904 (2019)]"
}

@article{doroshkevich1970spatial,
  title={Spatial structure of perturbations and origin of galactic rotation in fluctuation theory},
  author={Doroshkevich, AG302134},
  journal={Astrophysics},
  volume={6},
  number={4},
  pages={320--330},
  year={1970},
  publisher={Springer}
}

@book{Mukhanov:2005sc,
    author = "Mukhanov, V.",
    title = "{Physical Foundations of Cosmology}",
    doi = "10.1017/CBO9780511790553",
    isbn = "978-0-521-56398-7",
    publisher = "Cambridge University Press",
    address = "Oxford",
    year = "2005"
}

\end{document}